\documentclass{aa}  

\usepackage{graphicx}
\usepackage{txfonts}
\begin{document}

   \title{ \textit{\textbf{Gaia}} Data Release 1}

\subtitle{The Cepheid \& RR Lyrae star pipeline  and its application to the south ecliptic
pole region}
\titlerunning{{\it Gaia} Data Release 1, Cepheids and RR Lyrae stars}
\author{
G. Clementini\inst{\ref{inst1}},
V. Ripepi\inst{\ref{inst2}}, 
S. Leccia\inst{\ref{inst2}}, 
N. Mowlavi\inst{\ref{inst3}}, 
I. Lecoeur-Taibi\inst{\ref{inst4}},
M. Marconi\inst{\ref{inst2}}, 
L. Szabados\inst{\ref{inst5}},
L. Eyer\inst{\ref{inst3}},  
L.~P. Guy\inst{\ref{inst4}}, 
L. Rimoldini\inst{\ref{inst4}}, 
G. Jevardat de Fombelle\inst{\ref{inst6}},
B. Holl\inst{\ref{inst4}}, 
G. Busso\inst{\ref{inst7}},
J. Charnas\inst{\ref{inst4}}, 
J. Cuypers\inst{\ref{inst8}},
F. De Angeli\inst{\ref{inst7}}, 
J. De Ridder\inst{\ref{inst9}}, 
J. Debosscher\inst{\ref{inst9}}, 
D.~W. Evans\inst{\ref{inst7}},  
P. Klagyivik\inst{\ref{inst5}},
I. Musella\inst{\ref{inst2}}, 
K. Nienartowicz\inst{\ref{inst6}},
D. Ord\'o\~nez\inst{\ref{inst4}}, 
S. Regibo\inst{\ref{inst9}}, 
M. Riello\inst{\ref{inst7}},
L.~M. Sarro\inst{\ref{inst10}}, 
M. S\"{u}veges\inst{\ref{inst4}}
}
\institute{INAF-Osservatorio Astronomico di Bologna, Via Ranzani 1, I - 40127 Bologna, Italy\label{inst1}\\
\email{gisella.clementini@oabo.inaf.it}
\and
INAF-Osservatorio Astronomico di Capodimonte, Salita Moiariello 16, I - 80131 Napoli, Italy\label{inst2}
\and
Department of Astronomy, University of Geneva, Ch. des Maillettes 51, CH-1290 Versoix, Switzerland\label{inst3}
\and
Department of Astronomy, University of Geneva, Ch. d'Ecogia 16, CH-1290 Versoix, Switzerland\label{inst4}
\and 
Konkoly Observatory, Research Centre for Astronomy and Earth Sciences, Hungarian Academy of Sciences, H-1121 Budapest, Konkoly Thege M. ut 15-17, Hungary\label{inst5}
\and
SixSq, Rue du Bois-du-Lan 8, CH-1217 Geneva, Switzerland\label{inst6}
\and
Institute of Astronomy, University of Cambridge, Madingley Road, Cambridge CB3 0HA, UK\label{inst7}
\and
Royal Observatory of Belgium, Ringlaan 3, B-1180 Brussels, Belgium\label{inst8}
\and
Institute of Astronomy, KU Leuven, Celestijnenlaan 200D, 3001 Leuven, Belgium\label{inst9}
\and
Dpto. Inteligencia Artificial, UNED, c/ Juan del Rosal 16, 28040 Madrid, Spain\label{inst10} 
}
\authorrunning{Clementini et al.}
   \date{Received ... ; accepted .....}

 
  \abstract
 {The European Space Agency spacecraft {\it Gaia} is expected to observe about 10,000 Galactic Cepheids and over 100,000 Milky Way RR Lyrae stars (a large fraction of which will be new discoveries), during the five-year nominal lifetime spent scanning the whole sky to a faint limit of $G$ = 20.7 mag, sampling their light variation on average about 70 times.}
{We present an overview of the Specific Objects Study (SOS) pipeline developed within the Coordination Unit 7 (CU7) of the Data Processing and Analysis Consortium (DPAC), the  coordination unit charged with the  processing and analysis of variable sources observed by {\it Gaia}, to validate and fully characterise Cepheids and RR Lyrae stars observed by the spacecraft. 
The algorithms developed to classify and extract information such as the pulsation period, mode of pulsation, mean magnitude, peak-to-peak amplitude of the light variation, 
sub-classification in type, multiplicity, secondary periodicities,  light curve Fourier decomposition parameters,  as well as physical parameters such as mass, metallicity, 
reddening and, for classical Cepheids, age, are briefly described.}
{The full chain of the CU7 pipeline was run on the time-series photometry collected by {\it Gaia} during 28 days of Ecliptic Pole Scanning Law (EPSL) and over a year of
 Nominal Scanning Law (NSL), starting from the general Variability Detection, general Characterisation, proceeding through the global Classification and ending with the 
 detailed checks and typecasting of the SOS for Cepheids and RR Lyrae stars (SOS Cep\&RRL). 
We describe in more detail how the SOS Cep\&RRL pipeline was specifically tailored to analyse {\it Gaia}'s $G$-band  photometric time-series with a South Ecliptic Pole (SEP) 
footprint, which covers an external region of the Large Magellanic Cloud (LMC), and  to produce results for confirmed RR Lyrae stars and Cepheids  to be published in  {\it Gaia} 
Data Release 1 ({\it Gaia} DR1).} 
{$G$-band time-series photometry and characterization by the 
SOS Cep\&RRL pipeline (mean magnitude and pulsation characteristics)  are published in {\it Gaia} DR1   for a total 
sample 
of 3,194 variable stars, 599 Cepheids and  2,595 RR Lyrae stars, of which 386 (43 Cepheids and 343 RR Lyrae stars) are new discoveries by {\it Gaia}. 
All 3,194 stars are distributed over an area  extending 38 degrees on either side from a point offset from the centre of the LMC by about 
3 degrees to the north and 4 degrees to the east. 
The vast majority, but not all, are located within the LMC. 
The published sample also  includes a few bright RR Lyrae stars that  trace  the  outer halo of the Milky Way  in front of the LMC.}
{}


   \maketitle
%

\section{Introduction}
Easy to recognise thanks to their characteristic  light variation, Cepheids and RR Lyrae stars are radial pulsating variables that trace stellar populations with different 
age and chemical 
    composition: classical Cepheids (hereinafter, DCEPs) trace a young ($t \lesssim <$300 Myr) stellar component;  anomalous 
    Cepheids (ACEPs) can trace stars of intermediate age ($t \sim$1-5 Gyr)  and metal poor content ([Fe/H]$< -1.5$ dex), although it is still matter of debate whether they might 
also arise from coalescence of binary stars  as old as  about 10 Gyr;  finally, the RR Lyrae stars and the Type II Cepheids (T2CEPs) trace an  old    
     ($t >$10 Gyr) stellar population. 

They are 
primary standard candles 
in establishing the cosmic distance ladder, because Cepheids conform to period--luminosity ($PL$),  period--luminosity--colour ($PLC$)
and
period -- Wesenheit ($PW$)
relations, whereas the RR Lyrae stars follow a luminosity--metallicity
relation in the visual band ($M_V$--[Fe/H]) and a period--luminosity--metallicity ($PLZ$) relation in the infrared.

With its multi-epoch 
 monitoring of the full sky, {\it Gaia} will 
discover and measure position, parallax, proper motion and time-series photometry of thousands of Cepheids and RR~Lyrae stars 
 in the Milky Way (MW)  and its surroundings, down to a faint magnitude limit of 
$G\sim$ 20.7 mag.
The spacecraft  is expected to observe from 2,000 to 9,000 MW Cepheids,  
about 70,000 RR~Lyrae stars in the Galactic halo, 
from 15,000 to 40,000 RR~Lyrae in the MW bulge, (see table~3 in  \citealt{eyer12},  and references therein) and  according to the most recent estimates (\citealt{soszynski15b, soszynski16}) over 45,500 RR Lyrae stars and 9,500 Cepheids in the Magellanic Clouds. 
{\it Gaia} will revise upwards these statistics,  
as ongoing surveys 
such as OGLE-IV (\citealt{soszynski15a,soszynski15b,soszynski15c,soszynski16}), Catalina Sky Survey (CSS, \citealt{drake13,torrealba15}), Pan-STARRS (\citealt{hernitschek16}), LINEAR (\citealt{sesar13}) and PTF (\citealt{cohen15}), are constantly reporting 
new discoveries and increased numbers of RR Lyrae stars and Cepheids both in the MW 
and in its neighbour companions.

{\it Gaia}'s complete census of the Galactic Cepheids and RR~Lyrae will allow a breakthrough 
in our understanding of the MW structure by tracing  young and old variable stars  
all the way through from the Galactic bulge, to the disk, to the halo, and likely  revealing new streams and faint satellites that bear witness of the MW hierarchical 
build-up (see e.g. \citealt{clementini16}).
But most importantly, {\it Gaia} will measure the parallax of tens of thousands of Galactic Cepheids and RR Lyrae stars,  
 along with milli-mag optical 
spectrophotometry ($G$ broad-band white-light magnitude, blue and red spectro-photometry) and  radial velocities and chemistry for those within reach of the Radial Velocity 
Spectrometer  (RVS; $G\lesssim$17 mag).
The unprecedented accuracy of {\it Gaia} measurements  
for local Cepheids and RR~Lyrae stars 
will  allow the absolute 
calibration via parallax of the Cepheid $PL$, $PLC$, $PW$ and of the
$M_{V} -$[Fe/H]  and infrared $PLZ$ relations for RR Lyrae stars, along  
with a test of the metallicity effects through simultaneous abundance measurements. This will enable    
 re-calibration of  ``secondary'' distance indicators probing distances far into the unperturbed Hubble flow and  
 a total re-assessment of the whole cosmic distance ladder, from local to cosmological distances, in turn significantly improving our knowledge of the Hubble constant. 
The physical parameters of Cepheids and RR Lyrae stars will be constrained by {\it Gaia} photometry, parallax, metallicity and radial velocity (RV) 
measurements, that will also constrain  the input physics of theoretical pulsation models. This will further improve the use of Cepheids and RR Lyrae stars as standard 
candles and stellar population tracers.
  
In this paper we describe the Specific Objects Study (SOS)  pipeline developed within the  Coordination Unit 7 (CU7)
 of the Data Processing and Analysis Consortium (DPAC), the  coordination unit in charge of the  processing and analysis of variable sources 
 observed by {\it Gaia}, to validate and fully characterise Cepheids and RR Lyrae stars observed by the spacecraft.
 A detailed description of the {\it Gaia} mission, its scientific goals and performance,  as well as a comprehensive  illustration of the {\it Gaia} DPAC structure and activities
 can be found in \citet{gaiacol-prusti}. A summary of the astrometric, photometric and survey properties of {\it Gaia} Data Release 1 ({\it Gaia} DR1) and a description of scientific quality and 
 limitations of this first data release are provided in  \citet{gaiacol-brown}. 
The photometric data set and the processing of the  $G$-band photometry released in {\it Gaia} DR1 
are thoroughly discussed in \citet{vanleeuwen16}, \citet{carrasco16}, \citet{riello16}, and \citet{evans16}.
 
We note that a rather strict policy is adopted within DPAC for what concerns the processing and dissemination of {\it Gaia} data.
Specifically,  it was decided to be consistent in how DPAC does the processing and what it is published in {\it Gaia} 
releases. That is, we do not release results which are based on {\it Gaia} data that is
not published.  For instance, since no $G_\mathrm{BP}$, $G_\mathrm{RP}$ photometry, 
is released in {\it Gaia} DR1, only the $G$-band time series photometry was used for processing and 
classification of the variable sources released in {\it Gaia} DR1. 
Furthermore,  characterisation and classification of {\it Gaia} variable sources rely only on {\it Gaia} data. That is,  
we do not complement {\it Gaia}'s time series with external non-{\it Gaia} data to increase the number 
of data points  or the time-span of the {\it Gaia} observations.
Literature published data are used once the processing is  completed, only to validate
results (i.e. characterisation and classification of the variable sources) which are, 
however, purely and exclusively based on {\it Gaia} data. 
If and how this may have limited the efficiency of the {\it Gaia}  pipeline for
Cepheids and RR Lyrae stars is extensively discussed in the paper
and most specifically in Section~\ref{tailoring}. On the other hand,
we are also sure that  the next  {\it Gaia}  data
releases will significantly improve both census and results for variable
stars,
and will also emend misclassifications if/where they have occurred. 
 
The 
 SOS pipeline for Cepheids and RR Lyrae stars, hereinafter referred to as 
SOS Cep\&RRL pipeline,  is one of the latest stages of the general 
 variable star analysis pipeline. 
Steps of the processing prior the SOS Cep\&RRL pipeline are fully described in \citet{eyer16}, to which the 
reader is referred to for details.

Validation of the classification provided by the 
 general variable star analysis pipeline is  necessary, since Cepheids and RR Lyrae
 stars overlap in period with other types of variables (e.g. binary
systems, long period variables, etc.). SOS Cep\&RRL uses specific
 features  such as the parameters of the light curve Fourier decomposition 
 and diagnostic tools like the {\it Gaia} colour-magnitude diagram (CMD), the amplitude ratios,
the period-amplitude ($PA$), $PL$, $PLC$ and $PW$ relations,  and the Petersen diagram \citep{petersen73}, to check the
 classification, to  derive periods and pulsation modes, and to
 identify multimode pulsators.  RV measurements obtained by the RVS are also planned for use as soon as they become available,  
 to identify binary/multiple systems.
 
The main tasks of the SOS Cep\&RRL pipeline 
are: i) to validate and refine the detection and classification of Cepheids and RR Lyrae stars in the {\it Gaia} data base,
  provided by the 
  general variable star analysis pipeline, by cleaning  the sample from contaminating objects, i.e. other types
of variables falling into the same period domain; ii) to check and improve the period determination and  the light curve modelling;
 iii) to identify the pulsation modes and the objects with secondary and multiple periodicities; iv) to classify RR Lyrae stars and Cepheids  into sub-types 
  (fundamental mode -- RRab, first overtone -- RRc, and double mode -- RRd) according to the pulsation mode for the RR Lyrae stars,   and  DCEPs,
ACEPs and T2CEPs for the Cepheids, along with identification of pulsation modes for the former two,  and sub-classification into W Virginis (WVIR), BL Herculis (BLHER) and 
RV Tauri (RVTAU) types 
for  the latter; 
v) to identify and flag variables showing modulations of the light curve, due to a
binary companion or to the Blazhko effect \citep{blazhko}, that may falsify both the
star brightness and the derived trigonometric parallax; vi) to use the pulsation properties and derive physical parameters 
(luminosity, mass, radius, effective temperature, 
metallicity, reddening, etc.) of confirmed {\it bona fide} RR Lyrae stars and Cepheids to be
ingested into the {\it Gaia} main data base, by means of a variety of methods specifically tailored 
 to these types of variables. 
 
 The paper is organised as follows: Section~\ref{s2} provides a description of the whole architecture of the SOS Cep\&RRL pipeline,  its diagnostic tools and their definition 
 in the {\it Gaia}  pass-bands.  Section~\ref{s3new}  presents the dataset and source selection on which the SOS Cep\&RRL  pipeline was run and 
 describes how the pipeline was specifically  tailored and simplified to analyse the {\it Gaia} SEP $G$-band time series data of candidate Cepheids and RR Lyrae stars. Section~\ref{results2} presents results of the SOS Cep\&RRL analysis that are published in {\it Gaia} DR1. Finally, the main results  and future developments of the pipeline are summarised in Section~\ref{conclusions}. 

\section{The SOS Cep\&RRL pipeline
}\label{s2}
In this Section we present the SOS Cep\&RRL pipeline in its complete form, briefly describing all its algorithms and tools, only part of which could actually be applied due to 
the specific characteristics  
 of the dataset published in {\it Gaia} DR1 (see Section~\ref{tailoring} for details). 

In order to validate and refine the classification of Cepheids and RR Lyrae stars
provided by the 
general variable star analysis pipeline,  and be able 
 in the future to make comparisons with the parameters measured for these
stars by other 
DPAC processing tasks,  different procedures are implemented in the 
SOS Cep\&RRL  pipeline as described in the following sections. 
 The whole processing of the SOS Cep\&RRL  pipeline is presented schematically 
in Figs.~\ref{decomposition1}, \ref{decomposition2}, and  \ref{decomposition3} and described in detail in Sections~\ref{initialproc}, ~\ref{RRtrunk} and ~\ref{CepTrunk}. 
For each candidate confirmed as a bona fide RR Lyrae star or Cepheid, SOS Cep\&RRL  returns the 
  most probable period, pulsation mode, multiple periodicities, if
  any, associated amplitudes, 
  intensity-averaged magnitudes and mean radial velocities (if RVS data
  are available) as well as flagging of  binary or multiple systems, to be published in the {\it Gaia} main data base.

The main input of SOS Cep\&RRL  is the calibrated {\it Gaia} time-series photometry  
($G$-magnitudes and integrated $G_\mathrm{BP}$, $G_\mathrm{RP}$ photometry) 
processed  by CU5, the DPAC coordination  
unit charged with the photometric processing of {\it Gaia} data,
 for all sources  pre-classified as candidate 
Cepheids and RR Lyrae stars by the 
 general variability
{\it Characterization} and {\it Classification} processing, along with general information on the data, such as number of transits per source (after outlier removal), 
typical errors, time span of the observations, mean and median magnitude values, etc.,  computed by the   {\it Statistical Parameters} module in the general pipeline (\citealt{eyer16}) prior 
SOS Cep\&RRL.

Additional information will be progressively added in future releases  
 such as: time-series RVs for sources within the RVS 
magnitude limit, 
{\it Gaia}'s parallaxes (distances), 
  astrophysical parameters such as: effective temperature
($\log T_{\rm {eff}}$), gravity  ($\log g$) and absorption  ($A_{\lambda}$) inferred by combining {\it Gaia}'s astrometry, photometry and spectroscopy. They can be used to optimise the 
SOS Cep\&RRL  processing. 

\subsection{Initial processing: characterisation of the light curves  (Period search \& Fourier fitting)}\label{initialproc}  
\begin{figure*}
   \centering
   \includegraphics[trim= 0 40 10 80,width=12 cm,clip]{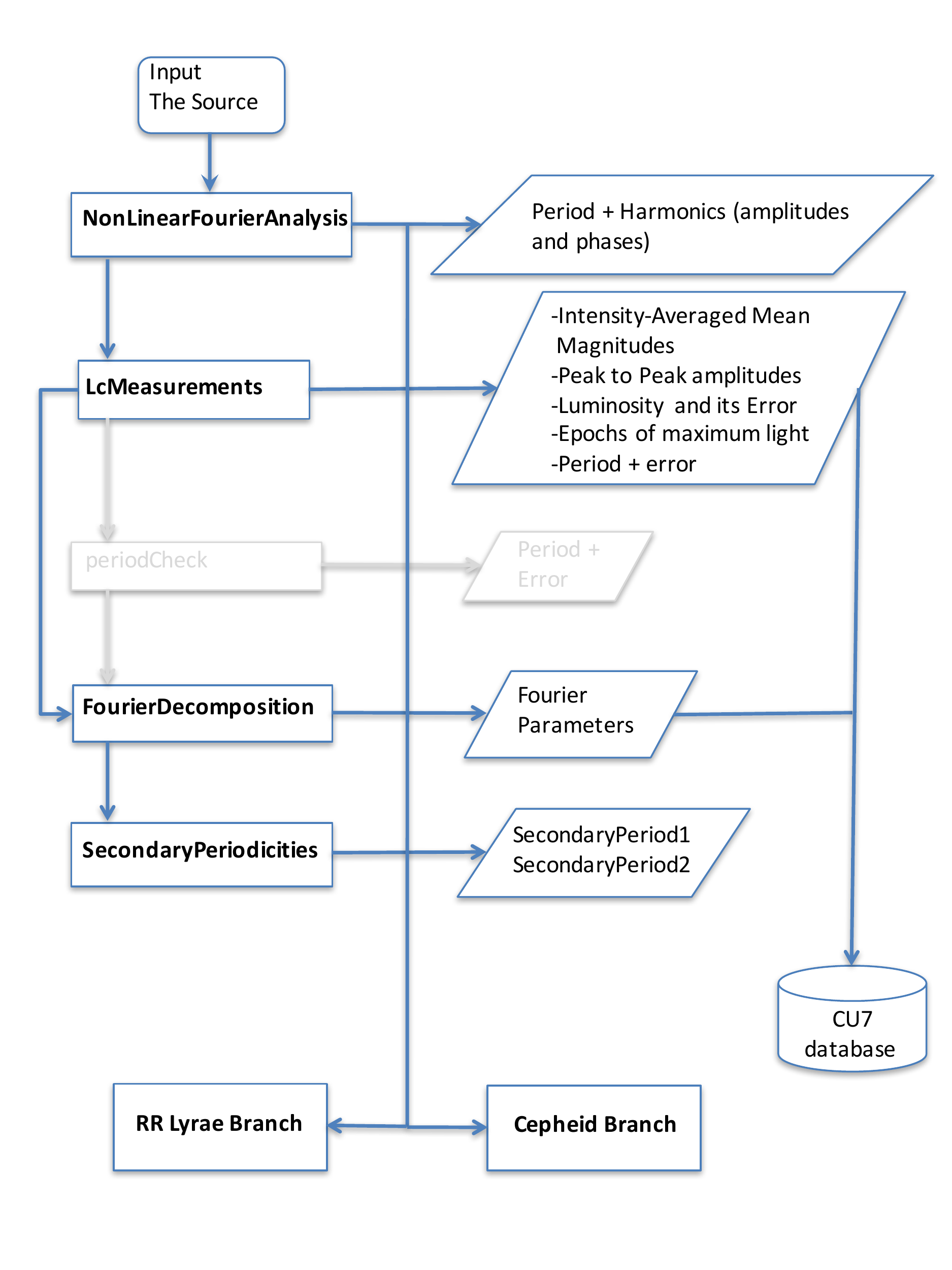}
   \caption{Flow-chart of the SOS Cep\&RRL pipeline that is common to both Cepheid  and RR Lyrae star processing. Boxes on the left-hand side show 
   different modules of this general trunk, their outputs  are indicated within rhombs on the right-hand side. Blue arrows connect modules activated 
   for the processing of {\it Gaia} DR1 data, their names are highlighted in bold-face. We have marked in light grey modules not operational for the {\it Gaia} DR1 processing and their connections to the rest of the pipeline.}
              \label{decomposition1}%
    \end{figure*}

The SOS Cep\&RRL processing starts by characterising the $G$-band light curve (as well as the $G_\mathrm{BP}$ and $G_\mathrm{RP}$  light curves and the RVS RV curve 
if/when available)  of  sources classified as candidate 
Cepheids and RR Lyrae stars by the {\it Supervised Classification} module of the general variable star analysis pipeline. 
This part of the  SOS Cep\&RRL  processing is common to both Cepheids and RR Lyrae stars and its main steps are  sketched in Fig.~\ref{decomposition1}.
The first step is the derivation of the source periodicity independently and with a different method to that used in the general pipeline {\it Characterisation} module. We used the Lomb-Scarge algorithm (\citealt{lomb}, \citealt{scargle}) 
as opposed to the least squares method used in  {\it Characterisation} (see \citealt{eyer16}). 
Tests performed on Cepheids and RR Lyrae stars  
in the Hipparcos catalogue showed that the Lomb-Scarge method reduces the number of large deviating period values 
for these specific types of variables.
  The period derived by SOS Cep\&RRL  is used to fold the light curve which is then 
  modelled with a truncated Fourier series in the form:
 \begin{equation}
  mag(t_j)=zp+\sum[A_i
  sin(i \times 2 \pi \nu_{max} t_j +\phi_i)]
 \end{equation} 
Zero-point ({\it zp}), period
  (1/$\nu_{max}$), harmonic number ($i$), amplitudes ($A_i$), and
  phases ($\phi_i$) of the harmonics, for the $G$-band light curve are first computed with a linear fitting procedure. They provide initial trial values for the non-linear 
  modelling of the light curve which is performed using the Levenberg-Marquardt (\citealt{Levenberg1944}, \citealt{Marquardt1963}) non-linear fitting
  algorithm (module {\it NonLinearFourierAnalysis} of Fig.~\ref{decomposition1}).  The non-linear fitting 
  refines both the period and model of the light curve. 
  In {\it Gaia} DR1 only the $G$-band time-series photometry is available for the sources, hence, the period resulting from the non-linear fitting of the $G$-band light curve is  
 the final period adopted for the source and  module {\it LcMeasurements} directly measures the source intensity-averaged $G$ 
 mean magnitude,  peak-to-peak  $G$-amplitude and
  epoch of maximum light from the $G$-band light curve  modelled with the non-linear fitting algorithm\footnote{An independent search for the best period and best-fit model 
  calculations will be performed on the $G_\mathrm{BP}$,  $G_\mathrm{RP}$ and RV data, when they become available. Peak-to-peak amplitudes and amplitude ratios for all photometric bands will be 
  computed and an internal consistency check among periods from different pass-bands will be performed by the  {\it periodCheck} module (see Fig.~\ref{decomposition1}). 
  This module was not activated for DR1.}.
The peak-to-peak $G$-amplitude [Amp($G$)] is calculated as:  (maximum) $-$ (minimum) values of the model best-fitting the folded light curve. An estimate of the error on this
quantity is provided by  $\sqrt{2}\times \sigma(zp)$.  
The epoch of maximum light is computed as the Baricentric Julian day (BJD) of the maximum value of the $G$-band light curve model which
is closer to the BJD of the first observations minus 3 times the period of
the source. This procedure ensures  that the time of maximum light precedes the time of the first observation.  The mentioned BJD is offset by JD 2455197.5 d (= J2010.0).
Adopting the results of the non-linear modelling  
 amplitude ratios ($R_{ij} =A_i/A_j$), phase differences ($\phi_{ij}=j \times\phi_i-i \times \phi_j$) and related errors,  of the Fourier decomposition of the $G$-band light curve,  
are computed by the module {\it FourierDecomposition}\footnote{Fourier parameters will be computed in the future also for the $G_\mathrm{BP}$, $G_\mathrm{RP}$ light curves and for the RV curve 
to perform
consistency checks.}. 
Errors for all the parameters (period, amplitude, {\it zp}, etc.) derived with the non-linear Fourier modelling are estimated via Monte Carlo simulations\footnote{Only exceptions are the  errors of the Fourier parameters $\phi_{ij}$ and  $R_{ij}$ which are currently computed by propagation of the errors in  $\phi_i$, $\phi_j$, $R_i$ and  $R_j$, the latter being computed via Monte Carlo simulations.}
 according to the 
following procedure: 
\begin{itemize}

\item a new light curve is generated where: (value of each phase point) =
  (value of the original phase  point) + (random number) $\times$ ($\pm$error of the original phase point value). Random numbers range between 0 and 1;
\item a new model is computed with the non-linear modelling procedure, using
  as trial values the model parameters of the original light curve;
\item these two first steps are iterated 100 times;
\item for each parameter of the model, average
  and standard deviation are computed over the 100 simulations; 
\item the standard deviations derived in the previous step are assigned as
  uncertainties of the parameters resulting from the non-linear modelling.
\end{itemize}   
 \noindent
 The final step in  the general part of the SOS Cep\&RRL pipeline 
 is the detection of possible secondary periodicities in the source periodogram. This is the task of module {\it SecondaryPeriodicities}. 
 For RR Lyrae stars we look for one
and  for Cepheids for two additional frequencies beyond the first
periodicity.  Following the procedure adopted in the {\it NonLinearFourierAnalysis} module, the residuals: observed $-$ model phase points,  are searched for
secondary periodicities. If a significant secondary frequency is found
the procedure is iterated twice.
The steps of the {\it SecondaryPeriodicities} algorithm are as follows:
\begin{itemize}
\item[1.] residuals are computed as:  (observed) $-$ (corresponding model) phase points;
\item[2.] the Lomb-Scargle period search method is run on the residual time-series;   
\end{itemize}

As it will be discussed in Section~\ref{conclusions}, this module will be revised in preparation for the {\it Gaia} second data release ({\it Gaia} DR2) in order to properly take into account the actual 
significance of the detected secondary periodicities.

\noindent After this common  part the analysis proceeds in two separate branches each specifically tailored to the 
processing  of RR Lyrae stars and Cepheids, respectively.  
This is summarised in the flow-charts in Figs.~\ref{decomposition2} and ~\ref{decomposition3}. 
Sources classified as candidate Cepheids by the general 
 {\it Classification} pipeline will be sent first to the Cepheid branch. 
Conversely, sources classified as candidate RR Lyrae stars will first be processed through the RR Lyrae branch. 
However, as it will be described in  Section~\ref{tailoring} for the processing of the {\it Gaia} DR1 data we followed a different approach to decide which branch a source should be sent to.

 \subsection{Conversions to {\it Gaia}'s passbands}
\begin{figure*}[h!]
   \centering
   \includegraphics[width=12 cm,clip]{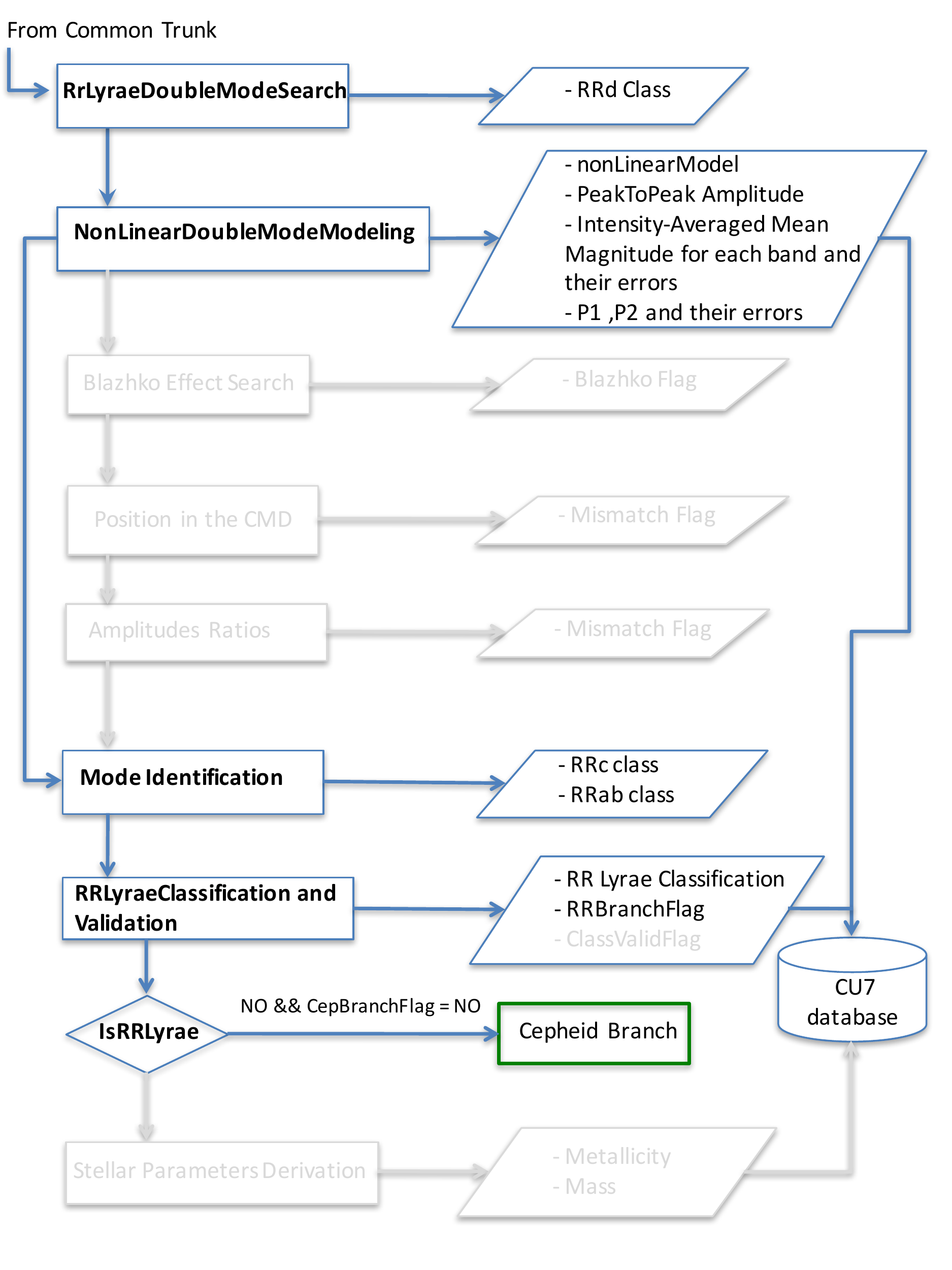}
   \caption{Flow-chart of the RR Lyrae branch in  the SOS Cep\&RRL pipeline. Boxes on the left-hand side show 
   the different modules, their outputs  are indicated within  rhombs on the right-hand side. Blue arrows connect modules activated 
   for the processing of the {\it Gaia} DR1 data, their names are highlighted in bold-face. 
 We have marked in light grey modules not operational for the {\it Gaia} DR1 processing, their outputs and their connections to the rest of the pipeline.  
  }
              \label{decomposition2}%
    \end{figure*}

\begin{figure*}[h!]
   \centering
   \includegraphics[width=12 cm,clip]{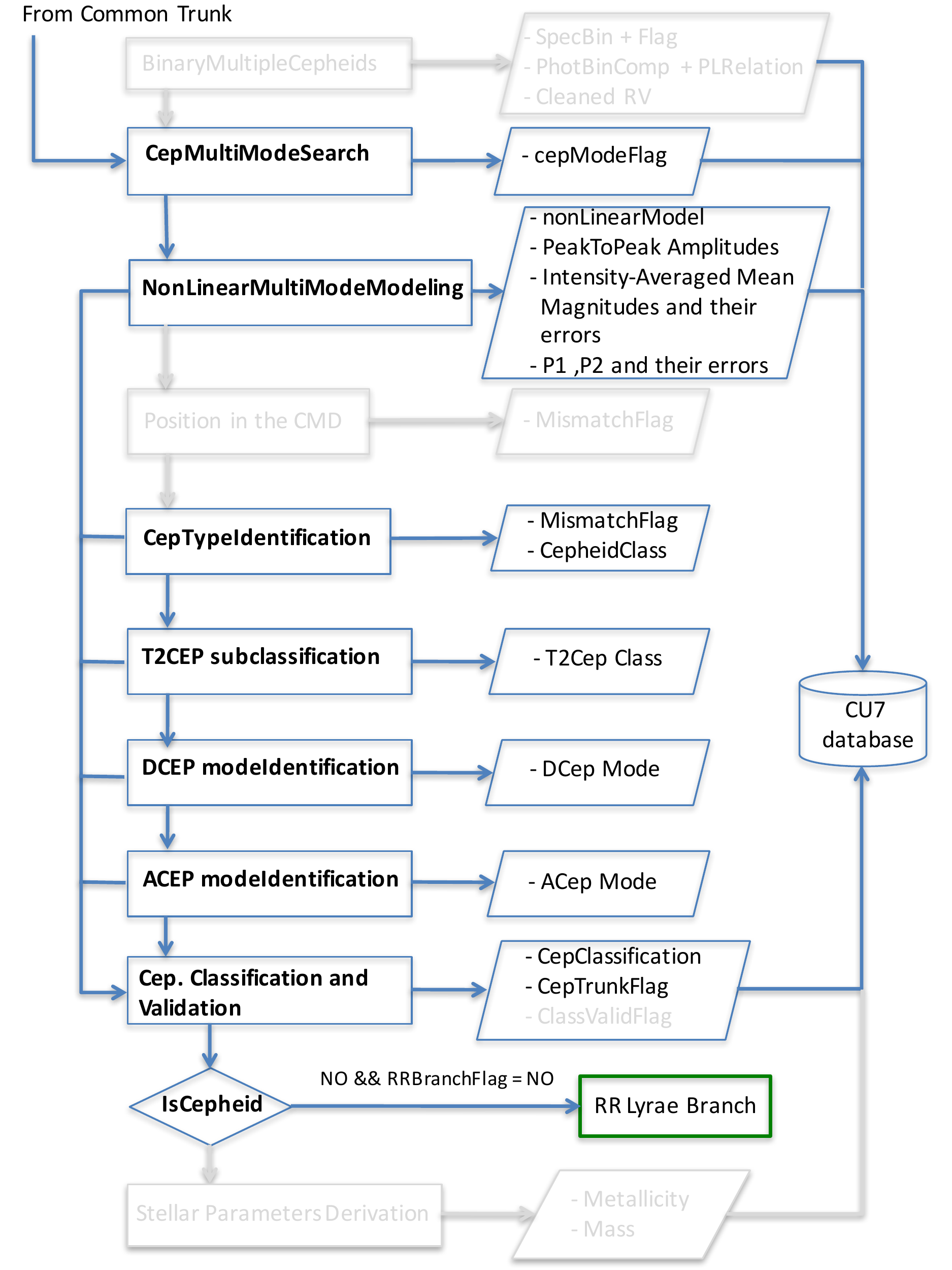}
   \caption{Flow-chart of the Cepheid branch in the SOS Cep\&RRL pipeline.  Layout and colour-coding are the same as in Fig. ~\ref{decomposition2}.
   }
              \label{decomposition3}%
    \end{figure*}

   A number of tools specifically applicable for the characterisation of RR Lyrae stars and Cepheids are adopted in the 
   different modules  of the Cepheid and RR Lyrae branches 
   (see Figs. ~\ref{decomposition2} and ~\ref{decomposition3}). 
They include: the CMD, the $G$-band 
 $PA$  diagram (also known, for RR Lyrae stars, as Bailey diagram, \citealt{bailey1902}), the $PL$, $PLC$ and $PW$ relations for Cepheids, the 
 different planes in the Fourier parameter space, and the Petersen diagram (\citealt{petersen73}) for double-mode Cepheids and RR Lyrae stars. All  limits and loci of these tools  are defined for the Johnson photometric system. We have converted them to the {\it Gaia}  photometric system in order to apply them directly to {\it Gaia} sources. We have used pass-band transformations provided in \citet{jordi10} and subsequent updates 
 (Jordi, personal communication) to compute conversion formulae appropriate for the colour and metallicity ranges of Cepheids and RR Lyrae stars   ($V-I <$ 2.5 mag and   $-2.5 <$ [Fe/H] $<$0.5 dex )   and transform the Johnson-Cousins $V$, $I$ to the {\it Gaia} $G$, $G_\mathrm{BP}$,  $G_\mathrm{RP}$ pass-bands. The conversion formula for the $G$ band is provided in Section~\ref{con-for} and shown in 
 Fig.~\ref{conversion}.
We then used a sample of 128 RR Lyrae stars 
belonging to the Galactic halo and bulge, to the Galactic globular clusters M3 and M68 and to the LMC and Small Magellanic Cloud (SMC) and 
77 classical Cepheids in the MW, LMC and SMC, for which  excellent light curves 
in the Johnson $V$ and $I$  passbands have been  
published by \citet{soszynski08a,soszynski09,soszynski10a,soszynski10c,soszynski11a}, \citet{cacciari05}, \citet{walker94}, 
  \citet{moffet84}, \citet{gieren81}, \citet{coulson85a,coulson85b}, and \citet{berdnikov95,berdnikov12,berdnikov14}. 
The two samples cover entirely the  parameter space of pulsation modes, 
periods and metallicities for RR Lyrae stars and Cepheids. 
We used Eq. (A.1) 
to transform the $V$, $I$ light curves of these selected samples  
to  {\it Gaia} $G$ 
passband and fitted both original $V$, $I$ and {\it Gaia}-transformed $G$ band  
light curves with truncated Fourier series obtaining $R_{21}$, $R_{31}$, $\phi_{21}$, $\phi_{31}$ parameters and peak-to-peak amplitude in each band. 
Finally, we used the quantities obtained with the above procedure to derive relationships and transform the Fourier parameters and peak-to-peak amplitudes from the Johnson-Cousins 
to the {\it Gaia} $G$-band.  The relations obtained with this procedure are described by Eqs. (2) to (19)\footnote{The $G$-$I$ conversions are particularly valuable since the OGLE team 
usually publishes results for RR Lyrae stars and Cepheids in the $I$ band.}.  

\begin{figure}
   \centering
   \includegraphics[trim=20 180 0 130, width=9.5 cm,clip]{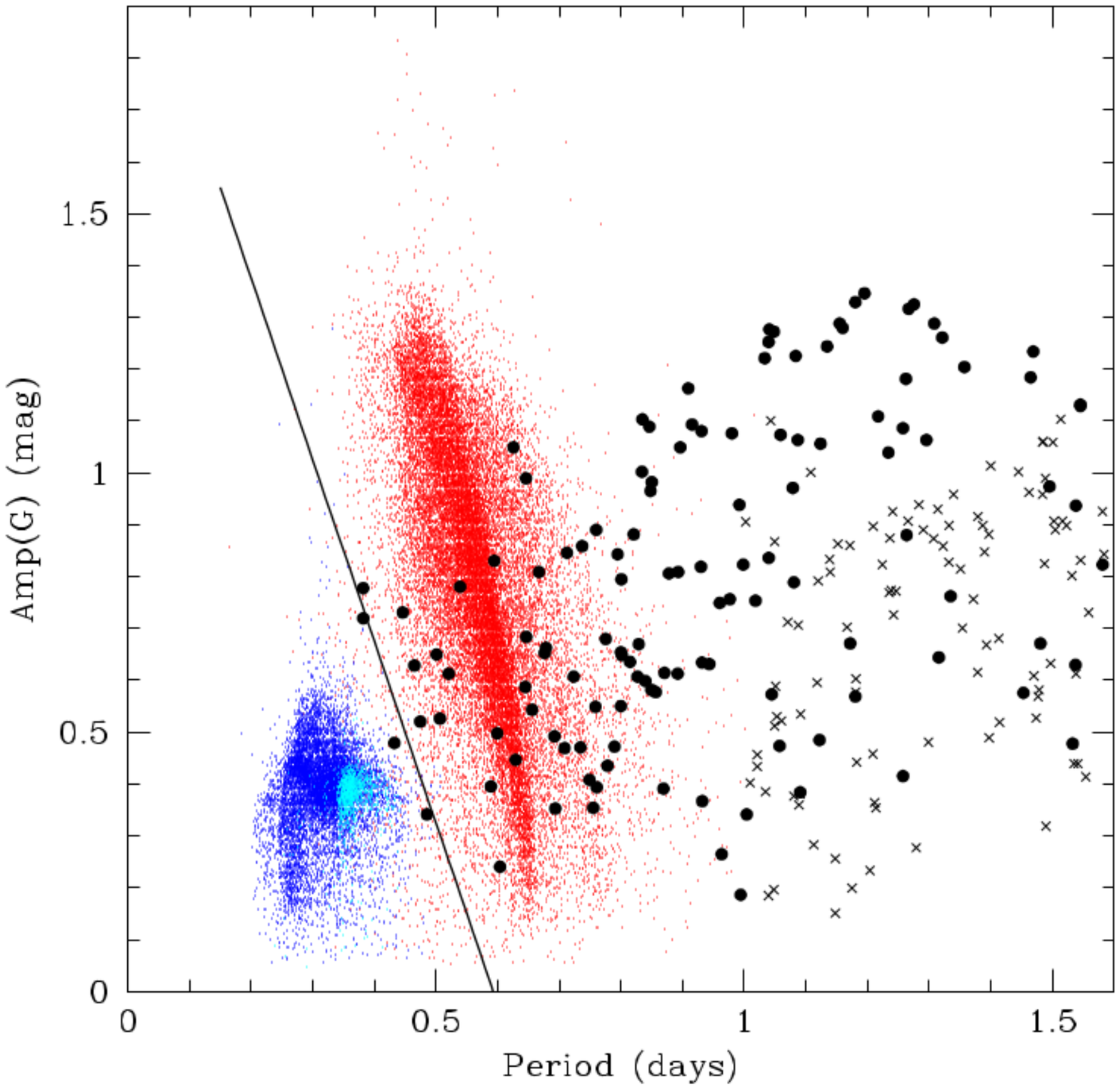}
   \caption{$G$-band $PA$ diagram for RR Lyrae stars (blue: RRc, cyan: RRd, red: RRab pulsators),  
ACEPs (black filled circles) and T2CEPs (crosses), obtained converting to $G$-band literature photometry from 
the catalogs of RR Lyrae stars, ACEPs and T2CEPs of the OGLE and ASAS surveys of  the LMC, SMC
 and Galactic bulge and halo, by \citet{soszynski8b,soszynski09,soszynski10b,soszynski10c,soszynski11a,soszynski11b}, \citet{poleski} and \citet{pojmanski97}. The black solid line separates 
RRc from RRab types. Contamination  between the two pulsation modes occurs close to the separation line, further contamination is also due to ACEPs.}
              \label{bailey}%
    \end{figure}

Formulae  to convert to the {\it Gaia} $G$-band the literature peak-to-peak amplitudes of RR Lyrae stars are provided by  Eqs. (2) and (3):  
\begin{equation}
\begin{split}
A(G)=(1.374 \pm 0.011)\times A(I)+(0.031\pm 0.006)\\
 rms=0.025~{\rm mag}\\
\end{split}
\end{equation}
\begin{equation}
\begin{split}
A(G)=(0.925 \pm 0.003)\times A(V)-(0.012\pm 0.003)\\
    rms=0.011~{\rm mag}\\
\end{split}
\end{equation}
We show in Fig.~\ref{bailey}  the  $PA$  diagram in the {\it Gaia} $G$ band of RR Lyrae stars in  the LMC, the SMC and the Galactic bulge and halo,  
 obtained converting the literature $V$ and $I$ amplitudes using eqs. (2) and (3).  
 We used literature amplitudes from \citet{soszynski09,soszynski10c,soszynski11a} for the RR Lyrae stars in the Magellanic Clouds, and from  \citet{pojmanski97} for the Galactic variables.  In the 
figure a black solid line is drawn to separate RRab from RRc types. Also shown in the figure are ACEPs (filled circles)
 and T2CEPs (crosses) in the LMC and SMC taken from   \citet{soszynski8b,soszynski15a}. Their amplitudes were transformed to the $G$-band also using eqs. (2) 
 and (3),  since colours and amplitudes of ACEPs and T2CEPs are similar to those of  RR Lyrae stars. The diagram in Fig.~\ref{bailey} is the main tool used  to separate RR Lyrae stars into 
RRab and RRc  
 types (see Section~\ref{s3new}). However,  since mixing of the pulsation modes  occurs close to the
 separation line and 
 further contamination is also caused by ACEPs,  we also used the Fourier parameters to separate  different pulsation modes and variable types.

Eqs. (4) to (11) provide formulae  to convert to the {\it Gaia} $G$-band the literature  values of the Fourier parameters $R_{21}$, $R_{31}$,
 $\phi_{21}$ and $\phi_{31}$ of  RR Lyrae stars:
\begin{equation}
\begin{split}
\phi_{21}(G)=(0.35 \pm 0.07)+(0.816\pm0.016)\times \phi_{21}(I)\\
-(0.64\pm 0.06)\times \log P{\rm (d)} ~~~~rms=0.006\\
\end{split}
\end{equation}
\begin{equation}
\begin{split}
\phi_{21}(G)=(0.13 \pm 0.03)+(0.996\pm0.007)\times \phi_{21}(V)\\
+(0.22\pm 0.03)\times \log P{\rm (d)} ~~~~rms=0.26\\
\end{split}
\end{equation}
\begin{equation}
\begin{split}
\phi_{31}(G)=(-0.50 \pm 0.05)+(0.875\pm0.018)\times \phi_{31}(I)\\
-(1.10\pm 0.11)\times \log P{\rm (d)} ~~~~rms=0.11\\
\end{split}
\end{equation}
\begin{equation}
\begin{split}
\phi_{31}(G)=(0.104 \pm 0.020)+(1.000\pm0.008)\times \phi_{31}(V)\\
~~~~rms=0.55\\
\end{split}
\end{equation}
\begin{equation}
\begin{split}
R_{21}(G)=(0.029 \pm 0.005)+(0.953\pm0.011)\times R_{21}(I)\\
-(0.05\pm 0.02)\times \log P{\rm (d)} ~~~~rms=0.017\\
\end{split}
\end{equation}
\begin{equation}
\begin{split}
R_{21}(G)=(0.000 \pm 0.002)+(1.015\pm0.004)\times R_{21}(V)\\
~~~~rms=0.006\\
\end{split}
\end{equation}
\begin{equation}
\begin{split}
R_{31}(G)=(0.034 \pm 0.004)+(0.935\pm0.011)\times R_{31}(I)\\
-(0.07\pm 0.02)\times \log P{\rm (d)} ~~~~rms=0.015\\
\end{split}
\end{equation}
\begin{equation}
\begin{split}
R_{31}(G)=(-0.001\pm 0.001)+(1.018\pm0.004)\times R_{31}(V)\\
~~~~rms=0.004\\
\end{split}
\end{equation}

\begin{figure}
   \centering
   \includegraphics[trim=10 180 -50 130, width=10.0 cm,clip]{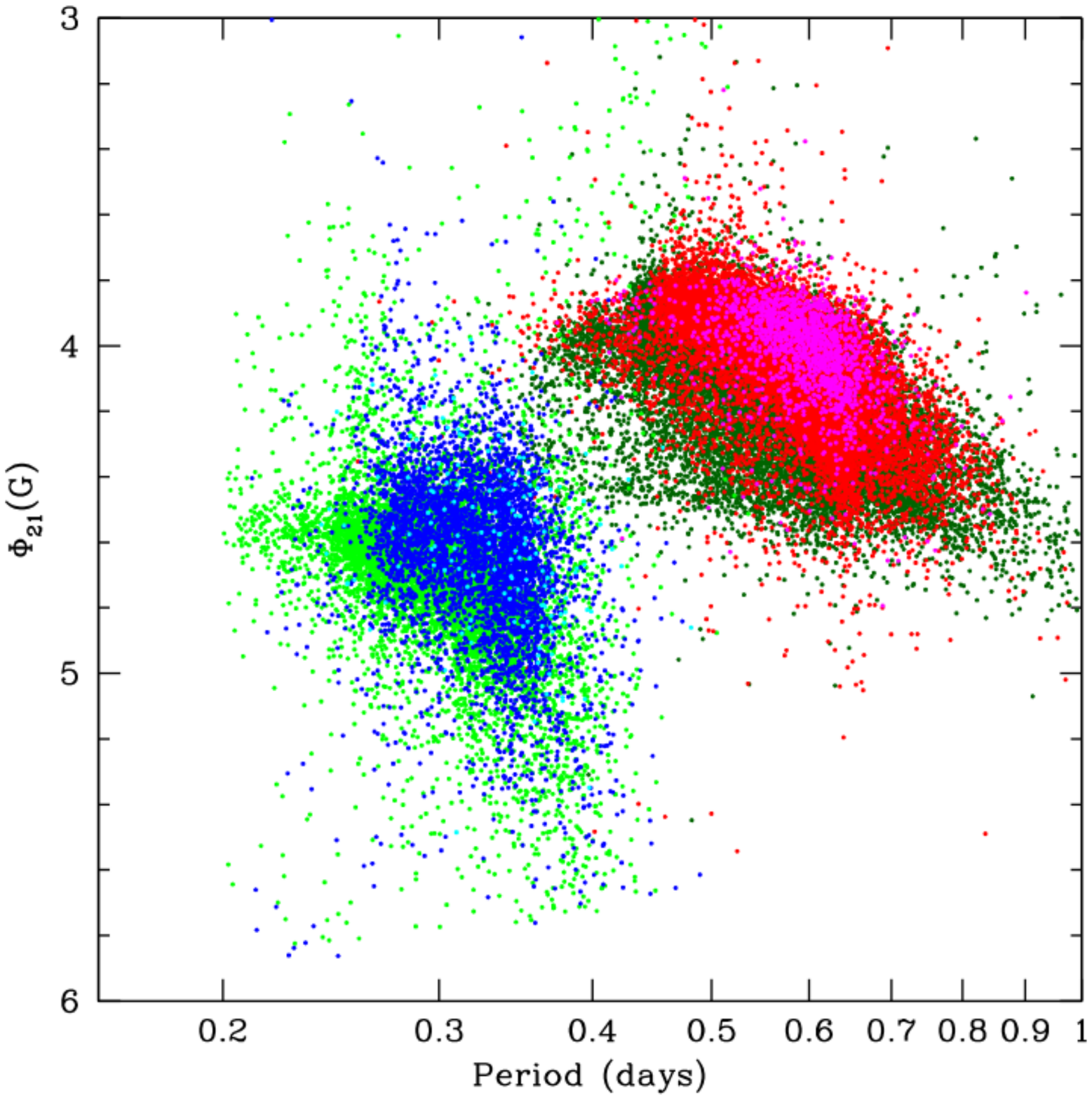}
   \caption{$G$-band $\phi_{21}$ versus period  diagram for RR Lyrae stars in the LMC (red: RRab, blue: RRc), SMC (magenta: RRab, cyan: RRc) and Galactic bulge (dark green: RRab, green: RRc), obtained converting to $G$-band the literature $I$-band $\phi_{21}$ values from OGLE (see text for details).}
              \label{RRL-F21}%
    \end{figure}

\begin{figure}
   \centering
   \includegraphics[trim=10 180 -50 130, width=10.2 cm,clip]{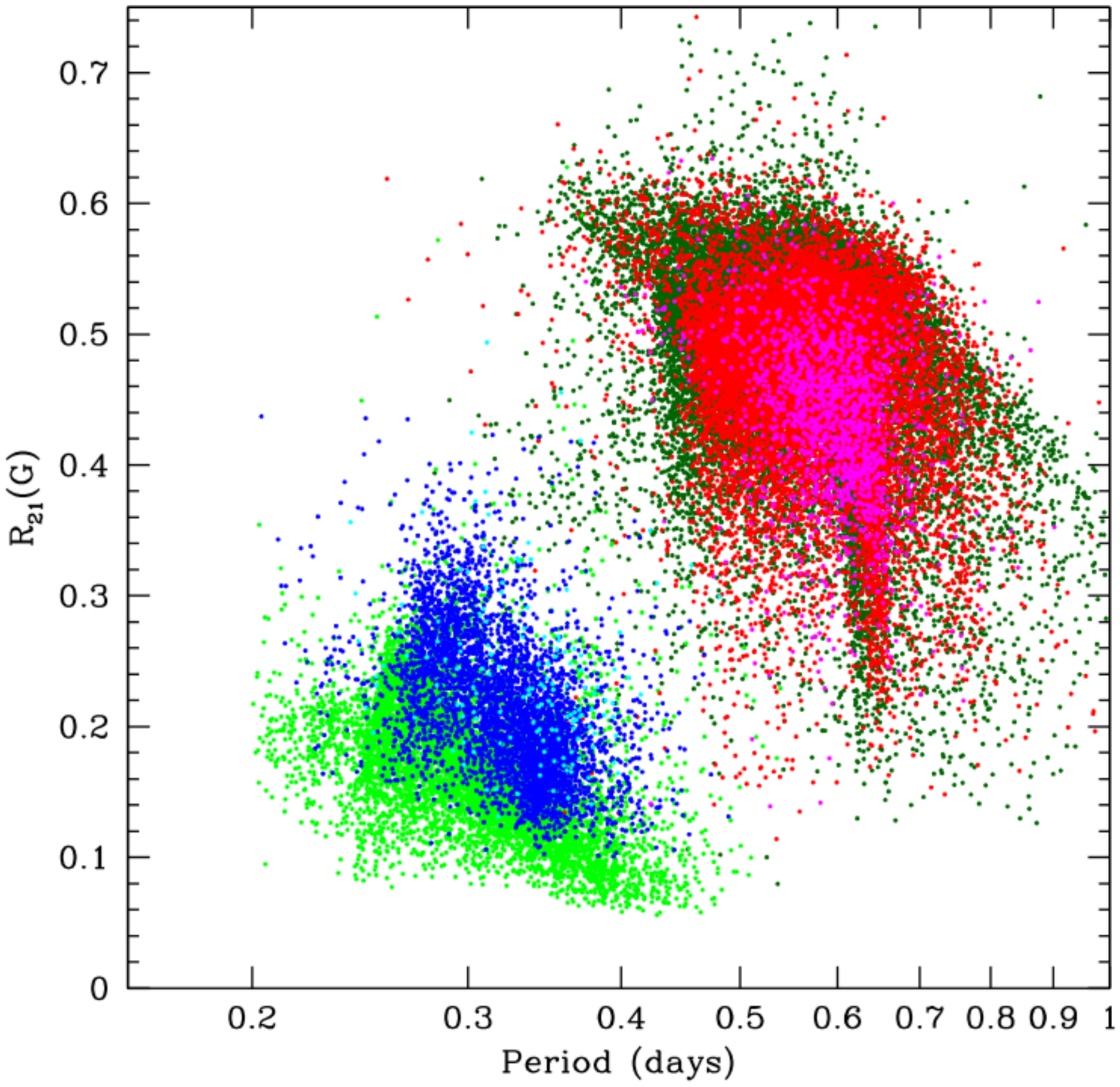}
   \caption{Same as in Fig.~\ref{RRL-F21}, for the $G$-band $R_{21}$ versus period diagram.}
              \label{RRL-R21}%
    \end{figure}

Figs.~\ref{RRL-F21} and ~\ref{RRL-R21} show the $G$-band Fourier parameters $\phi_{21}$ and $R_{21}$ of RR Lyrae stars computed using eqs.  (4), (5), (8) and (9) plotted versus period ($P$).
The $\phi_{21}$ {\it vs} $P$ plot was used along with the $PA$  diagram  to distinguish  RR Lyrae stars from ACEPs  and to better separate the RRab and RRc types (see Section~\ref{s3new}).

Finally. Eqs. (12) to (19) provide formulae  to convert to the {\it Gaia} $G$-band the literature  values of the Fourier parameters $R_{21}$, $R_{31}$,
 $\phi_{21}$ and $\phi_{31}$ of  Cepheids:
\begin{equation}
\begin{split}
\phi_{21}(G)=(0.76 \pm 0.14)+(0.81\pm0.03)\times \phi_{21}(I)\\
-(0.11\pm 0.04)\times \log P{\rm (d)} ~~~~rms=0.088\\
\end{split}
\end{equation}
\begin{equation}
\begin{split}
\phi_{21}(G)=(0.19 \pm 0.08)+(0.959\pm0.017)\times \phi_{21}(V)\\
+(0.10\pm 0.02)\times \log P{\rm (d)} ~~~~rms=0.063\\
\end{split}
\end{equation}
\begin{equation}
\begin{split}
\phi_{31}(G)=(0.00 \pm 0.04)+(1.006\pm0.011)\times \phi_{31}(I)\\
-(0.50\pm 0.11)\times \log P{\rm (d)} ~~~~rms=0.095\\
\end{split}
\end{equation}
\begin{equation}
\begin{split}
\phi_{31}(G)=(0.061 \pm 0.018)+(0.980\pm0.006)\times \phi_{31}(V)\\
~~~~rms=0.063\\
\end{split}
\end{equation}
\begin{equation}
\begin{split}
R_{21}(G)=(0.014 \pm 0.007)+(0.98\pm0.02)\times R_{21}(I)\\
 ~~~~rms=0.018\\
\end{split}
\end{equation}
\begin{equation}
\begin{split}
R_{21}(G)=(0.002 \pm 0.004)+(0.996\pm0.012)\times R_{21}(V)\\
~~~~rms=0.01\\
\end{split}
\end{equation}
\begin{equation}
\begin{split}
R_{31}(G)=(0.005 \pm 0.006)+(0.99\pm0.03)\times R_{31}(I)\\
~~~~rms=0.012\\
\end{split}
\end{equation}
\begin{equation}
\begin{split}
R_{31}(G)=(0.003\pm 0.003)+(1.002\pm0.017)\times R_{31}(V)\\
~~~~rms=0.005\\
\end{split}
\end{equation}
\noindent
 
Figs.~\ref{DCEP-F21}  and ~\ref{DCEP-R21} show the $G$-band Fourier parameters $\phi_{21}$ and $R_{21}$ of Cepheids computed using eqs. (12), (13), (16) and (17)   plotted versus period.
These diagrams were used to distinguish different Cepheid modes and to separate Cepheids  from RR Lyrae stars (see Section ~\ref{s3new}).


\subsection{RR Lyrae branch}\label{RRtrunk}
The different steps of the processing for sources classified as RR Lyrae stars by the general {\it Classification} pipeline are summarised in Fig.~\ref{decomposition2}. 
However, only some modules of the RR Lyrae branch were activated for the analysis of the {\it Gaia} DR1 data.

\subsubsection{Identification of double-mode RR Lyrae stars: {\it RRLyraeDoubleModeSearch} module}
\begin{figure}
   \centering
   \includegraphics[trim=20 180 20 120, width=9 cm,clip]{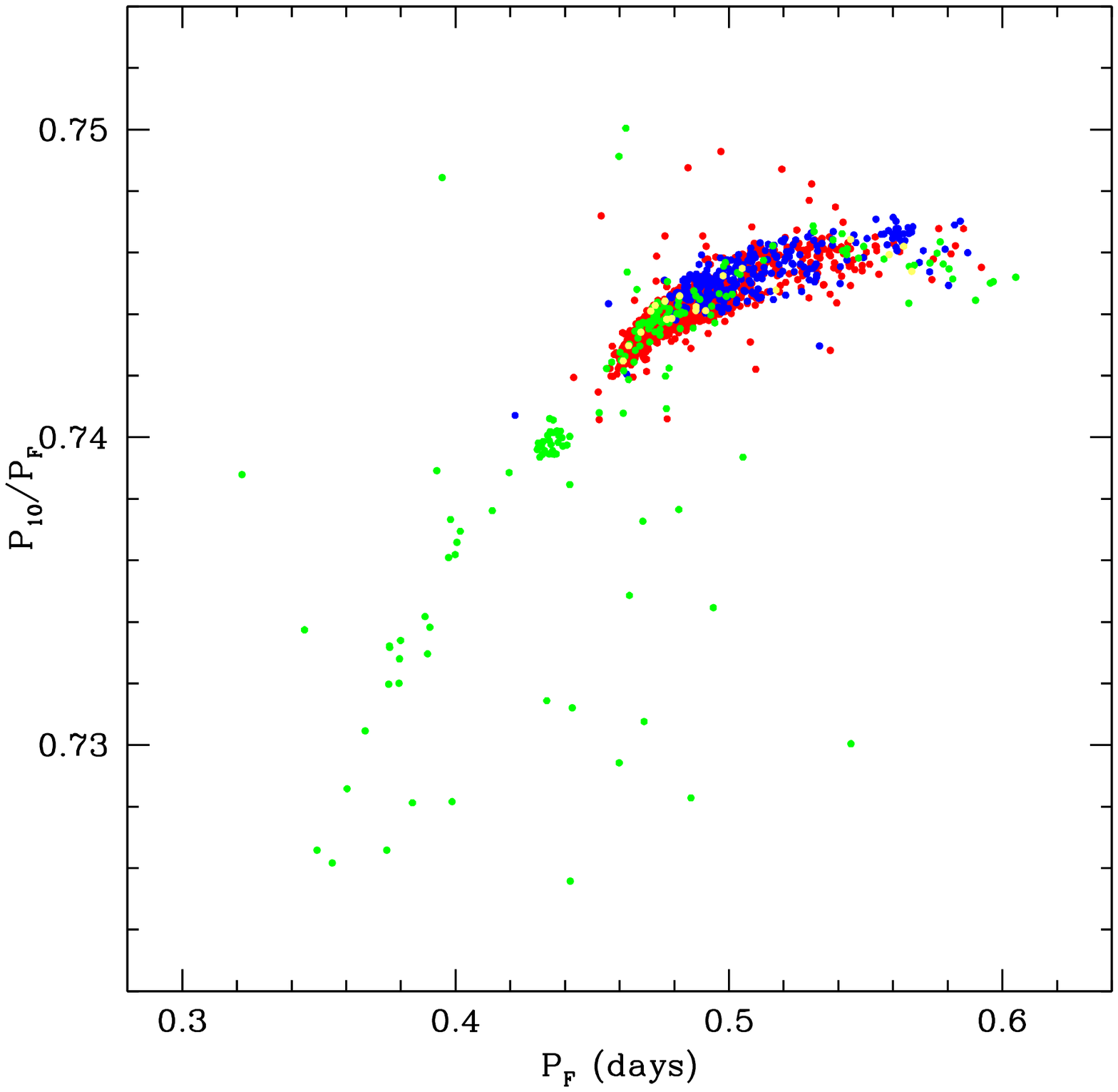}
   \caption{Petersen diagram of double-mode RR Lyrae stars observed by the OGLE and ASAS surveys in the
LMC (red filled circles), SMC (blue filled circles), and in the Galactic bulge (green filled circles) and halo (yellow filled circles). $P_{\rm F}$  and $P_{\rm 1O}$ are fundamental and first overtone pulsation mode, respectively (see text for details). 
}
              \label{petersen}%
    \end{figure}

\begin{figure}
   \centering
  \includegraphics[trim=10 180 20 120, width=9 cm,clip]{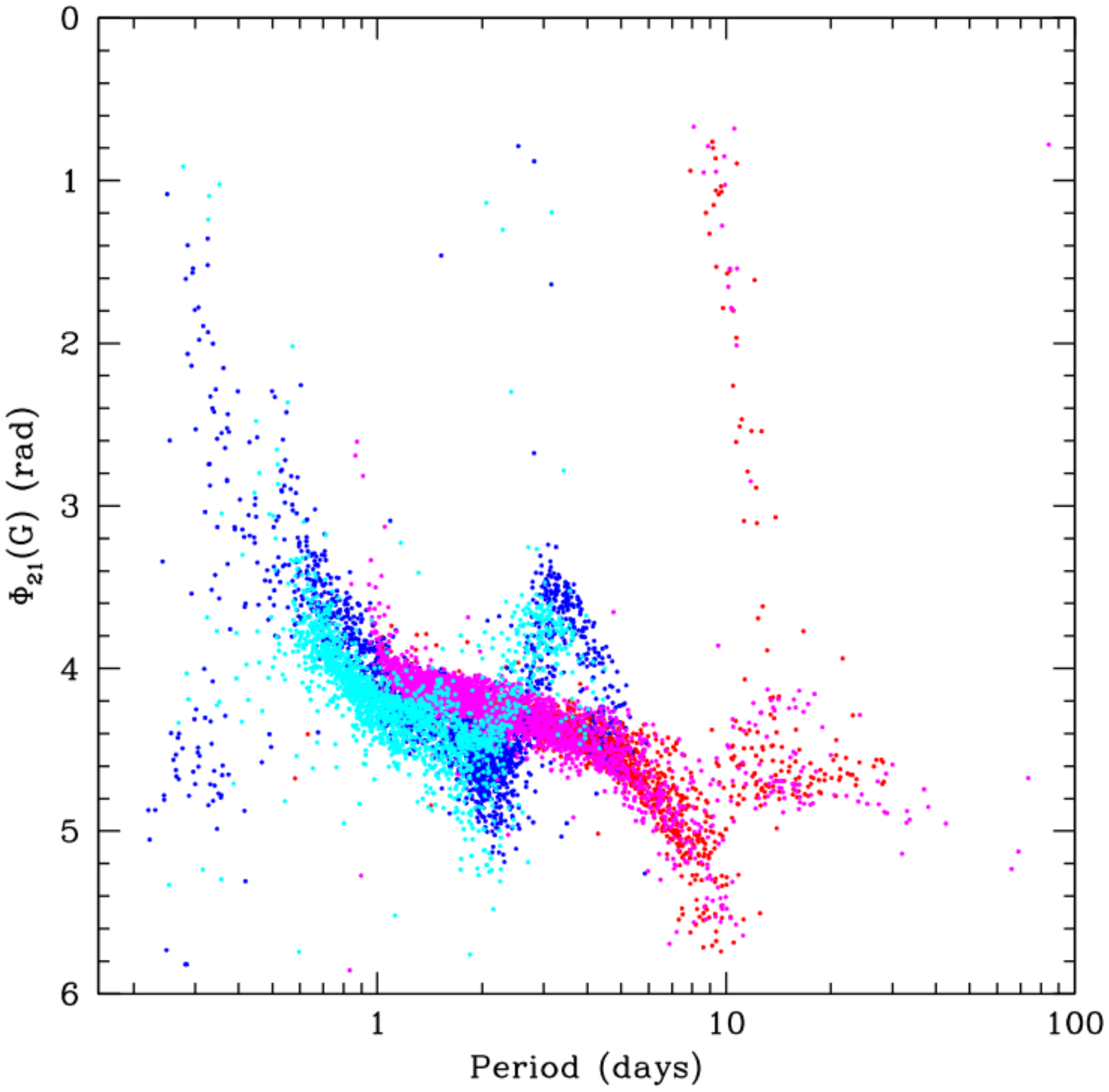}
   \caption{$G$-band $\phi_{21}$ versus period diagram for DCEPs in the Magellanic Clouds (red: F,  blue: 1O DCEPs in the LMC; magenta: F, cyan: 1O DCEPs  in the SMC).  The figure was obtained by transforming to the $G$-band the $I$-band  $\phi_{21}$ values in \citet{soszynski08a,soszynski10a,soszynski15b,soszynski15c} using eq. (12) (see text for details).} 
   \label{DCEP-F21}%
    \end{figure}

\begin{figure}
   \centering
   \includegraphics[trim=10 180 20 120, width=9 cm,clip]{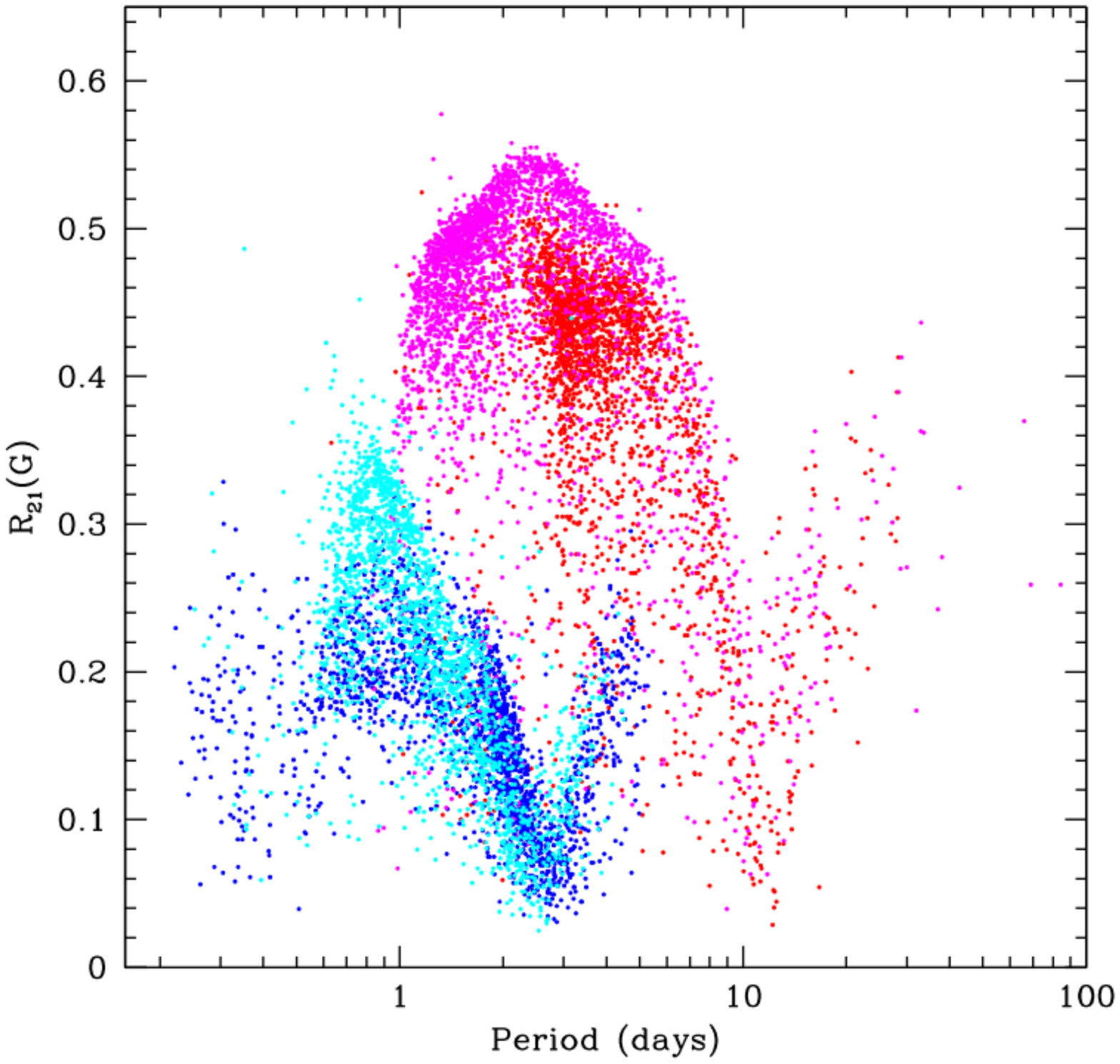}
   \caption{$G$-band $R_{21}$ versus period diagram for DCEPs in the Magellanic Clouds (red: F,  blue: 1O DCEPs in the LMC; magenta: F, cyan: 1O DCEPs  in the SMC). The figure was obtained by transforming to the $G$-band  the $I$-band $R_{21}$ values  in \citet{soszynski08a,soszynski10a,soszynski15b,soszynski15c} using eq. (16) (see text for details).}
   \label{DCEP-R21}%
    \end{figure}

For RR Lyrae stars for which a secondary periodicity was detected by module {\it SecondaryPeriodicities} of the initial SOS Cep\&RRL processing (see Fig.~\ref{decomposition1}) the {\it RRLyraeDoubleModeSearch} module sets to  
first overtone period ($P_{\rm 1O}$) and fundamental mode period ($P_{\rm F}$)  
the shorter and the longer periodicities, respectively. Then if both $P_{\rm 1O}$  
and  $P_{\rm F}$ 
 are longer than 
0.3 d, it checks whether
the star locates within the  regions of the  
$P_{\rm 1O}$/ $P_{\rm F}$ vs  $P_{\rm F}$ plane 
(generally known as Petersen diagram, \citealt{petersen73}), 
allowed for RR Lyrae  double-mode pulsation. The allowed loci in the Petersen diagram were defined using the   
$P_{\rm 1O}$,  $P_{\rm F}$ and  $P_{\rm 1O}$/ $P_{\rm F}$ values 
of 1,335 RRd variables observed by the OGLE and ASAS surveys of the
LMC, SMC, Galactic bulge and halo. Their Petersen diagram is shown in Fig.~\ref{petersen}. A lower limit of 0.3 d for the 
$P_{\rm 1O}$, $P_{\rm F}$ 
values was also inferred 
from Fig.~\ref{petersen}, which shows that no RRd pulsators are known with  
$P_{\rm F} < 0.3$ d.
Schematically, the {\it RRLyraeDoubleModeSearch} algorithm performs the following steps:

\begin{itemize}
\item computes the  $P_{\rm 1O}$/ $P_{\rm F}$ ratio; 
\item if the source falls in the region defined as follows:
  $0.724$$<$ $P_{\rm 1O}$/ $P_{\rm F}$$<$$0.752$ and 0.30$<$ $P_{\rm F}$$<$0.62 days, then the RR Lyrae star 
  is  identified as RRd.
\end{itemize}
\noindent
The light curve modelling  of a confirmed double-mode RR Lyrae star is then refined in module {\it NonLinearDoubleModeModeling} 
by applying the non-linear fitting procedure
with the proper truncated Fourier series and fitting simultaneously
the two pulsation modes.  In a similar way, the period, the epoch, the
peak-to-peak amplitudes and the Fourier decomposition are recomputed and refined.
The {\it RRLyraeDoubleModeSearch} module of SOS Cep\&RRL was tested but its results not yet included for {\it Gaia} DR1. 

\subsubsection{Blazhko effect search, position in the CMD \& amplitude ratios}
There are three modules in the RR Lyrae branch ({\it BlazhkoEffectSearch}, {\it PositionInTheCMD} and {\it AmplitudeRatios})  
that  could not be activated because in the {\it Gaia} DR1 we lack information to properly operate them. They are briefly described in the following.

The {\it BlazhkoEffectSearch} module searches for RR Lyrae stars affected by the Blazhko effect (\citealt{blazhko}),  a periodic modulation of the amplitude and/or the
phase of the main pulsation that occurs on timescales typically varying from a few
days to hundreds of days.  This phenomenon is shown by some 25-30\% of
the Galactic RRab and 5\% of the RRc stars. However, recent detections of
Blazhko stars with very small amplitudes suggest that these numbers may
be underestimated (an occurrence rate as high as 50\% follows from  
studies by Jurcsik et al., Jurcsik personal communication). Using data from future releases {\it Gaia} may  
provide  accurate and updated occurrence rates. RV 
measurements and multi-band photometric data, to be obtained by {\it Gaia} with the
foreseen long time base of observations, are fundamental to identify
irregular and Blazhko variables among the RR Lyrae stars, and can 
help to understand whether they are due to nonradial modes excited during pulsation.

The {\it PositionInTheCMD} module verifies that sources classified as RR Lyrae stars fall inside the RR Lyrae instability strip (IS) in the CMD, by 
checking whether the colours of the targets are compatible with the IS for RR Lyrae
stars.
{\it Gaia} CMD is a fundamental tool to clean the 
sample of RR Lyrae stars  by possible contaminating objects, since 
other types of variables falling in the same period domain, as for instance eclipsing binaries (ECLs), 
should lie predominantly outside the  RR Lyrae  IS, in this
diagram.  Knowledge of the source absolute magnitude (via parallax) and colour (or effective temperature) 
are needed to use this tool.  Parallaxes are published in {\it Gaia} in DR1 only for stars with $G \lesssim$  12 mag, as part of the Tycho-{\it Gaia} Astrometric Solution (TGAS; \citealt{lindegren16}). 
Cepheids and RR Lyrae stars for which results of the SOS Cep\&RRL processing are published in {\it Gaia} DR1 are fainter than this limit (see Section~\ref{s3}), hence, parallaxes are not yet 
available for them. 
 This limited the use of the CMD tool, as it is described in Section~\ref{s3new}.

The {\it AmplitudeRatios} module uses the peak-to-peak amplitude ratios 
between {\it Gaia}'s three different photometric bands to check
whether  the observed periodicity is due to a contact binary
system mimicking an RR Lyrae-like light curve (this may be the case
especially for RRc pulsators). In fact, amplitude ratios 
between different photometric bands are expected to be close to
unity in the case of ECLs. Lacking $G_\mathrm{BP}$,  $G_\mathrm{RP}$ time-series,   this tool could not be activated for {\it Gaia} DR1.

\subsubsection{$PA$ diagram and mode identification: {\it ModeIdentification} module}
RRab  and RRc types occupy
different locations in the $G$-band peak-to-peak amplitude  versus period diagram (see Fig.~\ref{bailey}) and in 
 the $P$ - $R_{21}$, and $P$ - $\phi_{21}$ planes  
(see Figs.~\ref{RRL-F21} and ~\ref{RRL-R21}). 
The  {\it ModeIdentification} module  combines results from these three diagnostics to identify the pulsation mode of
candidate RR Lyrae stars and/or recognise misclassified objects.  
According to Fig.~\ref{bailey}, 
RR Lyrae stars are confined within 0.2 d $\le P \le$ 1.0 d in period and within  0.05 mag $\le$ Amp($G$) $\le$ 2.0 mag in 
$G$-band amplitude. Sources outside these ranges will be flagged as possible misclassified
objects. Furthermore, RRc  and RRab types can then be separated with a  line  described by 
the equation:  Amp(G) = $-$4.0$\times  P$ +2.2.
Mixing of  RRc and RRab pulsators occurs close to the separation line, and significant contamination 
also exists in  Fig.~\ref{bailey}  between RR Lyrae stars and ACEPs. However, they can both be reduced by using the star position in the Fourier 
$R_{21}$ and $\phi_{21}$ versus $P$ planes (Figs.~\ref{RRL-F21} and ~\ref{RRL-R21}) where the two types locate in separated regions, of which we 
defined the borders as discussed in Section~\ref{results2}.


\subsubsection{RR Lyrae star classification}
This module performs the final assignment of a source to the RR Lyrae class, based on results of the previous SOS Cep\&RRL modules.
If the source is confirmed to be an RR Lyrae star, thus validating the initial class assignment provided by the general  {\it Classification} pipeline, 
the analysis proceeds with derivation of the stellar parameters\footnote{This occurs through the module {\it StellarParametersDerivation} still partially under development, 
by which stellar intrinsic parameters are derived through a variety of methods specifically appropriate
for RR Lyrae stars. These include:   radius estimates via Baade-Wesselink analysis (see, e.g., \citealt{cacciari89,cacciari92}, and references therein); metallicity estimates from the $\phi_{31}$ parameter of the Fourier 
light curve decomposition; reddening estimates from the light and colour curves; and mass estimates for double-mode pulsators through the Petersen diagram and the pulsation equation 
(\citealt{dicrisci04}) for single-mode pulsators. No specific tool to detect RR Lyrae  stars in binary systems has been implemented in the pipeline yet, as binary RR Lyrae stars  are an extremely rare event, only one has been firmly established  so far \citep{wade99},  less than two dozen candidates  are reported in total 
by \citet{hajdu15} and \citet{liska15}  but none of them has been spectroscopically confirmed yet.  
} and final export of the source  attributes (period, mean magnitude, peak-to-peak amplitude,  Fourier parameters, secondary periodicity, pulsation mode, and any other relevant quantities) to the Main Data Base. 
Conversely, if the source is not confirmed as an RR Lyrae star, it is sent for analysis through the Cepheid branch (this action is represented by the green box in Fig.~\ref{decomposition2}), unless it was already  analysed in that branch  and  
 not found to be a Cepheid either, in which case, it will be definitively rejected as an RR Lyrae star and/or Cepheid and fed back to the general variable star analysis pipeline for processing  
 through other SOS modules.

\subsection{Cepheid branch}\label{CepTrunk}  
The processing of sources classified as Cepheid candidates by the general {\it Classification} pipeline occurs in the Cepheid branch first, the  steps of which are schematically  shown in Fig.~\ref{decomposition3}. As with the RR Lyrae stars only some modules of the Cepheid branch could be used for the analysis of the Cepheids in {\it Gaia} DR1 .
In particular, the first step of the Cepheid branch processing would be the identification of Cepheids in binary or multiple systems, using RV measurements 
obtained by the RVS and the ratios of photometric amplitudes in the different passband. Since neither RVs nor $G_\mathrm{BP}$,  $G_\mathrm{RP}$ photometry are available in {\it Gaia} DR1  this module was not activated. Similarly, we could not use the {\it PositionInTheCMD} module to check whether the candidate Cepheids fell inside the DCEP,  ACEP and TCEP instability strips in the CMD. 
Analysis in the Cepheids branch starts from the search for multimode variables performed by the {\it CepMultModeSearch} module. 

\subsubsection{Identification of multi-mode Cepheids:  {\it CepMultiModeSearch} module}
The  {\it CepMultiModeSearch} module identifies Cepheids pulsating simultaneously in different modes  by checking the period ratios  of the different periodicities found by 
the {\it SecondaryPeriodicities} module of the 
 general trunk (Fig.~\ref{decomposition1}). The relevant period ratios are: 1O over F ($P_{\rm 1O}$/ $P_{\rm F}$), second-overtone (2O) over F ( $P_{\rm 2O}$/$P_{\rm F}$) and so on ( $P_{\rm 2O}$/ $P_{\rm 1O}$;  $P_{\rm 3O}$/$P_{\rm 1O}$;  $P_{\rm 3O}$/ $P_{\rm 2O}$).
The  typical values for these ratios are known both from empirical
and theoretical studies. Allowed periods and period ratios for multi-mode DCEPs, following \cite{soszynski08a}, 
are
reported in Table~\ref{tab:cepPeriodRatios}.

\begin{table}\label{table:1}
\caption{Permitted  periods and period ratios for multi-mode DCEPs.}
\label{tab:cepPeriodRatios}
\centering
\begin{tabular}{l}
\hline \hline
$0.7~~~~<P_{\rm 1O}/P_{\rm F}~~<0.76$~~~~~~\&~~~~$0.5~~<P_{\rm F}~~<8$ d\\
$0.79~~<P_{\rm 2O}/P_{\rm 1O}<0.81$~~~~~~\&~~~~$0.15<P_{\rm 1O}<1.8$ d\\
$0.67~~<P_{\rm 1O}/P_{\rm 3O}<0.68$~~~~~~\&~~~~$0.5~~<P_{\rm 1O}<0.6$ d\\
$0.58~~<P_{\rm 2O}/P_{\rm F}~~<0.68$~~~~~~\&~~~~$0.4~~<P_{\rm F}~~<1.6$  d\\
$0.655  <P_{\rm 3O}/P_{\rm 1O}<0.685$~~~~\&$~~~~0.1~~<P_{\rm 1O}<0.7$ d\\
$0.828  <P_{\rm 3O}/P_{\rm 2O}<0.844$~~~~\&$~~~~0.15 <P_{\rm 2O}<0.65$ d\\
\hline
\end{tabular}
\end{table}

\par\noindent
The following different cases are considered, depending on the number of secondary periodicities found by the {\it SecondaryPeriodicities} module of the SOS Cep\&RRL 
general trunk:  
\begin{itemize}

\item
{\it case 1: two periodicities are found}
\begin{itemize}
\item
The algorithm sets to $P_{\rm 1O}$ and  $P_{\rm F}$
the shorter and the longer periods, respectively. 
Then it checks whether the ratio $P_{\rm 1O}$/ $P_{\rm F}$ falls within the allowed ranges, as listed in Table~\ref{tab:cepPeriodRatios}.
\item
The algorithm sets to  $P_{\rm 2O}$ and  $P_{\rm F}$
the shorter and the longer periodicities, respectively. 
Then it checks whether the ratio  $P_{\rm 2O}$/ $P_{\rm F}$ is within the allowed ranges, as listed in Table~\ref{tab:cepPeriodRatios}.
\item
The algorithm sets to  $P_{\rm 2O}$ and  $P_{\rm 1O}$
the shorter and the longer periodicities, respectively. 
Then it checks whether the ratio  $P_{\rm 2O}$/ $P_{\rm 1O}$ is within the allowed ranges, as listed in Table~\ref{tab:cepPeriodRatios}.
\item
The algorithm sets to  $P_{\rm 3O}$ and  $P_{\rm 1O}$
the shorter and the longer periodicities, respectively. 
Then it checks whether the ratio  $P_{\rm 3O}$/ $P_{\rm 1O}$ is within the allowed ranges, as listed in Table~\ref{tab:cepPeriodRatios}.
\end{itemize}
If none of the above conditions is satisfied the source is  rejected as multimode Cepheid.

\item
{\it case 2: three periodicities are found}
\begin{itemize}
\item The algorithm sets to  $P_{\rm 2O}$,  $P_{\rm 1O}$ and  $P_{\rm F}$ the shorter, the
 intermediate, and the longer periodicity, respectively.  Then it
 checks whether the  $P_{\rm 1O}$/ $P_{\rm F}$,  $P_{\rm 2O}$/ $P_{\rm F}$ and  $P_{\rm 2O}$/ $P_{\rm 1O}$ ratios  are within the allowed ranges, as listed in Table~\ref{tab:cepPeriodRatios}.
\item The algorithm sets to  $P_{\rm 3O}$,  $P_{\rm 2O}$ and  $P_{\rm 1O}$ the shorter, the
 intermediate, and the longer periodicity, respectively.  Then it
 checks whether the $P_{\rm 3O}$/ $P_{\rm 1O}$,  $P_{\rm 3O}$/ $P_{\rm 2O}$ and  $P_{\rm 2O}$/ $P_{\rm 1O}$ ratios are within the allowed ranges, as listed in Table~\ref{tab:cepPeriodRatios}.
\end{itemize}
If none of the above conditions is satisfied the source is  rejected as multimode Cepheid.
\end{itemize}

The light curve modelling  of a confirmed multi-mode  Cepheid is refined in the {\it NonLinearDoubleModeModeling} module 
by applying the non-linear fitting procedure with the proper truncated Fourier series and fitting simultaneously all pulsation modes (generally two).  
Periodicities, epoch of maximum light, intensity-averaged mean magnitude, peak-to-peak amplitudes  of the two/three pulsation modes and the parameters of the Fourier decomposition 
are also recomputed and refined taking into account all periodicities.
The {\it CepMultiModeSearch} module of SOS Cep\&RRL was tested but its results not yet included for {\it Gaia} DR1.

\subsubsection{Identification of Cepheid type (DCEP, ACEP, T2CEP): {\it CepTypeIdentification} module}
\begin{figure}
   \centering
   \includegraphics[width=9.5 cm, trim= 20 420 0 130, clip]{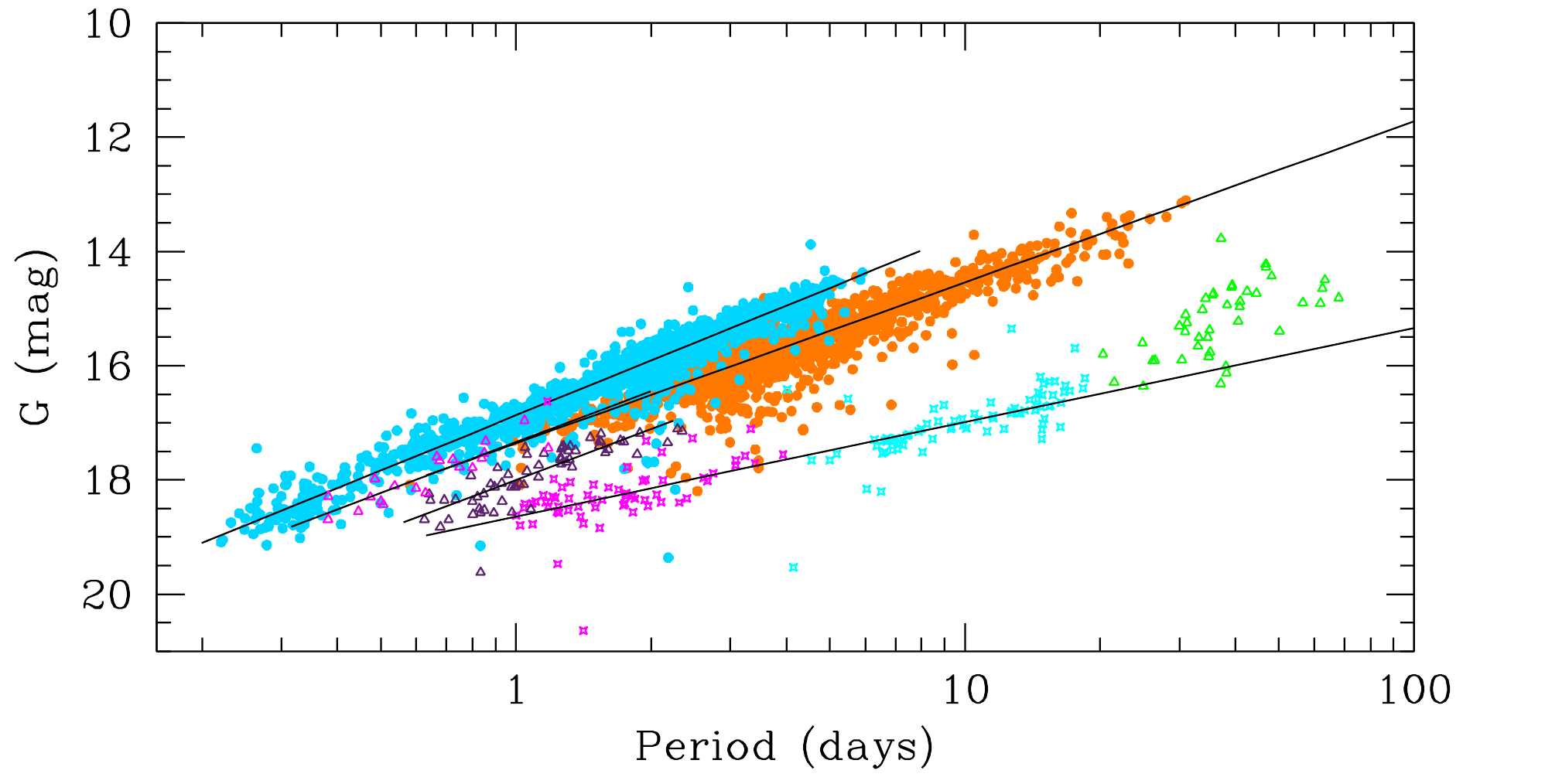}
   \caption{$G$-band $PL$ distribution of DCEPs,  ACEPs and T2CEPs in the LMC obtained transforming to the $G$-band the OGLE $V,I$ magnitudes by means of Eq. A.1.   Overlaid are the 
   $PL$ relations described by Eqs. (20) - (24).  Cyan filled circles: DECPs 1O; orange
filled circles: DCEPs F; cyan four starred symbols: ACEPs F; magenta
four starred symbols: ACEPs 1O; green open triangles: RVTAU; violet
open triangles: WVIR; magenta open triangles: BLHER.
}
              \label{plPlotPaper}%
    \end{figure}

The module {\it CepTypeIdentification} combines pulsation and photometric (in the future also spectroscopic) characteristics of the candidate Cepheids computed in the previous modules of the  
SOS Cep\&RRL processing (both in the general trunk and in the Cepheid branch)  to subdivide them into DCEP, ACEP and T2CEP types.  The module foresees  the use of two main 
diagnostics:  (i) the source metallicity  that will be derived by the DPAC processing of RVS measurements,  
as DCEP and T2CEP/ACEP types show in general different
metallicity content, being the DCEPs typically more metal-rich than the T2CEPs and ACEPs; (ii) the $PL$ and $PW$ relations, which are different for the three types of Cepheids.
However, for {\it Gaia} DR1 the identification of the Cepheid types could rely only on the use of the $G$-band $PL$ relations, for which we used the following set of equations:
\begin{equation}
\begin{split}
{\rm DCEP_F~~}:~M_G = (17.361 \pm 0.020) - DM_{\rm LMC} - (2.818\\ 
\pm 0.032) \times \log P~~~ \sigma=0.162~{\rm mag}
\end{split}
\end{equation}
\begin{equation}
\begin{split}
{\rm DCEP_{1O}}:~M_G  =  (16.872 \pm 0.010) - DM_{\rm LMC} - (3.195\\
\pm 0.030) \times \log P~~~ \sigma=0.181~{\rm mag}    
\end{split}
\end{equation}
\begin{equation}
\begin{split}
{\rm T2CEP~~}:~M_G  =  (18.640 \pm 0.085) - DM_{\rm LMC} - (1.650\\
\pm 0.109) \times  \log P~~~ \sigma=0.188~{\rm mag}   
\end{split}
\end{equation}
\begin{equation}
\begin{split}
{\rm ACEP_F~~}:~M_G  =  (18.00 \pm 0.04) - DM_{\rm LMC} - (2.95\\
\pm 0.27) \log P~~~ \sigma=0.22~{\rm mag}
\end{split}
\end{equation}
\begin{equation}
\begin{split}
{\rm ACEP_{1O} }:~M_G  =  (17.34 \pm 0.08) - DM_{\rm LMC} - (2.95\\
\pm 0.35) \times \log P~~~ \sigma=0.20~{\rm mag}    
\end{split}
\end{equation}
where $M_G$ is the absolute magnitude in the $G$ band. These relations were defined using Cepheids in the LMC for which light curves and pulsation characteristics have been published  by \citet{soszynski08a,soszynski8b},  and adopting an 
absolute de-reddened distance modulus for the LMC of $DM_{LMC}=18.49$ mag,  from \citet{pietrz13}. 
They are shown in Fig.~\ref{plPlotPaper}.
Candidate Cepheids  that fall within 4 $\sigma$ from any  of the above 
relations are assigned the Cepheid type  and pulsation mode of the closest $PL$.  
Conversely, candidate Cepheids falling beyond 4 $\sigma$ will be rejected as misclassified sources. 
T2CEPs are  further subdivided in three separate classes: the BLHER class, the WVIR class  and the RVTAU class, depending on the pulsation 
period.  The three different subclasses of the T2CEP group follow different $PL$ and $PW$ 
relations and populate different period ranges. The  module: {\it T2CEPSubclassification} identifies as BLHER the T2CEPs with period in the range  1$\leq P <$ 4 d; 
as WVIR  the T2CEPs with period in the range 4$\leq P <$ 20 d;  and as RVTAU the T2CEPs
with period equal to or longer than 20 days (\citealt{soszynski8b}).

Single-mode DCEPs are known to pulsate in the F, 1O and 2O modes.   Since F, 1O and 2O DCEP subclasses occupy different loci in
the $R_{21}$ versus Period diagram, the {\it DCEPModeIdentification} module assigns the
pulsation mode to a DCEP using the $R_{21}$ parameter of the light curve Fourier decomposition. 
Specifically, considering the Fourier decomposition of the $G$-band light
curve, and the  $R_{21}$ vs $P$ diagram, 
a DCEP will be pulsating in the 1O mode  if the following conditions are verified:\\

\noindent $R_{21}$ $< -0.13\times P{\rm (d)} + 0.53$\\
 and\\
 $P <2.34$ d or $R_{21}$ $<0.214$\\
 and\\
  0.234 d$< P < 7$ d\\
  
\noindent otherwise, it is assigned  the F mode. These limits were inferred from  the $PL$ relations  of DCEPs based on OGLE-III data. 

In  Section~\ref{results2} we discuss and illustrate graphycally the regions we specifically defined in the $\phi_{21}$ and $R_{21}$ versus Period diagrams
to separate Cepheids from RR Lyrae stars and to identify the Cepheid pulsation mode.

Similarly, ACEPs are known to pulsate in the F and 1O modes.   The {\it ACEPModeIdentification} module assigns the
pulsation mode to an ACEP by combining results from the classification in types based on the $PL$ relations (module {\it CepTypeIdentification}) and the source period, as 
1O ACEPs have periods in the range: 0.35 $<$P$\leq$ 1.20 d, whereas  F  ACEPs have periods in the range: 1.20 $<$P$\leq$ 2.5 d.
These limits were inferred from  $PL$ relations  of ACEPs based on OGLE-III data.

\subsubsection{Cepheid classification}
This module performs the final assignment of a source to the Cepheid class and its sub-classes, based on results of the previous SOS Cep\&RRL modules. 
 If the source is confirmed to be a Cepheid, thus validating the initial class assignment provided by the general  {\it Classification} pipeline, 
the analysis proceeds with derivation of the stellar parameters\footnote{This occurs in the module {\it StellarParametersDerivation}. This module is under development and includes: 
 radius estimates via Baade-Wesselink analysis (see, e.g. \citealt{barnes76}, \citealt{ripepi07}); metallicity estimates for DCEPs from the $R_{21}$ and $R_{31}$ parameters of the Fourier 
light curve decomposition; identification of binary Cepheids from astrometry and RVS measurements and determination of their orbital elements; DCEP mass and age estimates.} 
and the final export of the source  attributes (period, mean magnitude, peak-to-peak amplitude,  Fourier parameters, secondary periodicity, classification in Cepheid types, pulsation mode, and any other relevant quantities) to the Main Data Base. 
Conversely, if the source is not confirmed to be a Cepheid, it  will be sent for analysis through the RR Lyrae branch (this action is represented by the green box in Fig.~\ref{decomposition3}), unless it was already  analysed in that branch  and  
 not found to be an RR Lyrae star, in which case, it will be definitely rejected as a Cepheid  and/or RR Lyrae star and fed back to the general variable star analysis pipeline for processing  
 through other SOS modules.
 
 \section{Application of the SOS-Cep\&RRL pipeline to \textit{\textbf{Gaia}} DR1 dataset}\label{s3new}
Results on Cepheids and RR Lyrae stars published in {\it Gaia} DR1 are based only on the application of the SOS Cep\&RRL pipeline to 
{\it Gaia}  $G$-band time series photometry calibrated to the Vega magnitude system, as no $G_\mathrm{BP}$/$G_\mathrm{RP}$ 
photometry is published in the  {\it Gaia} DR1 and parallax and RV information are not yet available for these Cepheids and RR Lyrae stars which are fainter than the TGAS sample (\citealt{lindegren16}). 
This made it necessary  to re-arrange and specifically tailor the SOS Cep\&RRL  pipeline to cope with the limited information available for the analysis of the Cepheid and RR Lyrae star candidates.

 Processing of {\it Gaia} photometry 
 takes place in an iterative manner organized through cycles (\citealt{evans16}). 
 As detailed in \citet{eyer16}, two datasets with different photometric calibration, time extent, and source number were used exceptionally, in order to be able to publish Cepheids and RR Lyrae stars in advance of schedule for the {\it Gaia} DR1. 
The first dataset included Cycle 0 (C0) photometry of millions of objects observed in the first 3 months after commissioning (from July 25, 2014), which were reduced by the 
general variable star processing to a sample of under 20K potential candidate Cepheids and RR Lyrae stars.
This selection of sources was further analysed with the {\it Gaia} DR1 calibration photometry of Cycle 1 (C1), which spanned almost 14 months of observations. 
 
\begin{figure}
   \centering
   \includegraphics[width=9 cm,trim=128 145 150 148, clip]{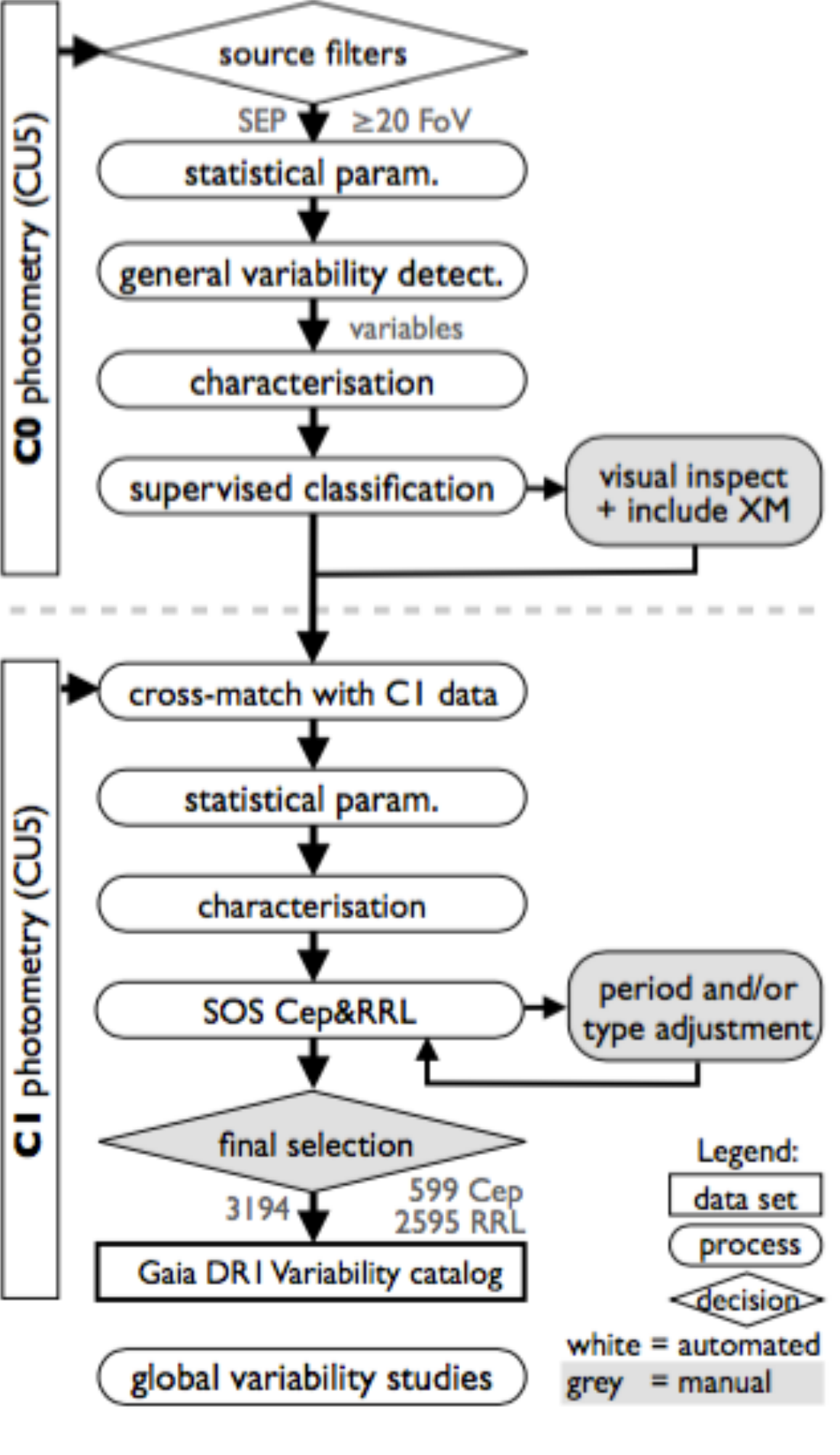}
    \caption{Flow-chart summarising the global  
    variable star analysis pipeline adopted for {\it Gaia} first data release. See fig.~10 in  \citet{eyer16} for a more detailed 
   version of this figure. 
       }
              \label{flow-chart-p3}
              \end{figure}

\begin{figure}
   \centering
   \includegraphics[width=9 cm, trim=100 170 110 164, clip]{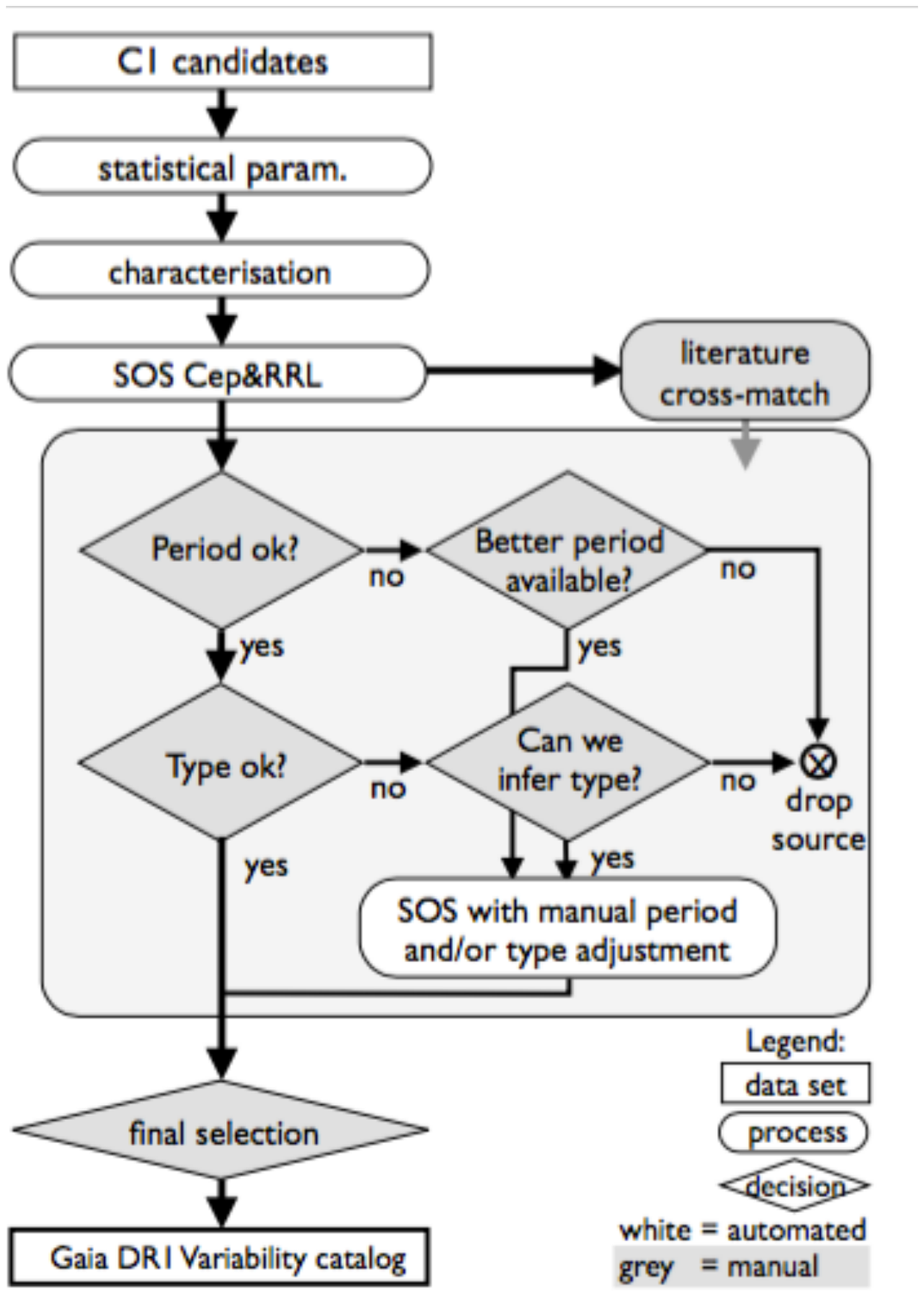}
   \caption{Flow-chart detailing the different steps of the SOS Cep\&RRL pipeline actual processing applied to produce results for {\it Gaia} first data release.
   }
              \label{flow-chart-p4}%
    \end{figure}

The global variable star analysis performed for the {\it Gaia} first data release is schematically summarised 
 in Fig.~\ref{flow-chart-p3} which shows both the initial processing on the C0 dataset and the final processing of the C1 photometry released in {\it Gaia} DR1. The SOS Cep\&RRL actual 
 processing is detailed in Fig.~\ref{flow-chart-p4} and specifically discussed in Section~\ref{tailoring}.

\subsection{Dataset, source selection and initial processing}\label{s3}
\begin{figure}
   \centering
   \includegraphics[width=10.5 cm,trim=20 355 -60 70, clip]{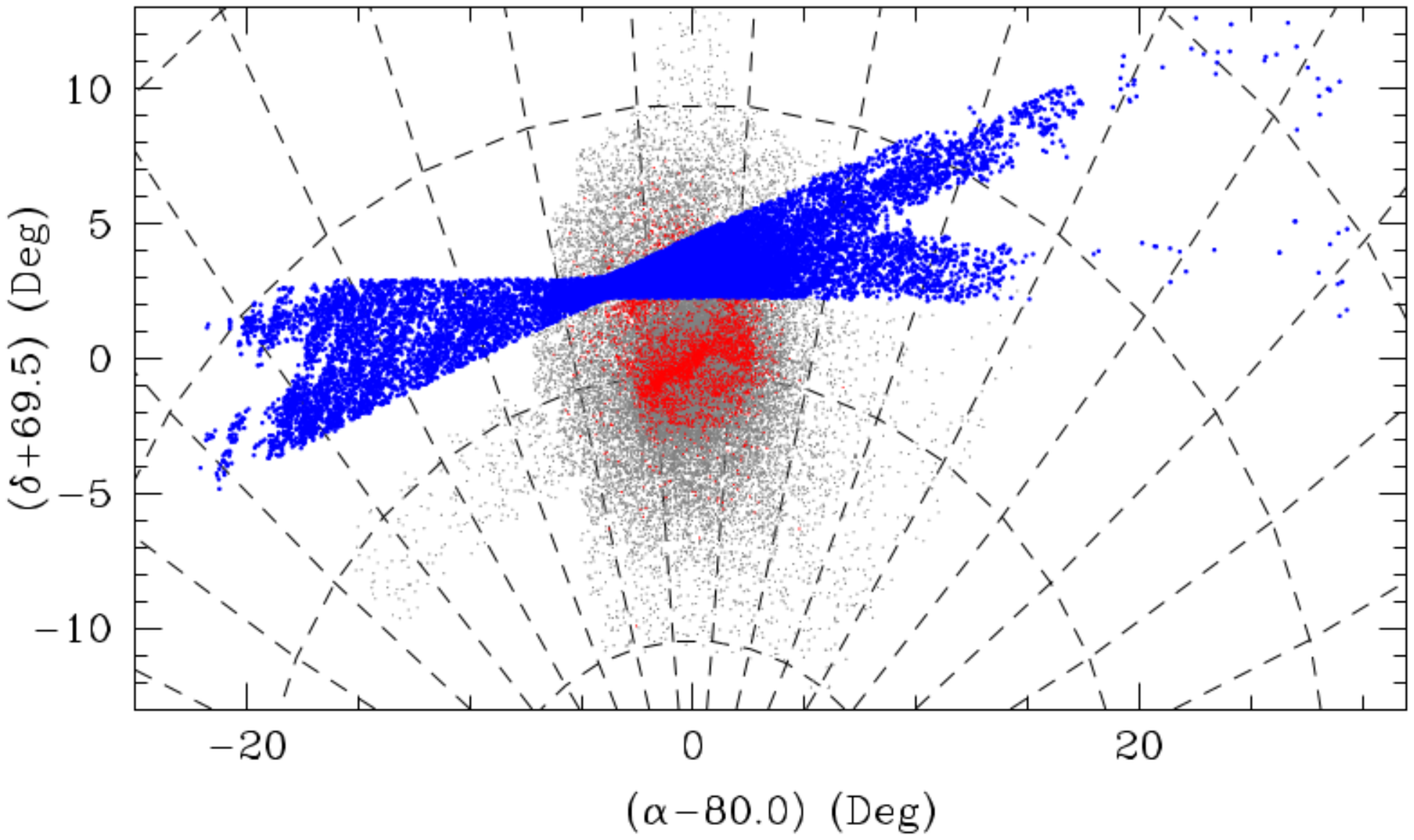}
   \caption{Blue filled circles: distribution on sky of 19,923 sources 
    in the {\it Gaia} SEP footprint, containing the enlarged set of Cepheid and RR 
Lyrae star candidates by the {\it Classification} workpackage of the general variable star analysis 
pipeline (9,347 candidates + 10,576 candidate outliers). {\it Gaia} SEP 
intercepts an external region of the LMC. Grey and red filled circles are, 
respectively, RR Lyrae stars and Cepheids observed in the LMC by the OGLE 
survey (see text for details).
 }
              \label{SEP-coverage}%
    \end{figure}

\begin{figure}
   \centering
   \includegraphics[width=10.5 cm,trim=20 355 -60 70, clip]{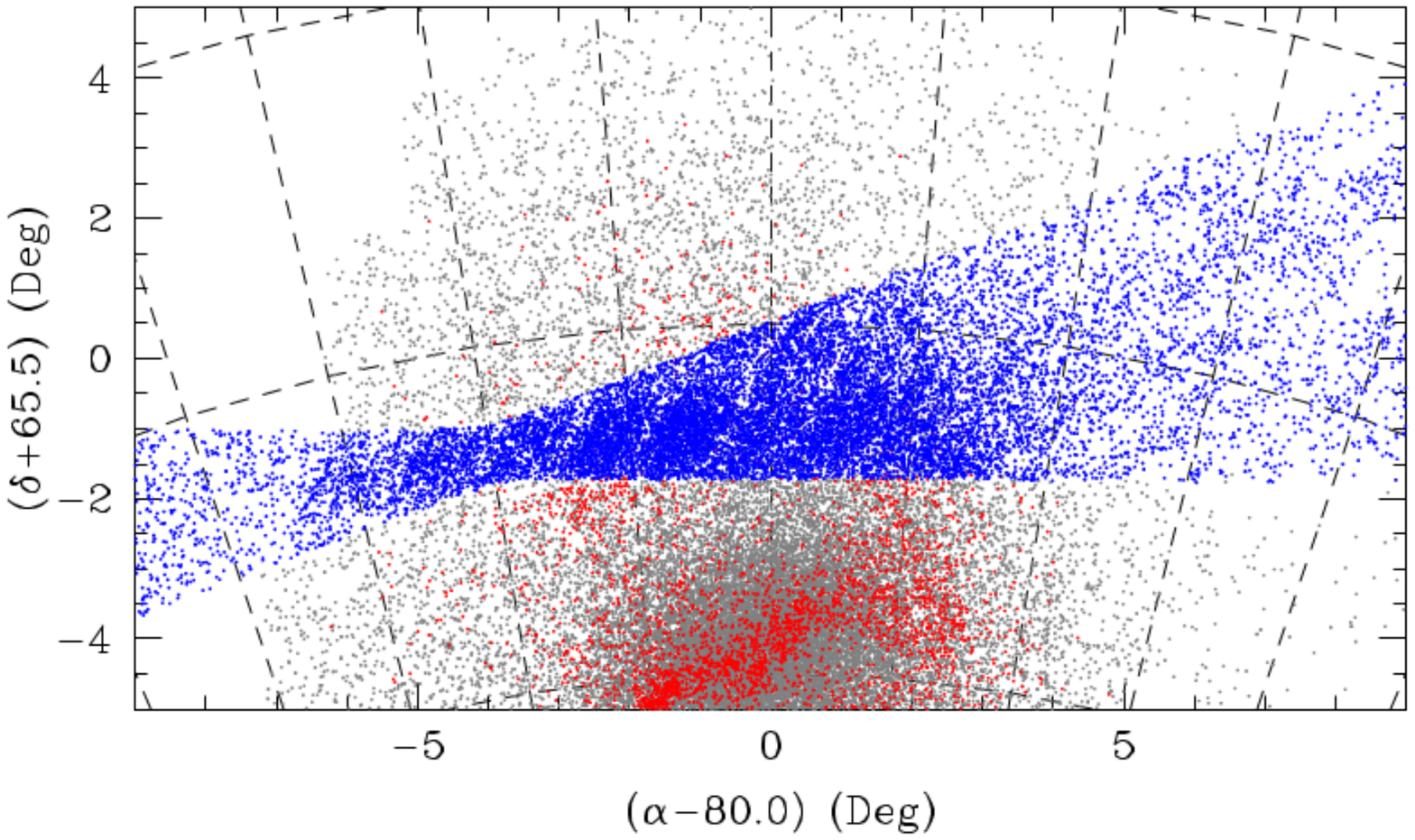}
   \caption{Enlargement  of Fig.~\ref{SEP-coverage} better showing the region of the LMC covered by the {\it Gaia} SEP. Symbols and colour-coding are as in Fig.~\ref{SEP-coverage}.}
              \label{SEP-coverage-bis}%
    \end{figure}

The dataset  processed by the SOS Cep\&RRL pipeline to produce results published in {\it Gaia} DR1 consists of $G$-band time-series photometry\footnote{Each point in the $G$-band
time-series is the mean of the 9 CCD measurements
collected during one observation of a source by {\it Gaia}.}
 of candidate Cepheids and RR Lyrae stars, observed by {\it Gaia} during 28-days Ecliptic Pole Scanning Law (EPSL)  from the end of July to the end of August 2014, followed by over a year of Nominal Scanning Law (NSL). We refer the reader to Section~5.2 in \citet{gaiacol-prusti} for a detailed description of {\it Gaia}'s scanning law. 
Source selection and steps of the preliminary analysis carried out on the C0 photometry 
(\citealt{evans16}) 
 are  summarized in the upper portion of Fig.~\ref{flow-chart-p3}. 
Sources were then reprocessed using  the 
C1 photometry, producing the final results for Cepheids and RR Lyrae stars in  the South Ecliptic Pole (SEP) footprint, that are published with {\it Gaia} first  data release.
Only sources having  20 or more data-points 
in the $G$-band time-series were analyzed, as this was deemed to be the minimum number of epochs allowing a reliable estimate of the period and other (pulsation) characteristics of the confirmed variables. However, the actual number of epochs per source is $<$ 20 in some cases due to subsequent removal of outliers.
As detailed in Eyer et al. (2016) and schematically summarised in Fig.~\ref{flow-chart-p3}, after applying the initial cut according to the number of epochs,  the {\it Statistical Parameters} of the 
 general variability pipeline computes the statistics of the time-series data  
for all sources, without any prior information on variability.   Sources showing variability are  then identified  by the {\it General Variability Detection} module, characterised in terms of periodicity and modelling of the light variation  by the {\it Characterization} module,  
and finally sent to the {\it Supervised Classification} module to determine the variability type. 
The SOS Cep\&RRL pipeline received sources classified as candidate Cepheids and RR Lyrae stars  by the three classifiers (Gaussian
Mixtures $-$GMs; Bayesian Networks $-$BNs, and Random
Forests $-$RFs) operated  by the {\it Supervised Classification} module of the general variable star analysis 
 pipeline (see \citealt{eyer16} and references therein, for details). 
A sample of 19,923 source were initially classified as 
 candidate RR Lyrae stars and Cepheids by applying the BN, RF and GM classifiers to the C0 photometry of {\it Gaia} SEP region. 
 Their  distribution  on sky  is shown in 
Figs.~\ref{SEP-coverage} and ~\ref{SEP-coverage-bis}. 
This rather
large number of candidates included all probability levels as well as candidates
flagged as class outliers, in order to maintain a high level of  completeness and not lose potentially valid candidates.

The characteristic geometric shape of the area covered by {\it Gaia} SEP observations is mainly due to the way the spacecraft scanned the sky during the first 28 days of science operation in EPSL  when {\it Gaia} was kept to repeatedly monitor the two ecliptic poles. 
Therefore,  sources very close ($\sim1$~deg) to the ecliptic poles 
 may have time-series data with more than 200 
 observations, though quickly dropping off at distances further away. 
 {\it Gaia} SEP footprint intercepts  a peripheral portion of the LMC, offset by about 3 degrees to the North and 4 degrees to the East of the LMC center, that  contains  several  Cepheids and RR Lyrae stars (as well as other types of  variable stars) with characteristics known from the studies by 
 the OGLE (\citealt{
 soszynski12,soszynski15a,soszynski15b,soszynski15c,soszynski16} and references therein) and EROS-2  
(according to \citealt{kim14}) surveys. 
They are shown as grey (RR Lyrae stars) and red (Cepheids) filled circles in Figs.~\ref{SEP-coverage} and 
~\ref{SEP-coverage-bis}.
 
Candidate Cepheids and RR Lyrae stars in the {\it Gaia} SEP most likely belong to the LMC, hence, to a first approximation, they are all the same distance from  us. 
Due to the lack of parallaxes/distances to support the 
variability analysis of these sources at this early stage of the {\it Gaia} mission, this occurrence, along with the high cadence of the EPSL observations, were important assets that helped the SOS Cep\& RRL 
processing for the first {\it Gaia} data release, which specifically focused on   
 the 19,923  
   Cepheids and RR Lyrae star candidates 
within 
 the {\it Gaia} SEP marked in blue in Figs.~\ref{SEP-coverage} and \ref{SEP-coverage-bis}. 

\begin{figure}
   \centering
   \includegraphics[width=9.5 cm, trim=30 150 0 100,clip]{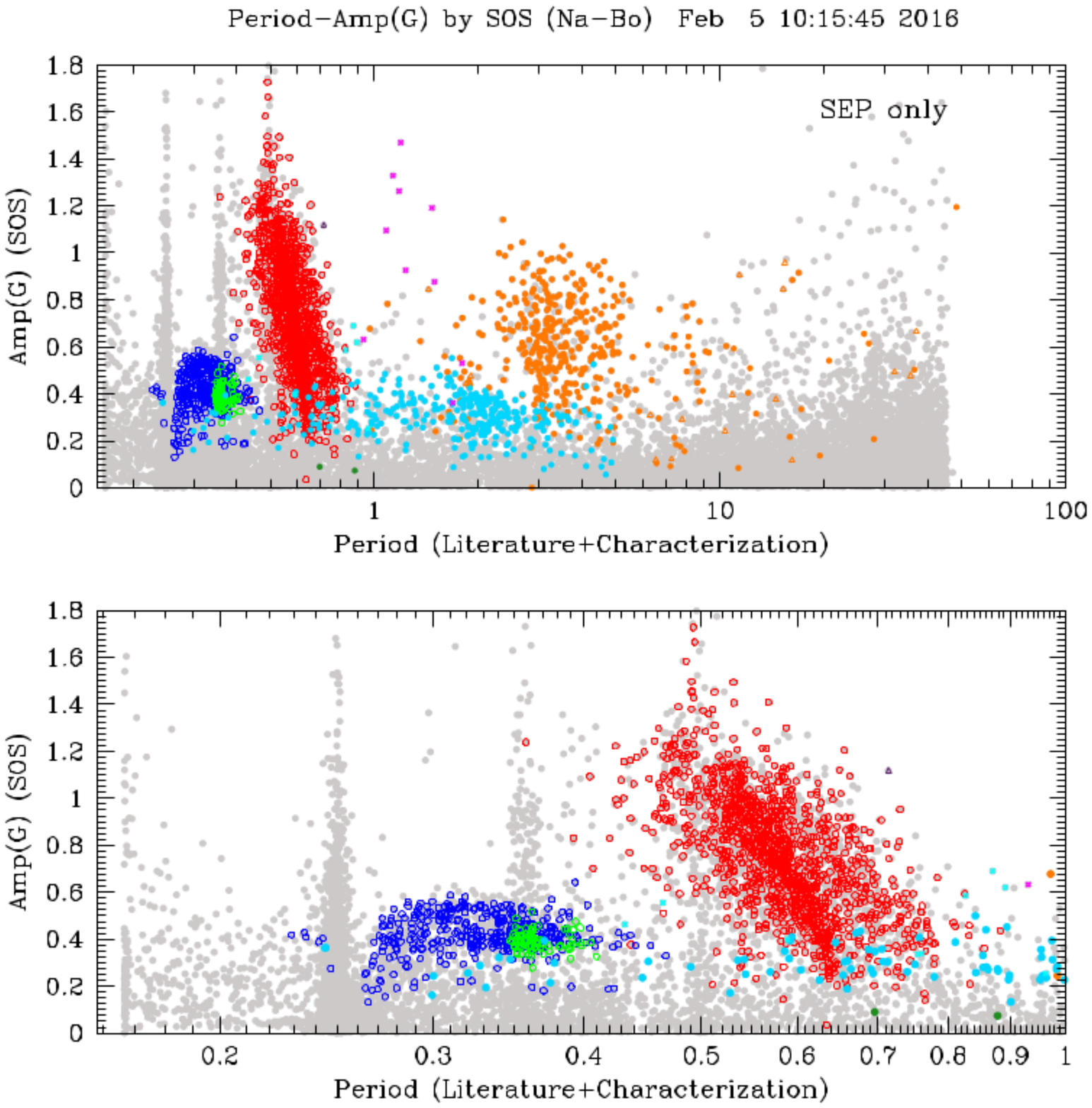}
   \caption{$P$ - $G$-band amplitude distribution of 19,923 
    candidate Cepheids and RR Lyrae stars (grey filled cicles) identified by 
    {\it Classification} pipeline of the general variability processing in the C0 
   photometry of the {\it Gaia} SEP. Units are, days and magnitude on the X and Y axis, respectively. The bottom panel is an expansion of the top panel for the region of $P \le $1 d. Orange and cyan filled circles  mark  F and 1O Cepheids, red, blue and green open circles are RRab, RRc and RRd pulsators known from OGLE-III and 
   OGLE-IV GSEP catalogues. Amplitudes are  values computed by running the SOS Cep\&RRL pipeline on C0 photometry. The strong peak around  $P$=0.25 d is an alias  
due to the  rotation period of {\it Gaia} around its axis. 
  }
              \label{pampSEP}%
    \end{figure}

The  scanning law determines the cadence of {\it Gaia} multi-epoch observations and, 
in turn, has a bearing on the alias patterns we may expect to show up in the 
power spectrum of {\it Gaia} time-series data.  
Particularly strong 
 is the 6 hour alias 
caused by the  
 spacecraft rotation around its axis, that during EPSL was always kept in the ecliptic plane and fixed with respect to the {\it Gaia}-Sun  axis (see section~3 
  in Eyer et al. 2016).
 This is clearly seen in the two panels of Fig.~\ref{pampSEP} that show the period - $G$-band amplitude distribution  of the 19,923 
  candidate RR Lyrae stars and Cepheids in the {\it Gaia} SEP.  
    Sources are plotted using the period computed by the {\it Characterization} workpackage of the general variability  pipeline for Cepheid and RR Lyrae star candidates and OGLE's period for the known variables. 
   The figure also shows that only very few of the known 
   Cepheids and RR Lyrae stars in this region of the LMC have $G$-band amplitude smaller than 0.1 mag, hence suggesting that the large majority of candidates with such small amplitudes 
  are misclassifications, lending support to the choice to not consider them any further (see Section~\ref{tailoring}). 
    Test runs of the SOS Cep\&RRL pipeline 
    along with visual inspection of a randomly selected sample of the 19,923 SEP Cepheid and RR Lyrae star candidates during this initial stage of the processing 
     helped to get a general overview of the dataset,  improve the global  analysis and fine tune the classifiers' predictions. The visual inspection also confirmed a few (5) candidates that were later 
     added to the final counts. 

By excluding sources flagged as RR Lyrae  star and Cepheid outliers in the predictions of 
the used classifiers  the sample of automatically identified Cepheid and RR Lyrae star candidates 
reduces to 9,347 sources.  
 Then further 
rejecting sources with $G$-band amplitudes smaller than 0.1 mag (more than 3,800) but including 
the aforementioned 5 candidates identified by visual inspection, 
the number of SEP Cepheid and RR Lyrae star candidates drops to 6,714 sources.
Of them 6,053 were recovered by cross-matching  the C0 and C1 photometries.
%
This forms  the  sample finally processed through the SOS Cep\&RRL pipeline.

Statistical properties (time sampling, number of observations, time-series duration, mean/median $G$ magnitudes, magnitude uncertainties, etc.)  of the $G$-band C0 time-series photometry 
out of which the initial 19,923 enlarged sample of Cepheid and RR Lyrae star candidates  analyzed by the SOS Cep\&RRL pipeline was extracted are described in great detail in section~5.2 and   figs.14, 15, 16, 17 of \citet{eyer16}.  Whereas, the characteristics of  the $G$-band C1 time-series photometry of the  sample of 6,053  sources finally processed through the SOS Cep\&RRL pipeline and turned into the  3,194 confirmed Cepheids and RR Lyrae stars released in {\it Gaia} DR1 are described in section~6.2, figs. 27, 28, 30, 31 and table~3 of \citet{eyer16}.

We briefly remind here that this latter sample spans the $G$-band magnitude range 12 $\lesssim \langle G \rangle \lesssim $ 20.3 mag, with typical errors on the order of  a few milli-mag and $\gtrsim$ 0.02 mag per individual datapoint at bright and faint ends, respectively, $G$-bands amplitudes in the range from $\sim$ 0.1 and 1.5 mag, and between 15 and over 200 phase-point sampling of the light curves (see text and figures in Section~\ref{results2} for details).

\begin{figure}
   \centering
   \includegraphics[width=9 cm, trim= 10 190 30 130, clip]{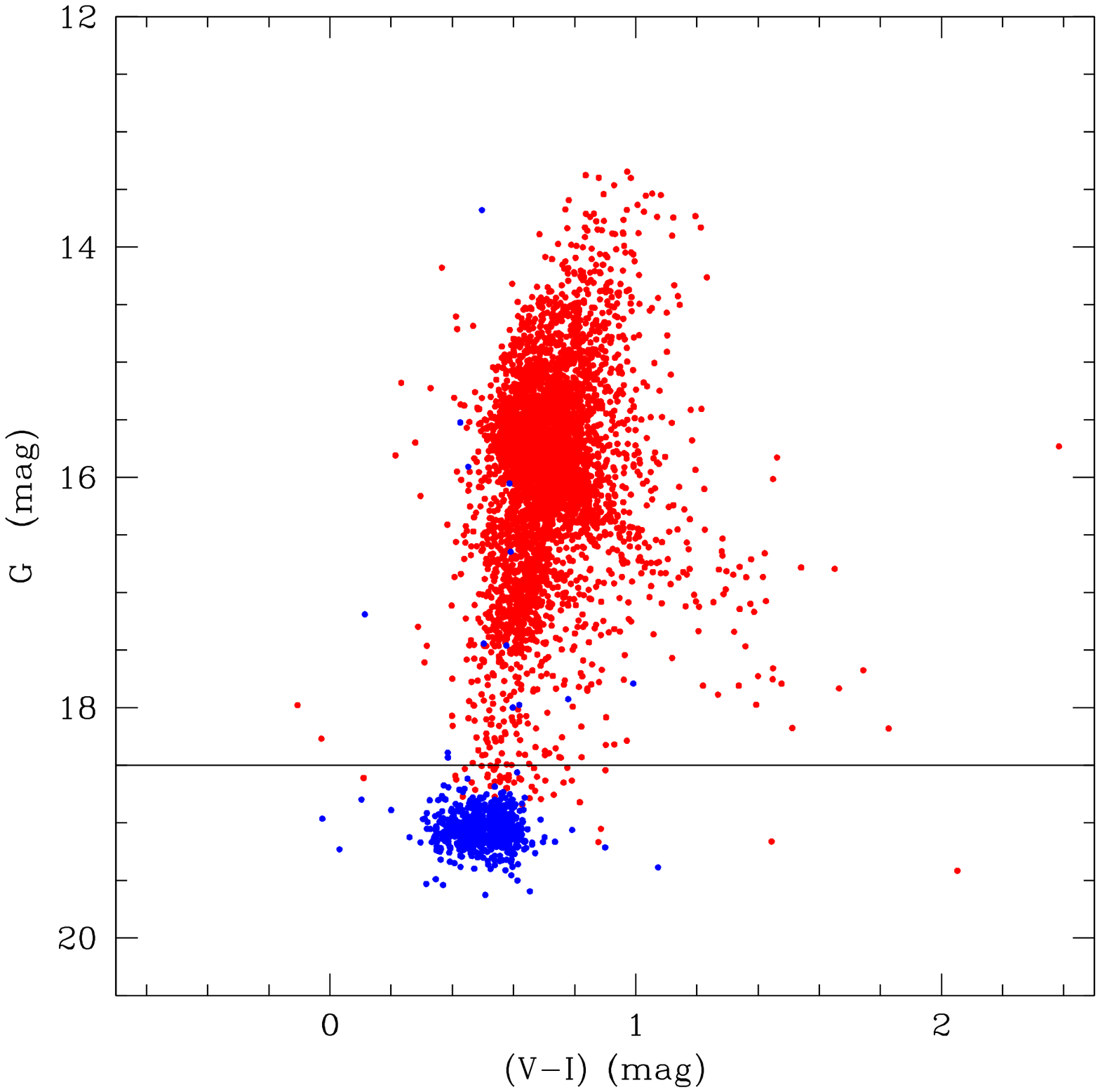}
   \caption{$G$-band versus $V -I$ colour-magnitude diagram of RR Lyrae stars (blue filled circles) and  Cepheids (red filled circles) in the LMC field, using the OGLE-IV GSEP catalogue for the RR Lyrae stars (\citealt{soszynski12}) and 
OGLE-IV LMC Cepheids  according to \cite{soszynski15a,soszynski15b,soszynski15c}.}
              \label{cmdSEP}%
    \end{figure}

Integrated blue ($G_\mathrm{BP}$) and red ($G_\mathrm{RP}$) spectrophotometry 
from the blue and red photometers on board {\it Gaia} is not released in {\it Gaia} DR1. 
  This additional limitation of the {\it Gaia} DR1 dataset means we could not build the {\it Gaia} CMD  of the SEP 
Cepheid and RR Lyrae star candidates.  
However, we built a $G$-band versus $V - I$ CMD, shown in Fig.~\ref{cmdSEP}, using OGLE-IV's GSEP catalogue of LMC RR Lyrae stars (\citealt{soszynski12}) and OGLE-IV's catalogues of  LMC Cepheids published by  \cite{soszynski15b,soszynski15c}. The mean $V$ magnitudes published by OGLE were transformed to  the {\it Gaia} $G$-band using eq.~(A.1). This gave us some insight on the  range 
in $G$ magnitude spanned by RR Lyrae stars and Cepheids  in the {\it Gaia} SEP (see Section~\ref{tailoring}). 
  
The C1 time-series photometry of the  6,053 Cepheid and RR Lyrae star candidates, selected as described in Section~\ref{s3}, 
 was fed into the SOS Cep\&RRL pipeline. 
 As a first step sources were  cross-matched against catalogues of known RR Lyrae stars and Cepheids existing in the literature. We primarily used 
 the OGLE-III (\citealt{soszynski09}) catalogue for RR Lyrae stars\footnote{The OGLE-IV catalogue of the LMC RR Lyrae stars (\citealt{soszynski16}) became available after 
 the SOS Cep\&RRL processing and been finished with discovery of over 1,000 new LMC RR Lyrae stars, of which 2/3 turned out to be in common with \citet{soszynski16}.}
 OGLE-IV catalogue (\citealt{soszynski15a,soszynski15b,soszynski15c}) for Cepheids and EROS-2 catalogue of \cite{kim14}. 
 Then we also used  RR Lyrae stars and Cepheids identified in the SEP region by the Catalina (\citealt{torrealba15},) and ASAS (\citealt{pojmanski97}) surveys and also checked the online version of the GVCS (http://www.sai.msu.su/gcvs/gcvs/ ) and the Simbad catalogue 
 (http://simbad.u-strasbg.fr/simbad/). 

\subsection{Tailoring the SOS Cep\&RRL pipeline for the analysis of the SEP data}\label{tailoring} 
Given the limitations of the first {\it Gaia} data, the SOS Cep\&RRL pipeline had to be  properly tailored to successfully process the {\it Gaia} DR1 data. This occurred through a number of assumptions and simplifications which impacted significantly the pipeline effectiveness and enhanced the need for visual inspections and manual intervention to check and adjust results (see Fig.~\ref{flow-chart-p4}). 
Specifically,
\begin{itemize}
\item Lacking parallaxes (distances) for the stars in the sample we assumed that the SEP sources were all the same distance from us, which is an acceptable first approximation given that they are located primarily in the LMC, and we adopted  a cut in apparent magnitude to initially separate Cepheids from RR Lyrae stars. A threshold was set using the CMD in Fig.~\ref{cmdSEP} by which we assumed that candidates with $\langle G \rangle >$ 18.5 mag are more likely RR Lyrae stars and after the general trunk (Fig.~\ref{decomposition1}) they were sent to the RR Lyrae branch (Fig.~\ref{decomposition2}) of the
SOS Cep\&RRL pipeline. Conversely, candidates with $\langle G \rangle <$ 18.5 mag were sent to the Cepheid branch (Fig.~\ref{decomposition3}). However, since the SEP sample might  include also Galactic RR Lyrae stars projected against the LMC, candidates brighter than 18.5 mag,  with $P <1$ d and $G$-band amplitude larger than 0.5 mag were analyzed in both Cepheid and RR Lyrae branches, as the two types overlap in this region of the parameter space. A fine tuning of the Fourier parameters and definition of strict loci in the $\phi_{21}$ {\it vs} $P$ and $R_{21}$ {\it vs} $P$ planes 
 ( see Section~\ref{results2})  were adopted to distinguish bright MW  RR Lyrae  stars from Cepheids.\\ 

\item Lacking source colours, the classification of Cepheids in different types and  pulsation modes could only rely on the $PL$ relations, complemented by visual inspection of the 
Fourier planes. Furthermore,  without colours, the contaminations by  ECLs (both for Cepheids and RRc) and by $\delta$ Scuti (particularly at the faint end of the 
LMC RR Lyrae star distribution, $\langle G \rangle \gtrsim$ 20.0 mag) were issues  that only the visual inspection of the light curves could alleviate.\\

\item The limited time span  covered by the {\it Gaia} DR1 data, whose bulk spans roughly 28 d, made the measure of period most reliable only for Cepheids with periods shorter than about 10 d.  In addition, the strong aliases  affecting the SEP data  required   visual inspection of the light curves and manual intervention (period or type tagging of sources) to solve ambiguous cases. 
\end{itemize}

\noindent When $G_\mathrm{BP}$, $G_\mathrm{RP}$ colours, parallaxes and an increasingly complete dataset  will become available starting already with {\it Gaia} DR2   in 2017, the need for visual inspection and manual intervention will definitely decrease. 

\begin{figure}
   \centering
 \includegraphics[width=9 cm, trim= 0 190 30 365, clip]{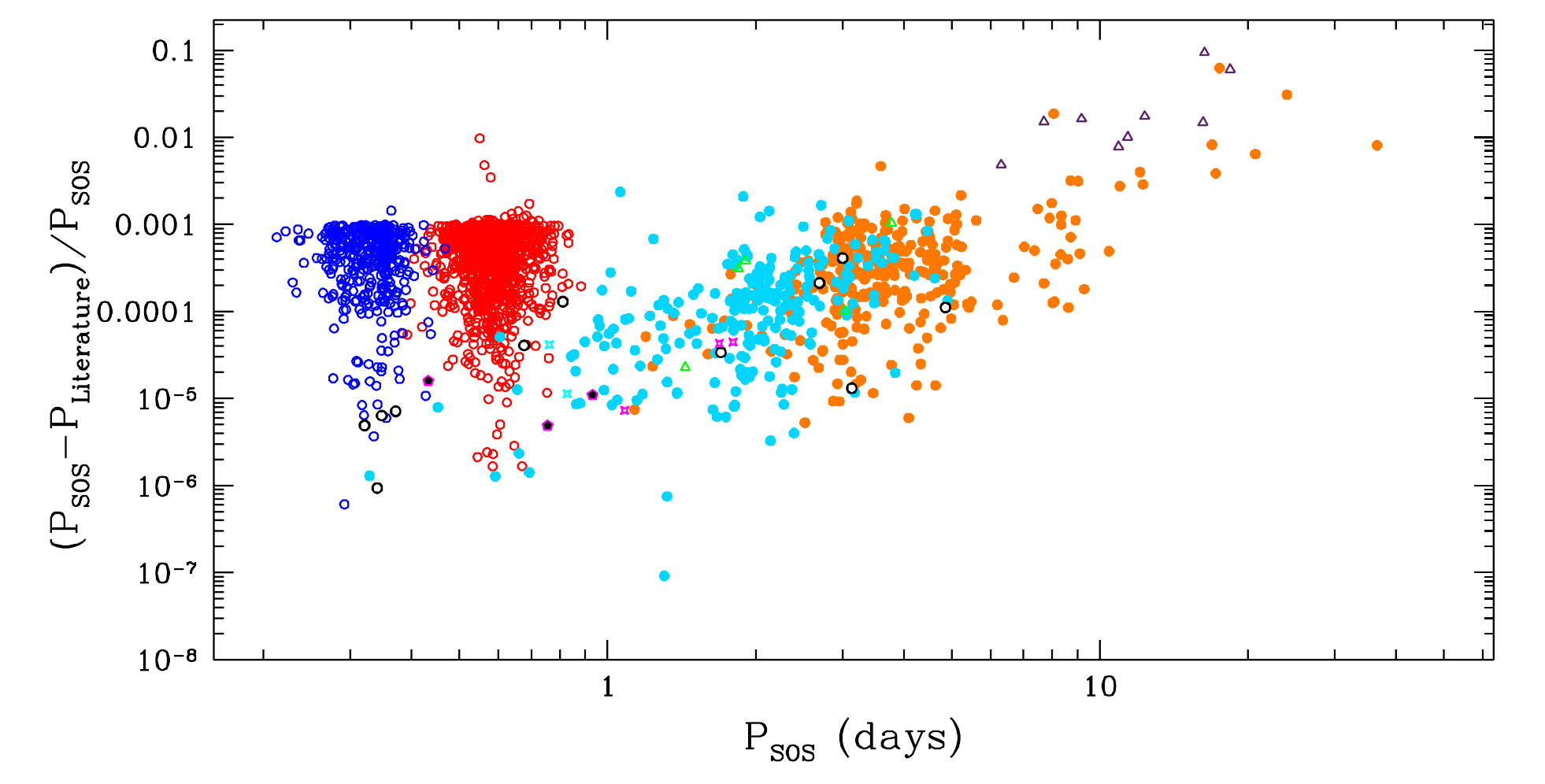}
   \caption{Difference between the period computed by the SOS Cep\&RRL pipeline (P$_{\rm SOS}$) and the literature period for known RR Lyrae stars and Cepheids in the SEP region, plotted versus 
 P$_{\rm SOS}$. Blue open circles: RRc; red open circles: RRab; cyan filled circles:  DECPs 1O; orange filled circles: DCEPs F; cyan four starred symbols: ACEPs F;  magenta four starred symbols: ACEPs 1O; magenta pentagons filled in black: ACEPs without mode identification; green open triangles: BLHER;  violet open triangles: WVIR; magenta open triangles: RVTAU; black open circles: DCEPs and T2CEPs without mode or subtype identification.}
              \label{dp}%
    \end{figure}

For  known Cepheids and RR Lyrae stars the  periods measured by the SOS Cep\&RRL pipeline (hereinafter, $P_{\rm SOS}$) were compared with the literature values.  Agreement was deemed satisfactory if differences 
 were within 0.01 d for the Cepheids and  0.001 d for RR Lyrae stars. 
When differences exceede these limits visual inspection of the light curve folded according to the two different periods helped to decide  which 
periodicity to finally adopt\footnote{For new RR Lyrae stars and Cepheids discovered by {\it Gaia}, such a  check was performed by  comparing P$_{\rm SOS}$ and the period computed by  
 {\it Characterization}  
 in the general variable star analysis pipeline.}.  
Fig.~\ref{dp} shows the  comparison with the literature of the periods finally adopted  
for  the  known RR Lyrae stars and Cepheids released in {\it Gaia} DR1. The flattening of RR Lyrae stars/Cepheids around  $\Delta P$=0.001/0.01 d 
is because we have assumed that $P_{\rm SOS}$ and literature period are in agreement if
they differed by less than 0.001/0.01 d for the RR Lyrae stars and  Cepheids, respectively.  A few sources above these limits in Fig.~\ref{dp}  are variables for which $\Delta P>$0.001/0.01 d and P$_{\rm SOS}$  produces a better folding of the light curve than the literature period.  
 Visual inspection of all light curves during final assessment of the SOS Cep\&RRL results  confirms that the comparison procedure described above worked well for the present dataset which does not contain many long period (above 10 d) Cepheids.  
 
Type and mode classification  of the known RR Lyrae stars and Cepheids  assigned by the  SOS Cep\&RRL pipeline were also cross-checked against literature. A number of obvious 
type misclassifications were emended by manual adjustment.  

After processing through the SOS Cep\&RRL pipeline the light curves of all  6,063 candidates  
were visually inspected for data quality assurance, validation of the modelled light curve and quantities thereby derived 
(period, peak-to-peak amplitude, mean magnitude, etc.)  and final assessment of the Cepheids and RR Lyrae stars to be published in {\it Gaia} DR1.

The run of the SOS Cep\&RRL pipeline on the 6,053 candidates combined with visual inspection of the resulting light curves produced a final sample of 3,602 confirmed 
Cepheids and RR Lyrae stars and rejection of a total number of 2,451 sources. 
This quite large number of rejections is due to various reasons among which are:  
candidates identified in C0 which are not present in C1 photometry,  photometry issues in the time series data among which very large outliers and/or systematically noisy light curves. However, a major and most important  reason for rejection was incomplete sampling of the light curves resulting in unreliable values for the source period, peak-to-peak amplitude and mean magnitude. Finally,  further rejections were also due to  contamination of the sample by ECLs, and other types of variables ($\delta$ Scuti, LPVs, etc.). We explicitly note that a large fraction of these rejections (over 50\%) are in fact bona fide 
RR Lyrae stars and Cepheids  for which unambiguous 
determinations of period and/or type classification could be obtained,  or are  RR Lyrae stars and Cepheids that did not pass the strict quality control criteria we  have adopted 
 to validate  
 these first  variability results from {\it Gaia}. 
During quality assessment,  it was also decided not to release results  
for double/multi-mode sources (239  in total between RR Lyrae stars and Cepheids)  as they were not fully compliant with 
 our internal validation criteria. These are quite strict to ensure 
a high reliability of  the information  for variable sources released  in {\it Gaia} DR1, notwithstanding the actual limitations of the available dataset.
By rejecting double-mode RR Lyrae  stars and multi-mode Cepheids the sample reduced to 3,363 sources. 
Finally, 169  additional sources, out of the 3,363 sample, were rejected  during the final 
validation process 
 which was meant to achieve consistency on the total number of sources 
finally released by different tasks of the DPAC processing contributing results for {\it Gaia} DR1.
 This further trimming lead to a final sample of 3,194 Cepheids and RR Lyrae stars.

\section{Results}\label{results2}
Position, $G$-band time series photometry and final results of the 
SOS Cep\&RRL pipeline are published in {\it Gaia} DR1  for a total  of 3,194 sources. 
They comprise 599 Cepheids (of which 43 are new from {\it Gaia}) and 2,595 RR Lyrae stars (of which  343 are new discoveries).
The subdivision of the 3,194 sources according to type, subtype and pulsation mode is  summarized in Table~\ref{table:2}. 
\begin{table}[h!]
\caption{Number and type/mode classification of RR Lyrae stars and Cepheids published in {\it Gaia} DR1}             
\label{table:2}     
\centering                         
\begin{tabular}{l r r r }       
\hline\hline                 
\noalign{\smallskip}
Type & Total & New  & Known \\ 
\hline                          
\noalign{\smallskip}
RRab & 1910 & 228 & 1682\\  
RRc & 685 & 115 & 570\\  
\hline                        
\noalign{\smallskip}
RR Lyrae Total &   2595   &  343    &  2252   \\
\hline                        
\noalign{\smallskip}
DCEP F       & 313 & 12 & 301 \\ 
DCEP 1O       & 230 & 10 & 220 \\ 
DCEP NoMode\tablefootmark{\rm (a)}   & 15 & 4 & 11 \\ 
\hline                          
\noalign{\smallskip}
DCEP Total   & 558  & 26 & 532 \\ 
\hline                          
\noalign{\smallskip}
ACEP F       & 6 & 3 & 3 \\  
ACEP 1O       & 7 & 5 & 2 \\  
ACEP NoMode\tablefootmark{\rm (a)}   & 3 & 0 & 3 \\  
\hline                          
\noalign{\smallskip}
ACEP Total   & 16  & 8 & 8 \\ 
\hline                       
\noalign{\smallskip}
T2CEP BLHER       & 11 & 6 & 5 \\  
T2CEP WVIR      & 10 & 1 & 9 \\   
T2CEP RVTAU      & 2 & 2 & 0 \\   
T2CEP NoSubType\tablefootmark{\rm (a)}   & 2 & 0 & 2 \\  
\hline                          
\noalign{\smallskip}
T2CEP Total   & 25  & 9 & 16 \\ 
\hline
\noalign{\smallskip}
Cepheid Total & 599 & 43 & 556\\
\hline                                  
\end{tabular}
\tablefoot{$^{\rm (a)}$ With the terms NoMode and NoSubType, we have flagged sources for which the pulsation mode (for DCEPs and ACEPS) or the subtype (for T2CEP) could not be automatically and univocally 
identified by the SOS Cep\&RRL pipeline.}\\
\end{table}

\noindent For these 3,194 sources the  following parameters, computed by the SOS Cep\&RRL  pipeline,  are released in {\it Gaia} DR1  along with the related errors:\\ 

\noindent 
- source pulsation period\\
- intensity-averaged mean $G$ magnitude\\
- epoch of maximum light\\
- $\phi_{21}$ and $R_{21}$ Fourier parameters\\
- peak-to-peak $G$-band amplitude [Amp($G$)]\\
- RR Lyrae star subclassification into RRab and RRc types\\
- Cepheid classification into DCEP, ACEP, T2CEP types\\
- DCEPs and ACEPs pulsation mode  (F, 1O)\\
- T2CEPs sub-classification into BLHER, WVIR, RVTAU types.\\

Gaia's sourceids, coordinates, values of the above quantities and associated statistics, along with the $G$-band time series for each of the 3,104 sources can be retrieved from  {\it Gaia} first data release archive:
\texttt{http://archives.esac.esa.int/gaia/}
 and other distribution nodes. The archives also provide tools for queries and to crossmatch Gaia data with other catalogues available in the literature.
 
 We provide in Table~\ref{table:3a} specific links to the archive tables as well as summarise names of the parameters, computed by
SOS Cep\&RRL that can be retrieved from the archive tables. In Section~\ref{app:queries} we provide queries to retrieve the various quantities and parameters listed in Table~\ref{table:3a}.

\begin{figure}
   \centering
   \includegraphics[width=9 cm, trim= 0 190 30 130, clip]{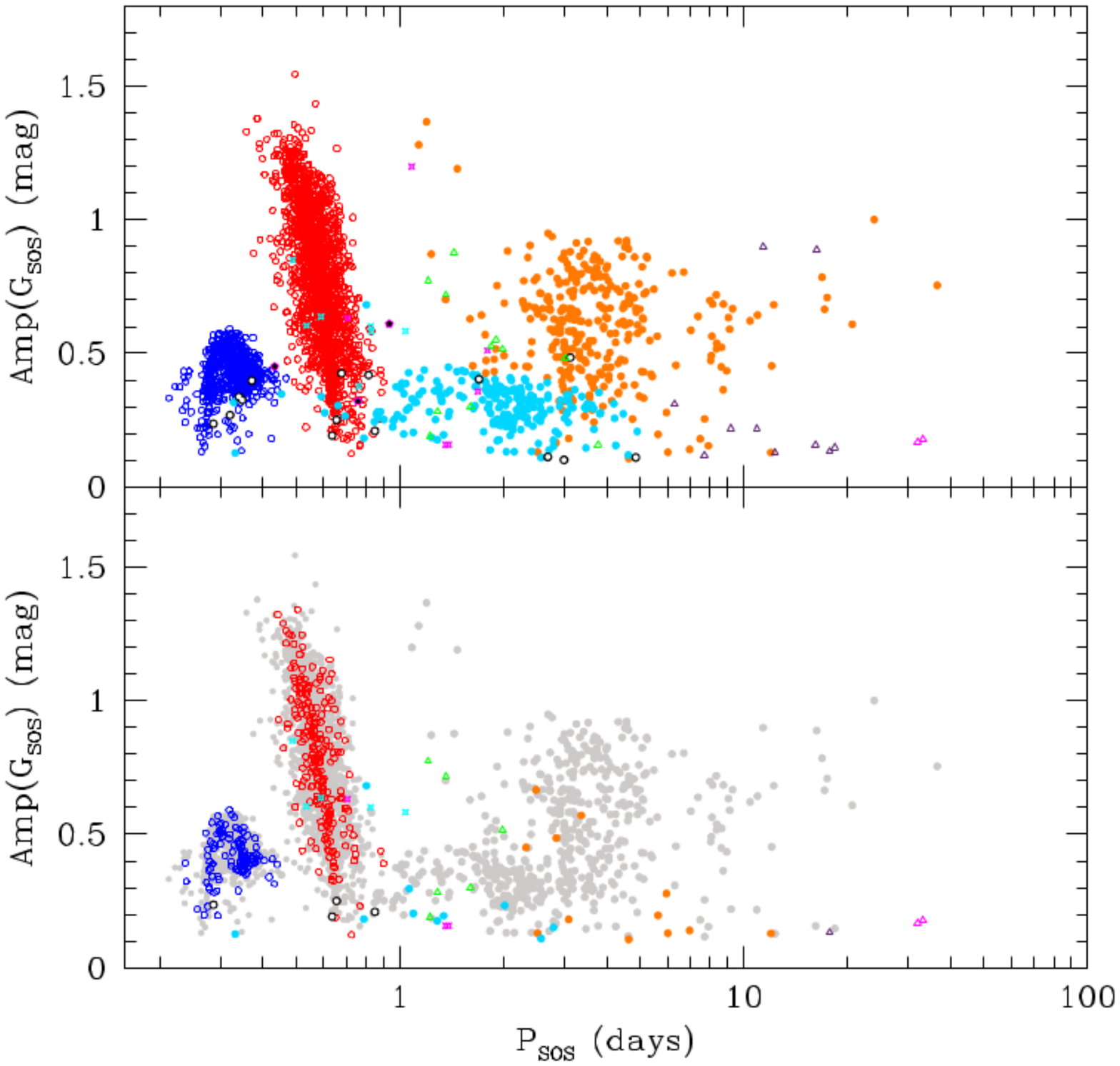}
   \caption{Period-$G$-band amplitude diagram of the Cepheids and RR Lyrae stars published in {\it Gaia} DR1.  The upper panel shows all 3,194 sources, symbols and colour coding are as in Fig.~\ref{dp}. In the lower panel  new discoveries by {\it Gaia} are plotted in colour, known variables in grey.}
              \label{pamp}%
    \end{figure}
\begin{figure}
   \centering
   \includegraphics[width=9 cm, trim= 10 190 30 130, clip]{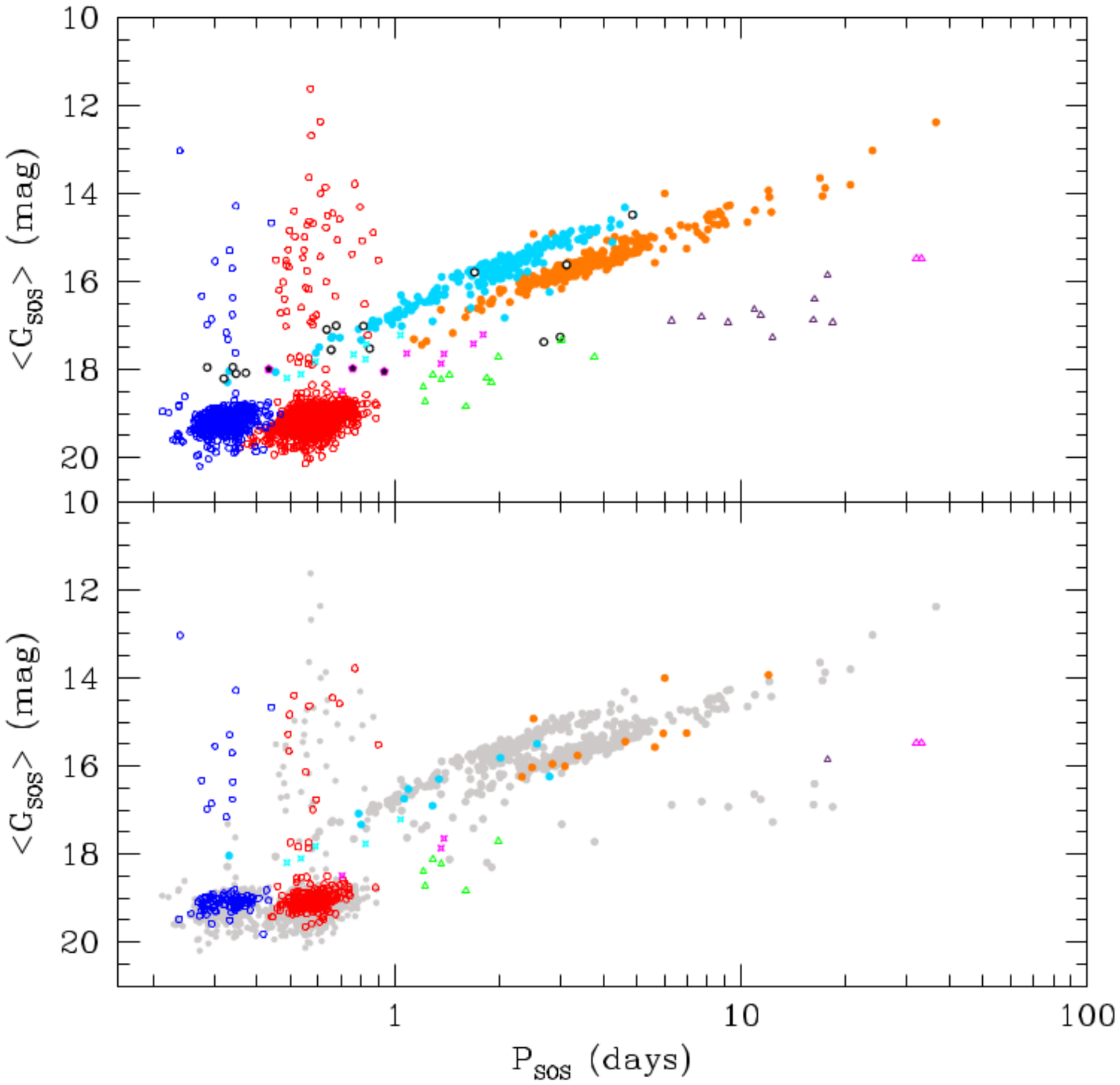}
   \caption{$G$-band $PL$ distribution of the 3,194 Cepheids and RR Lyrae published in {\it Gaia} DR1. 
   The upper panel shows all 3,194
sources, symbols and colour coding are as in Fig.~\ref{dp}. In the lower panel
new discoveries by Gaia are plotted in colour, known variables in grey.}
              \label{ppl_cep1}%
    \end{figure}
   \begin{table}
\tiny
\setlength\tabcolsep{5pt}
\caption{Links to {\it Gaia} archive tables to retrieve the $G$-band time series photometry and the pulsation characteristics: period, epoch of maximum light (E), peak-to-peak $G$ amplitude [Amp(G)], intensity-averaged mean $G$ magnitude, $\phi_{21}$, $R_{21}$ Fourier parameters and related uncertainties, computed by the SOS Cep\&RRL pipeline for the 599  Cepheids and 2,595 RR Lyrae stars published in {\it Gaia} DR1.
To ease the data retrieval, we also provide the  correspondence between parameter (e.g. period, E, etc.) and the name of the parameter in the {\it Gaia} archive table.}
\label{table:3a}     
\centering                         
\begin{tabular}{ll}       
\hline\hline                 
\noalign{\smallskip}
Table URL & \texttt{http://archives.esac.esa.int/gaia/}\\
\noalign{\smallskip}
\hline
\noalign{\smallskip}
\multicolumn{2}{c}{$G$-band time series data}\\
\hline
\noalign{\smallskip}
Table Name &  \texttt{\textbf{phot\_variable\_time\_series\_gfov}}\\
Table Content & \textit{G}-band FoV averaged photometry:  \texttt{observation\_time}, \\
 & \texttt{g\_flux}, \texttt{g\_flux\_error}, \texttt{g\_magnitude}, \\
 & \texttt{rejected\_by\_variability\_processing} \\
\noalign{\smallskip}
\hline
\noalign{\smallskip}
Table Name & phot\_variable\_time\_series\_gfov\_statisti\linebreak cal\_parameters\\
$N_{\rm obs}$ & \texttt{num\_observations\_processed}\\
\noalign{\smallskip}
\hline
\noalign{\smallskip}
\multicolumn{2}{c}{Cepheid parameters computed by the SOS Cep\&RRL pipeline}\\
\hline
\noalign{\smallskip}
Table Name&  \texttt{\textbf{cepheid}}\\
Source ID & \texttt{source\_id}\\
Type& \texttt{type\_best\_classification} \\
&  (one of \texttt{DCEP}, \texttt{T2CEP} or \texttt{ACEP})\\
Mode& \texttt{mode\_best\_classification} \\
 & (one of \texttt{FUNDAMENTAL}, \texttt{FIRST\_OVERTONE}, \\
 & \texttt{SECOND\_OVERTONE}, or \texttt{UNDEFINED}, only for \texttt{DCEP} or \texttt{ACEP})\\
Subtype&\texttt{type2\_best\_sub\_classification} \\
& (\texttt{BL\_HER}, \texttt{W\_VIR}, or \texttt{RV\_TAU}, only for \texttt{T2CEP})\\
$P$ &\texttt{p1} \\
$\sigma (P)$&\texttt{p1\_error} \\
$N_{\rm harm}$&\texttt{num\_harmonics\_for\_p1}\\
E\tablefootmark{\rm (a)}&\texttt{epoch\_g} \\
$\sigma {\rm E}$ &\texttt{epoch\_g\_error}\\
$\langle G \rangle$&\texttt{int\_average\_g}\\
$\sigma \langle G \rangle$ &\texttt{int\_average\_g\_error} \\
Amp($G$)&\texttt{peak\_to\_peak\_g}\\
$\sigma {\rm [Amp(}G{\rm )]}$&\texttt{peak\_to\_peak\_\linebreak g\_error}\\
$\phi_{21}$ &\texttt{phi21\_g}\\
$\sigma (\phi_{21})$ &\texttt{phi21\_\linebreak g\_error} \\
$R_{21}$ &\texttt{r21\_g}\\
$\sigma (R_{21})$&\texttt{r21\_g\_error}\\ 
\noalign{\smallskip}
\hline
\noalign{\smallskip}
\multicolumn{2}{c}{RR Lyrae star parameters computed by the SOS Cep\&RRL pipeline}\\
\hline
\noalign{\smallskip}
Table Name&  \texttt{\textbf{rrlyrae}}\\
Source ID & \texttt{source\_id}\\
Type& \texttt{best\_classification} (one of \texttt{RRC} or \texttt{RRAB})\\
$P$ &\texttt{p1} \\
$\sigma (P)$&\texttt{p1\_error} \\
$N_{\rm harm}$&\texttt{num\_harmonics\_for\_p1}\\
E\tablefootmark{\rm (a)}&\texttt{epoch\_g} \\
$\sigma {\rm E}$ &\texttt{epoch\_g\_error}\\
$\langle G \rangle$&\texttt{int\_average\_g}\\
$\sigma \langle G \rangle$ &\texttt{int\_average\_g\_error} \\
Amp($G$)&\texttt{peak\_to\_peak\_g}\\
$\sigma {\rm [Amp(}G{\rm )]}$&\texttt{peak\_to\_peak\_\linebreak g\_error}\\
$\phi_{21}$ &\texttt{phi21\_g}\\
$\sigma (\phi_{21})$ &\texttt{phi21\_\linebreak g\_error} \\
$R_{21}$ &\texttt{r21\_g}\\
$\sigma (R_{21})$&\texttt{r21\_g\_error}\\ 
\noalign{\smallskip}
\hline                                  
\end{tabular}
\tablefoot{$^{\rm (a)}$The BJD of the epoch of maximum light is offset by JD 2455197.5 d (= J2010.0).}\\
\end{table}

\begin{figure}
   \centering
  \includegraphics[width=9 cm, clip]{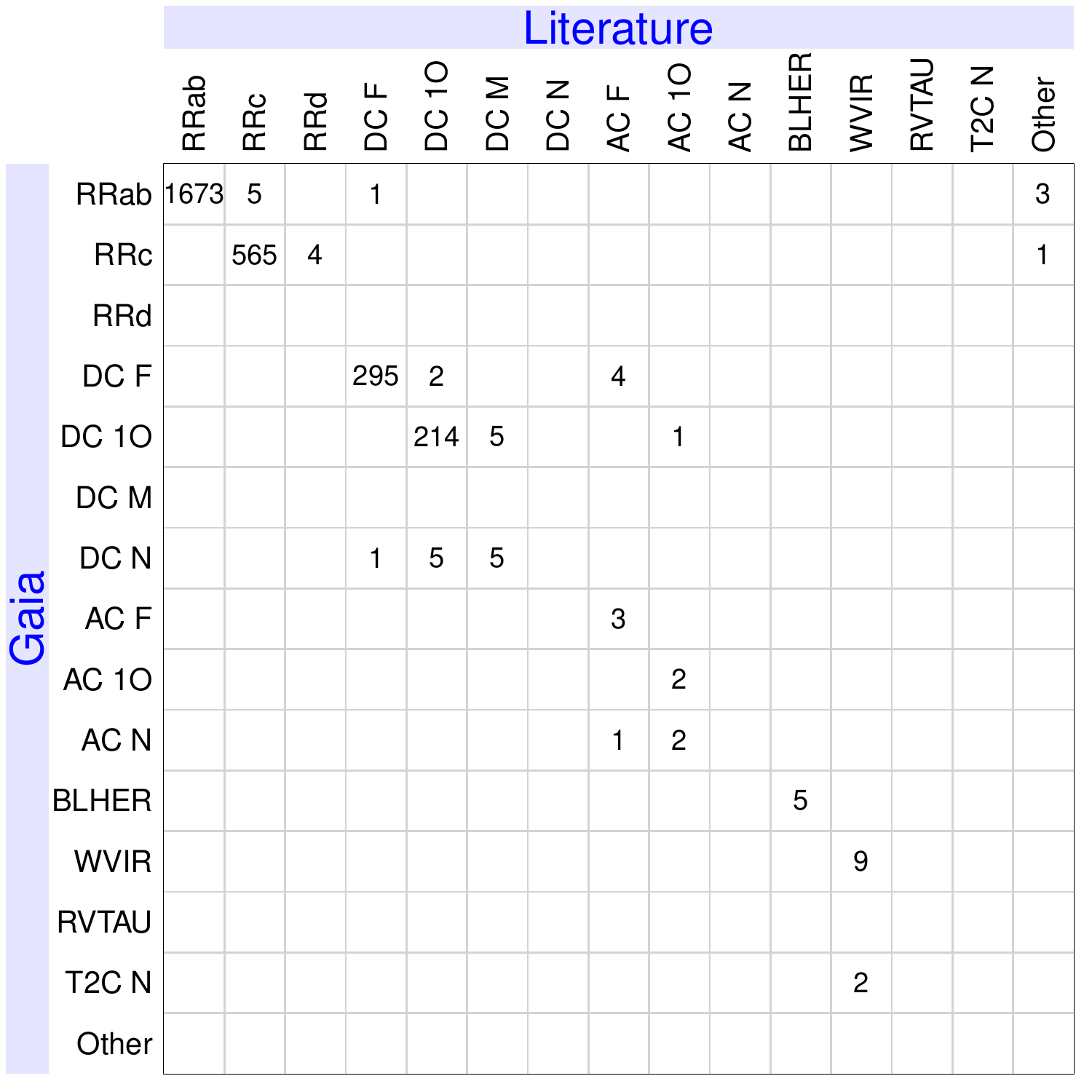}
   \caption{Confusion matrix of the comparison with the literature for 2,808 known Cepheids and RR Lyrae stars published in {\it Gaia} DR1. 
Rows refer to results of the SOS Cep\&RRL pipeline and columns to literature results. DC, AC, T2C indicate, respectively,  DCEPs, 
ACEPs  and T2CEPs; DC M indicates  multimode DCEPs.  DC N,  AC N and  T2C N indicate, respectively, DCEPs and ACEPs for which the 
pulsation mode was not identified and T2CEPs for which the subtype was not identified.}
\label{conf-matrix}%
\end{figure}

\begin{figure}
 \centering
    \includegraphics[width=9 cm, trim= 20 180 30 130, clip]{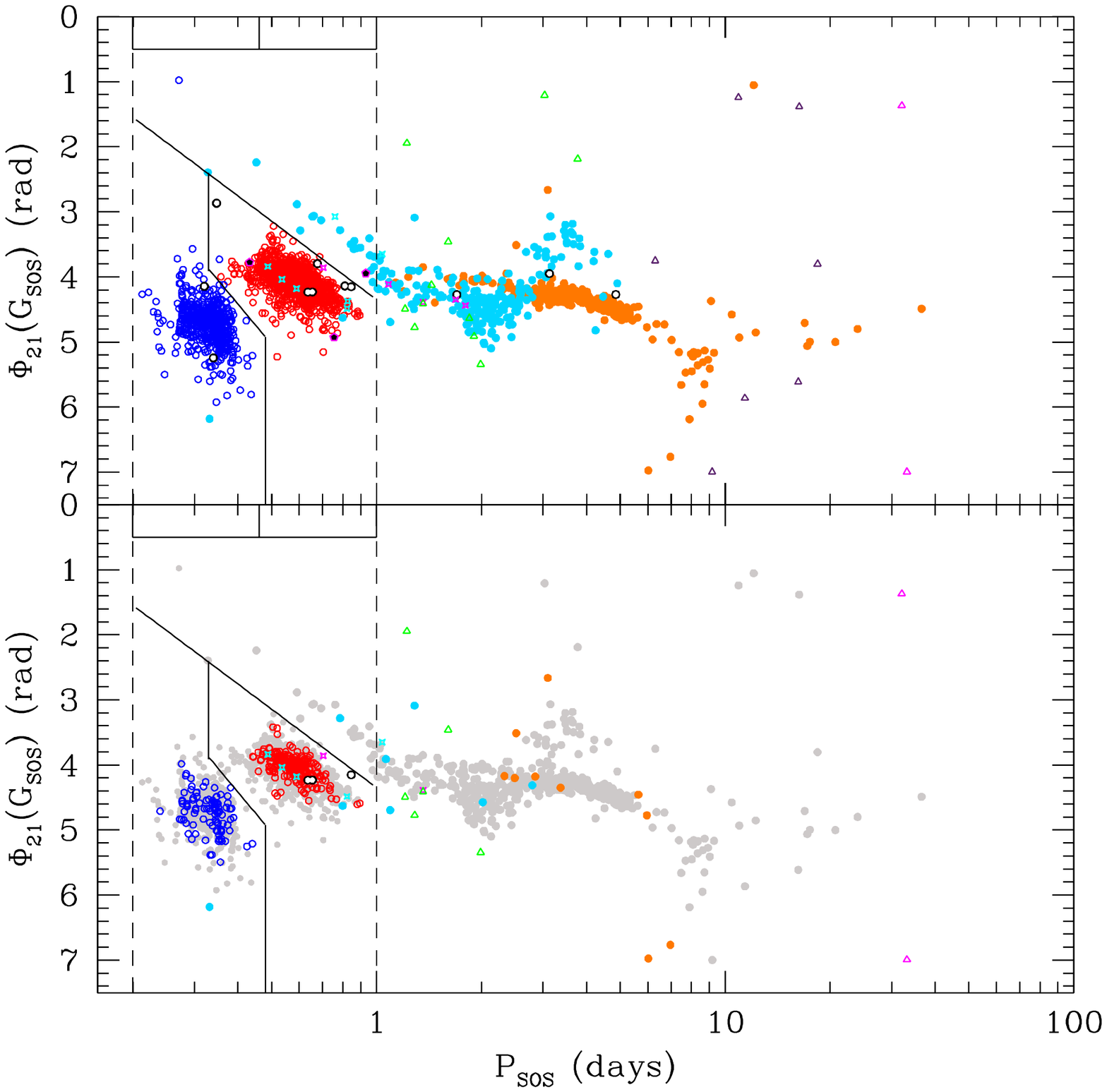}
   \caption{$\phi_{21}$ versus P$_{\rm SOS}$  distribution of the 3,194 Cepheids and RR Lyrae stars published in {\it Gaia} DR1. 
   Solid and dashed line show the loci that were set in the SOS 
   Cep\&RRL pipeline to separate  RR Lyrae stars from Cepheids and RRc from RRab pulsators. 
    The upper panel shows all 3,194
sources, symbols and colour coding are as in Fig.~\ref{dp}. In the lower panel
new discoveries by Gaia are plotted in colour, known variables in grey.
         }
              \label{p21cep}%
    \end{figure}

\begin{figure}
   \centering
   \includegraphics[width=9 cm, trim= 20 180 30 130, clip]{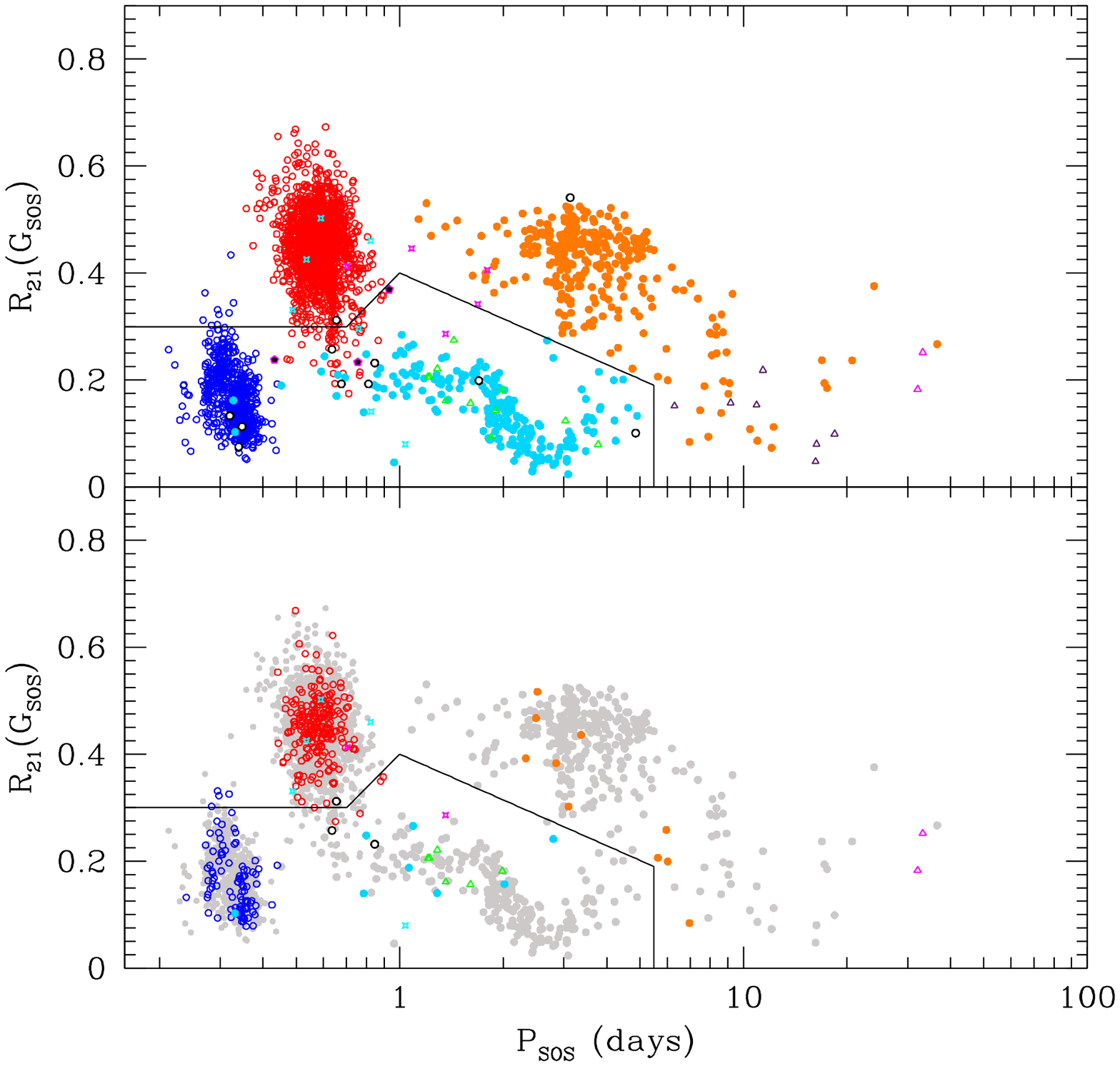}
   \caption{$R_{21}$ versus P$_{\rm SOS}$  distribution of the 3,194 Cepheids and RR Lyrae stars published in {\it Gaia} DR1. 
     Solid and dashed line show the loci that were set in the SOS Cep\&RRL pipeline to separate  RR Lyrae stars from Cepheids and to identify the Cepheid pulsation mode. 
     The upper panel shows all 3,194
sources, symbols and colour coding are as in Fig.~\ref{dp}. In the lower panel
new discoveries by Gaia are plotted in colour, known variables in grey.
}
              \label{r21cep}%
    \end{figure}

$PA$ and $PL$ distributions of the total sample of 3,194 sources are shown in Figs.~\ref{pamp} and ~\ref{ppl_cep1}, whereas  
Figs.~\ref{p21cep} and ~\ref{r21cep} show their distribution in the Fourier $\phi_{21}$ {\it vs} $P$ and  $R_{21}$ {\it vs} $P$ planes, respectively, according to the period
determined by the SOS Cep\&RRL pipeline ($P_{\rm SOS}$).  
The MW RR Lyrae stars are clearly recognized in Fig.~\ref{ppl_cep1} for $P <$ 1 d. This figure also shows the overlap between RR Lyrae stars, DCEPs and ACEPs, 
occurring in the short period regime ($P<$ 1d). 
 Of the 3,194 Cepheids and RR Lyrae stars published in {\it Gaia} DR1  2,808 are already known. Fig.~\ref{conf-matrix}  
shows the  comparison 
with the literature. The comparison  is done primarily with OGLE, for  which there is the largest number of sources in common 
(2,659 out of the 2,808 known RR Lyrae stars and Cepheids). Other surveys, among those listed at the end of Section~\ref{s3} were considered in case an OGLE 
classification was not available. Since contrasting classifications have been manually checked during the processing 
the number of sources in 
Fig.~\ref{conf-matrix}  
for which the SOS Cep\&RRL classification differs from the literature is limited and reflects actual 
differences resulting from the analysis based on {\it Gaia} data.

\begin{figure*}[h!]
\centering
\includegraphics[width=16.5 cm, trim= 10 130 10 130, clip]{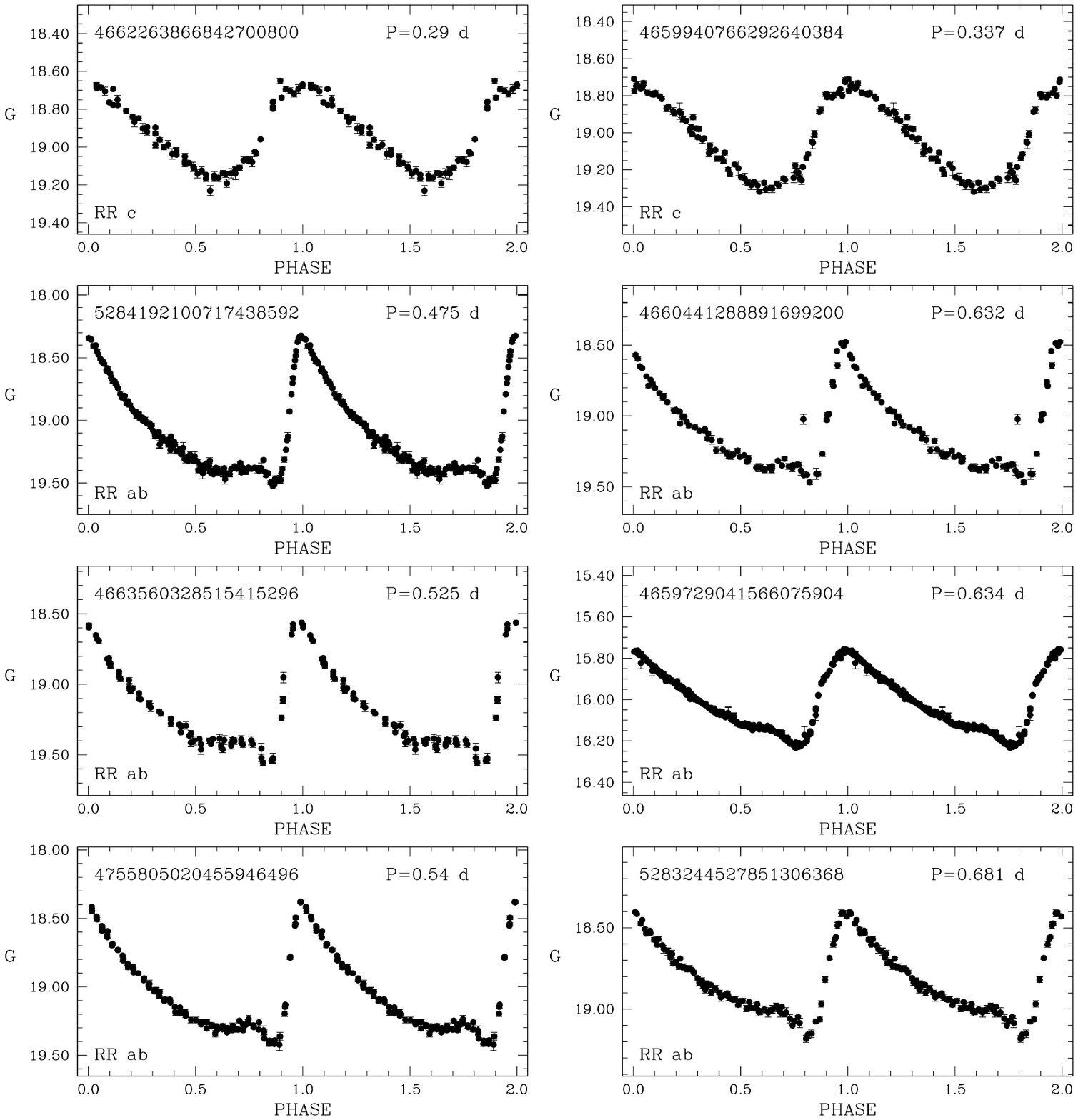}
  \caption{Examples of $G$-band  light curves of RRc (top two panels)  and RRab stars in the LMC and in the MW halo (third panel on the right).   Typical errors of individual measures at the magnitude level of the LMC RR Lyrae stars are on the order of $\gtrsim$ 0.02 mag. A few measurements  with errors  larger than 0.05 mag are not displayed.   
  }   
    \label{LC-RRL}%
    \end{figure*}

\begin{figure*}[h!]
   \centering
\includegraphics[width=16.5 cm, trim= 10 130 10 130, clip]{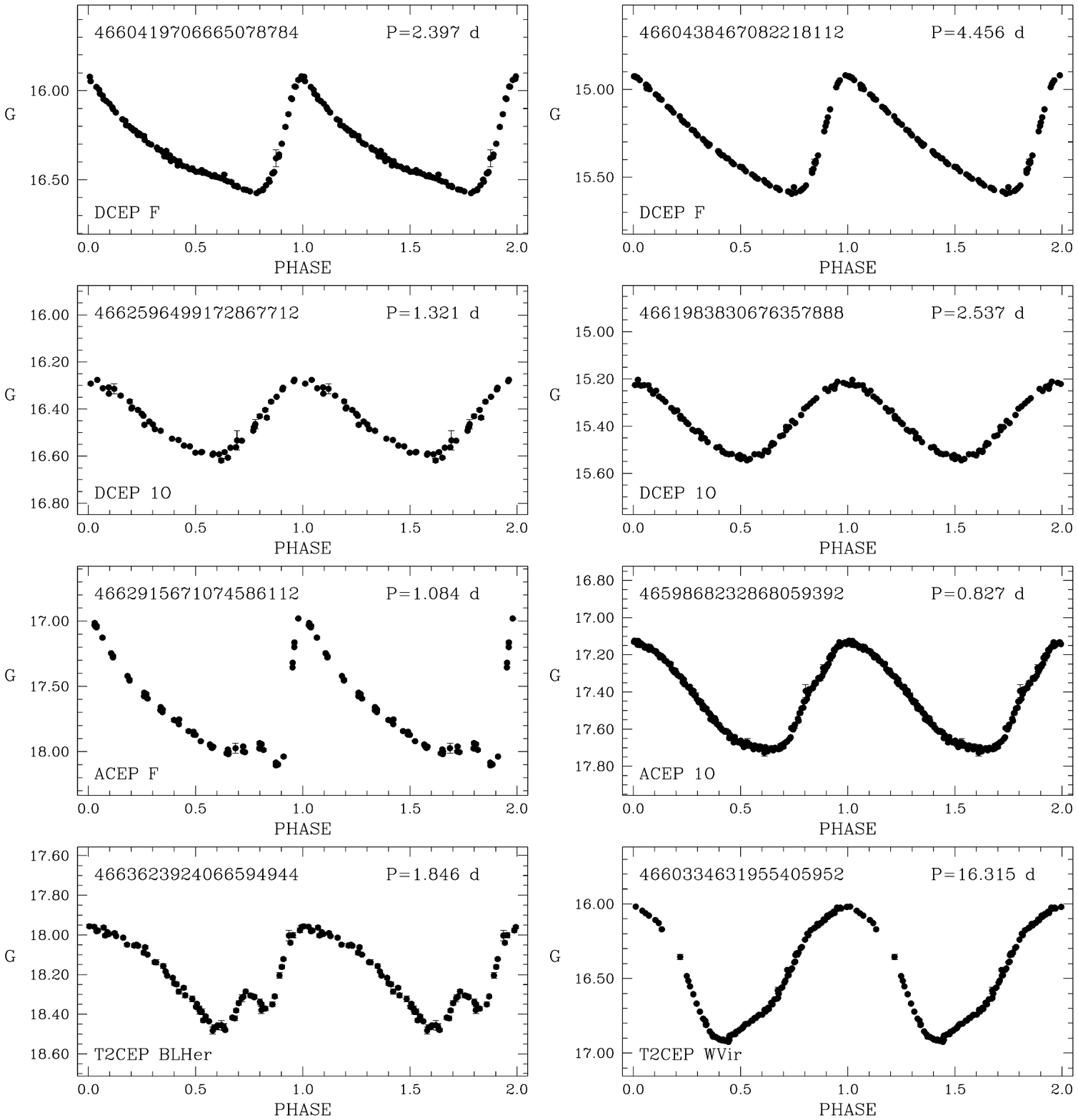}
   \caption{$G$-band light curves of Cepheids of different type and pulsation mode released in {\it Gaia} DR1. 
     Top panels: DCEPs F  (top two panels); DCEPs 1O (mid-upper two panels);
   ACEP F  (mid-lower-left panel), ACEP 1O (mid-lower-right panel);  T2CEP BLHER (lower-left panel) and  T2CEP WVIR (lower-right panel). Error bars are comparable or smaller than symbol size.}
              \label{LC-CEP}%
    \end{figure*}

Examples of light curves for RR Lyrae stars and Cepheids released in {\it Gaia} DR1  are presented in Figs.~\ref{LC-RRL} and  Figs.~\ref{LC-CEP}. Light curves are folded according to period and epoch of maximum light determined by the SOS Cep\&RRL  pipeline.  The full atlas of light curves is available in the electronic edition of the journal.  Samplers are shown in Sections~\ref{atlas-cep} (Figs.~\ref{atlas-DcepFO}, ~\ref{atlas-DcepF},  ~\ref{atlas-DcepTagged}, ~\ref{atlas-Acep}, ~\ref{atlas-T2cep-p1}, ~\ref{atlas-T2cep-p2}) and ~\ref{atlas-rrl} (Figs.~\ref{atlas-RRc}, ~\ref{atlas-RRab}), for Cepheids and RR Lyrae stars, respectively.  New discoveries by Gaia are labelled.

DCEPs with periods ranging from about  
6 to 16 d show a secondary maximum (bump) in their light and RV curves, which position varies with the pulsation 
period (see \citealt{bono00} for details). This phenomenon is called 
Hertzsprung progression (\citealt{hertzsprung1926}; \citealt{ledoux1958}).
\begin{figure*}
 \centering
\includegraphics[width=16.5 cm, trim= 10 130 10 130, clip]{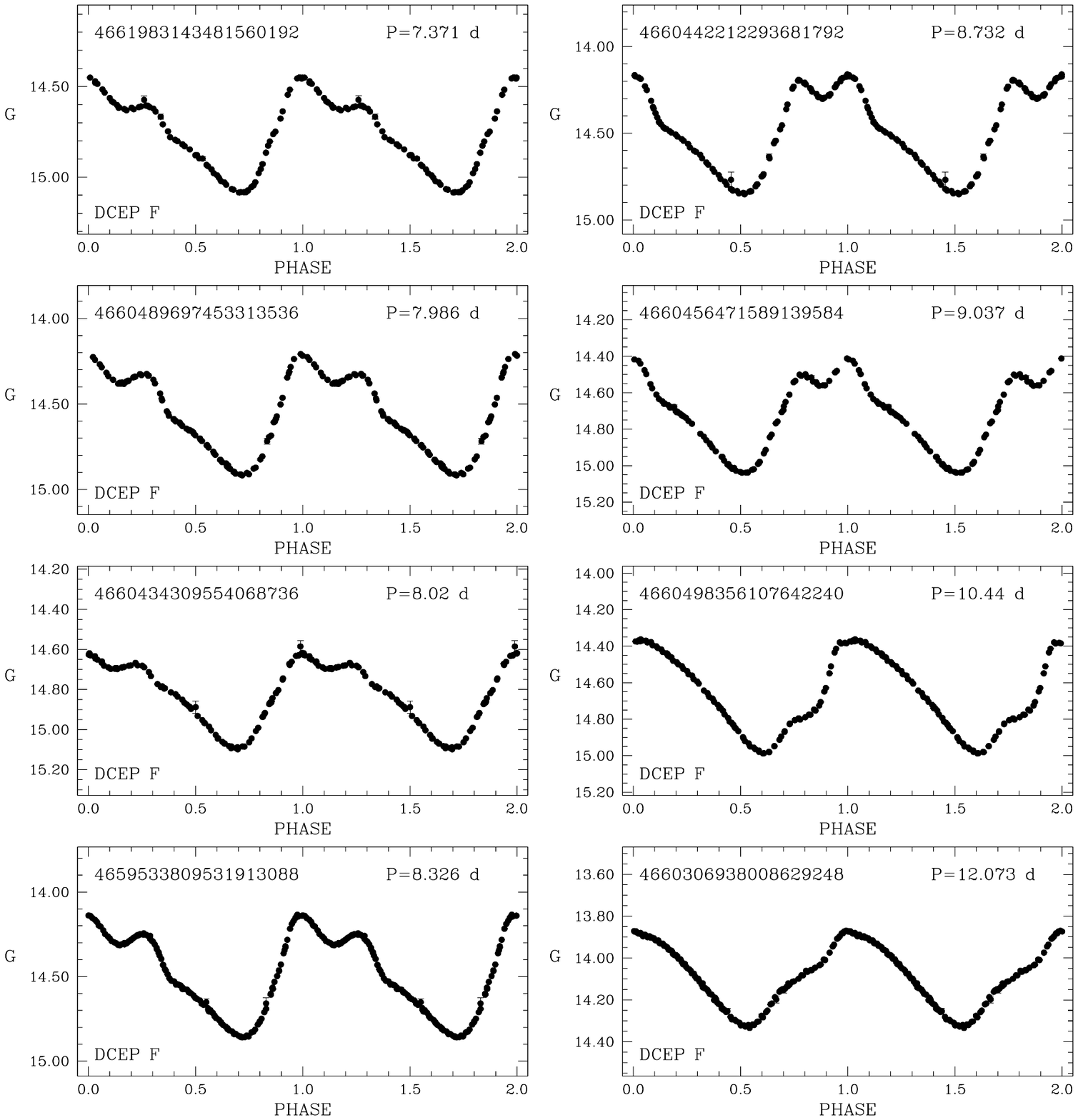}
   \caption{$G$-band light curves of a subset of LMC DCEPs  that exhibit the Hertzsprung progression. 
     Error bars are comparable or smaller than symbol size.}
               \label{HR-prog}%
    \end{figure*}
In the case of Galactic DCEPs the bump appears on the descending branch of the light and RV curves for $P< 9$ d, close to maximum light 
for $9 < P < 12$ d, and along the rising
branch for longer periods. The period at the center of the Hertzsprung progression varies with metallicity moving to  
 longer periods as metallicity decreases (\citealt{marconi05,soszynski11a}).
Fig.~\ref{HR-prog} shows the $G$-band light curves for a  subset of  DCEPs
that exhibit the Hertzsprung progression. The center of the progression for  these 
DCEPs is close to 9 days, as with the MW DCEPs, thus possibly suggesting that their metallicity is higher than for the bulk of the LMC Cepheids.

\begin{figure}
   \centering
   \includegraphics[width=9 cm, trim= 30 430 10 110, clip]{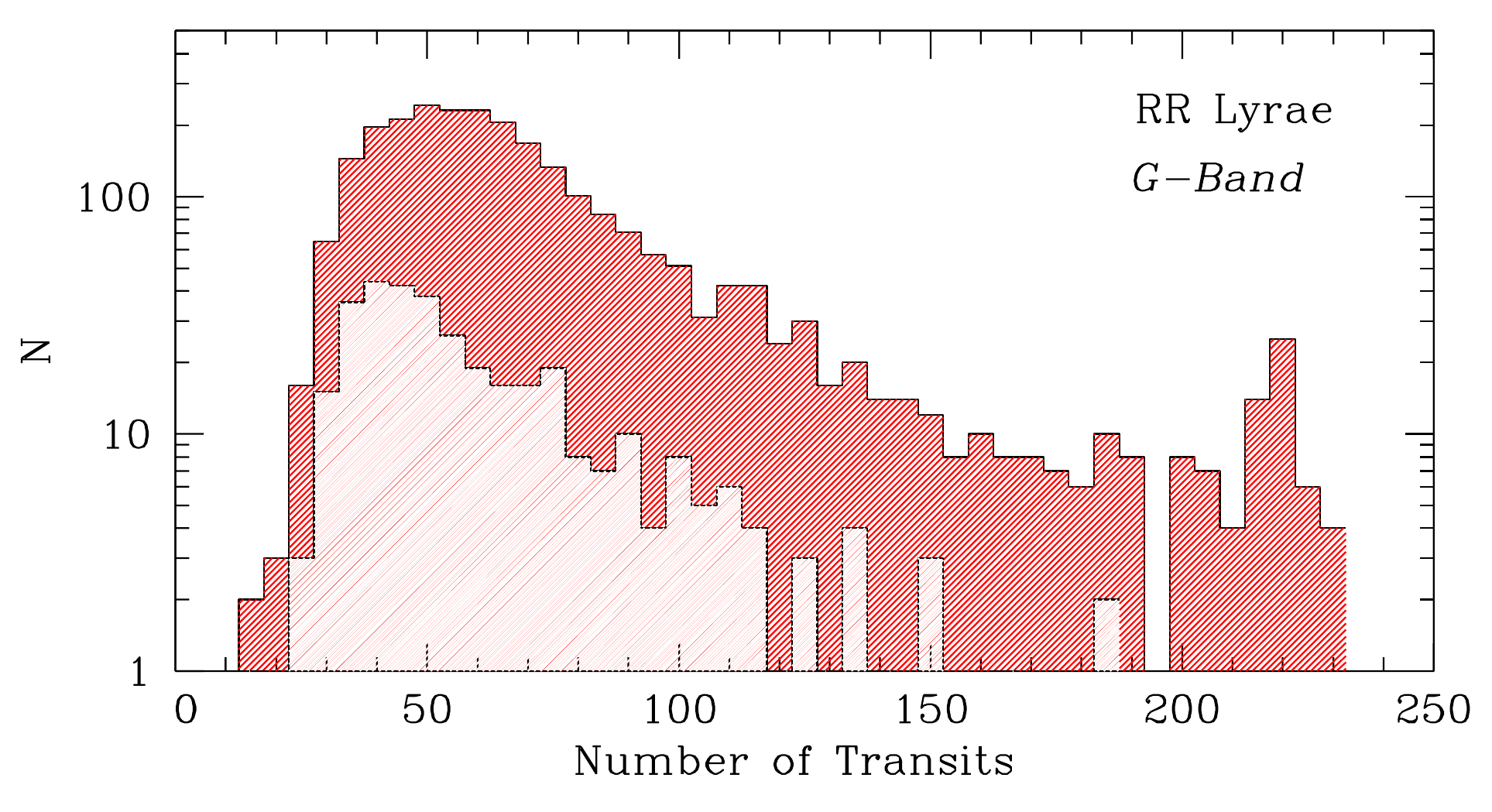}
  \caption{Histogram showing the number of observations available for the 2,595 SEP RR Lyrae stars published in {\it Gaia} DR1. 
    The pink area shows the number of observations available for 
   343 new RR Lyrae stars in the sample.}
              \label{NhistoRR}%
    \end{figure}

\begin{figure}
   \centering
   \includegraphics[width=9 cm, trim= 30 430 10 110, clip]{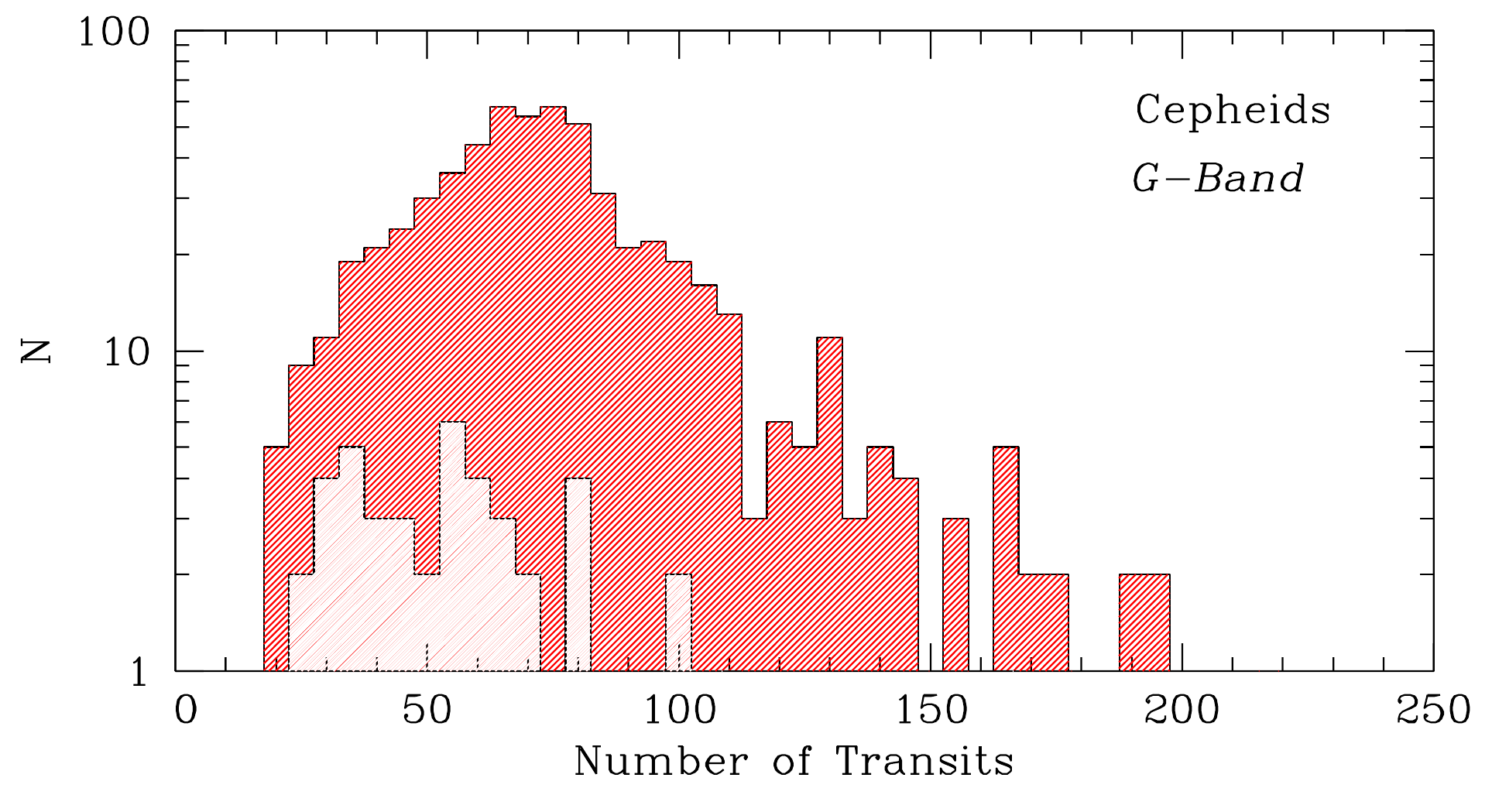}
   \caption{Histogram showing the number of observations available for the 599 Cepheids published in {\it Gaia} DR1. 
     The pink area shows the number of observations available for 
   43 new Cepheids in the sample.}
              \label{NhistoCC}%
    \end{figure}

Figs.~\ref{NhistoRR} and ~\ref{NhistoCC} show, respectively, the number of observations available for the 2,595 SEP RR Lyrae stars and the 599 Cepheids  in {\it Gaia} DR1. The distribution peaks  between 40 and 70  with maximum around 50 transits for the RR Lyrae stars and between 60 and 80 with maximum around 70 transits for the Cepheids. 
Hence, these SEP sources provide a reasonably realistic representation of the typical number of observations per source that will be available on average  at end of the five-year  
nominal life-time of the {\it Gaia} mission,  although with a totally different cadence. However, we note that  for the specific case of regular, periodic objects with typical periods of RR Lyrae stars  and Cepheids and over a large time interval (e.g. the five-year time-span),  
 the sparser NSL cadence being less prone to aliasing than  the EPSL cadence, which sampling is too regular, may  turn into an advantage for period derivation.


\begin{figure}
   \centering
   \includegraphics[width=9 cm, trim= 30 430 10 110, clip]{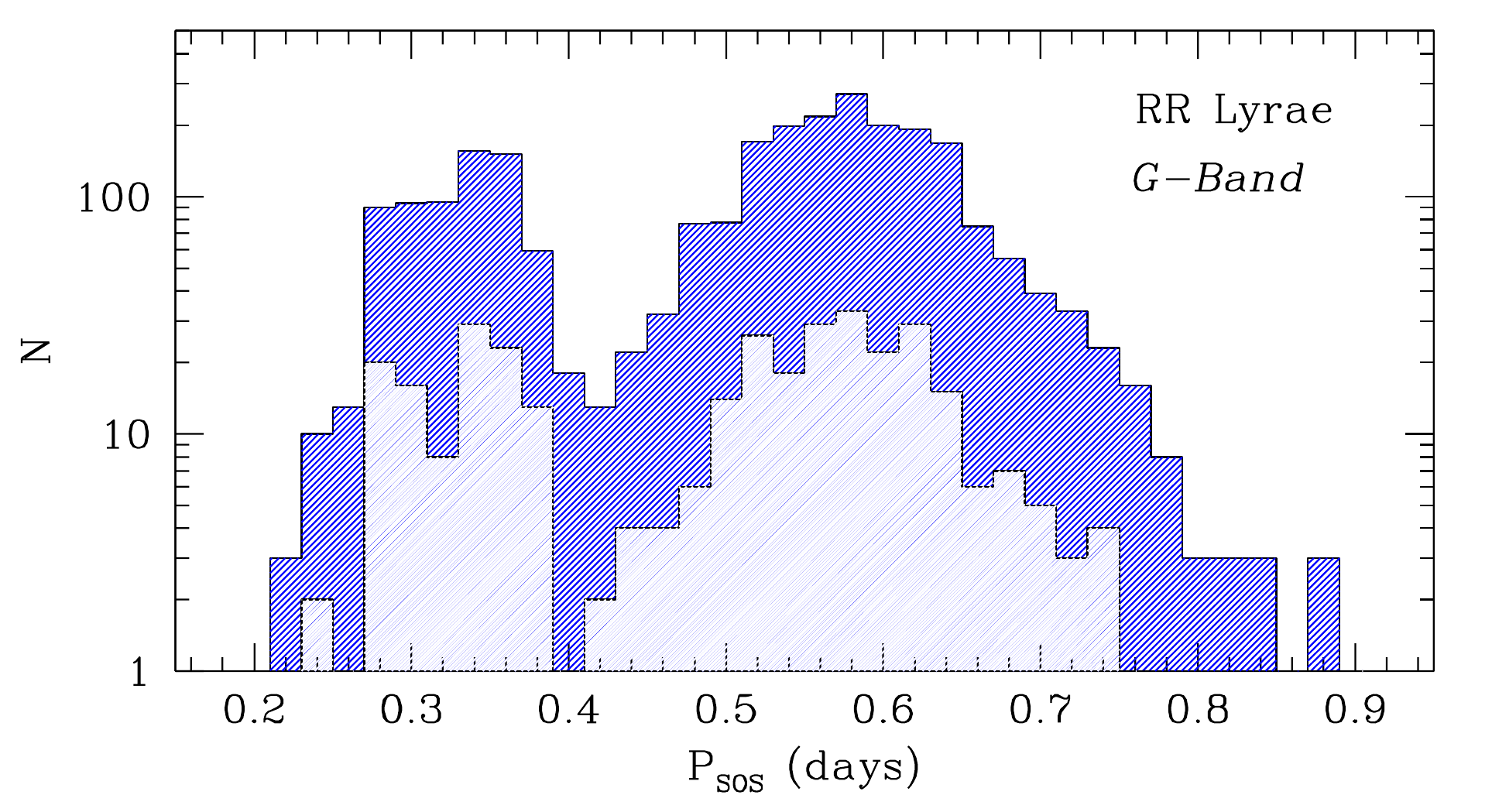}
   \caption{Histogram showing the period distribution of the 2,595 SEP RR Lyrae stars published in {\it Gaia} DR1. 
    The light-blue area shows the period distribution of 343  RR Lyrae stars that are new discoveries by {\it Gaia}.}
              \label{PhistoRR}%
    \end{figure}

\begin{figure}
   \centering
   \includegraphics[width=9 cm, trim= 30 430 10 110, clip]{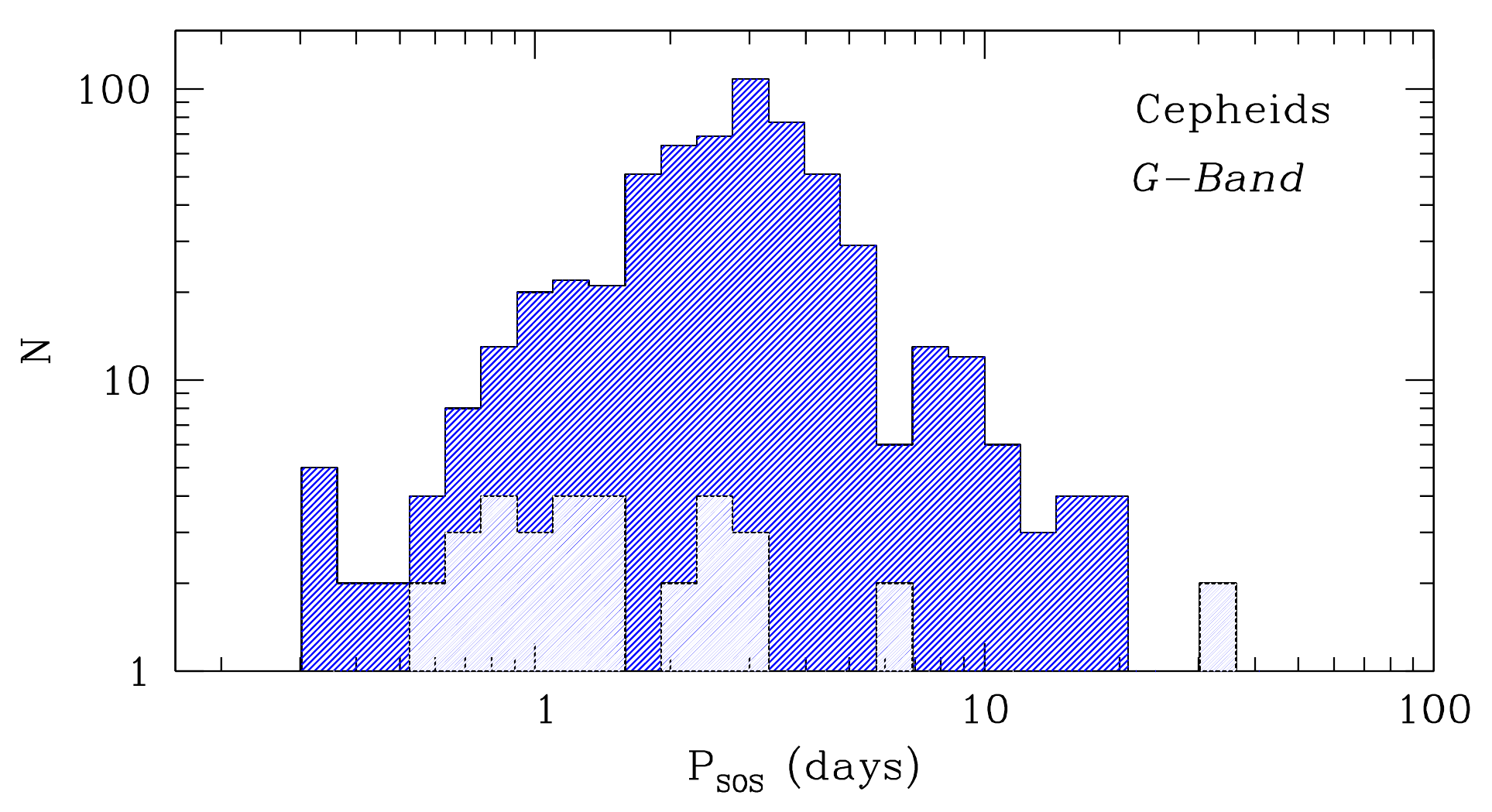}
   \caption{Histogram showing the period distribution of the 599 Cepheids published in {\it Gaia} DR1. 
    The light-blue area shows the period distribution of 43  Cepheids  in the sample that are new discoveries by {\it Gaia}.
   }
              \label{PhistoCC}%
    \end{figure}

\begin{figure}
   \centering
   \includegraphics[width=9 cm, trim= 10 190 30 110, clip]{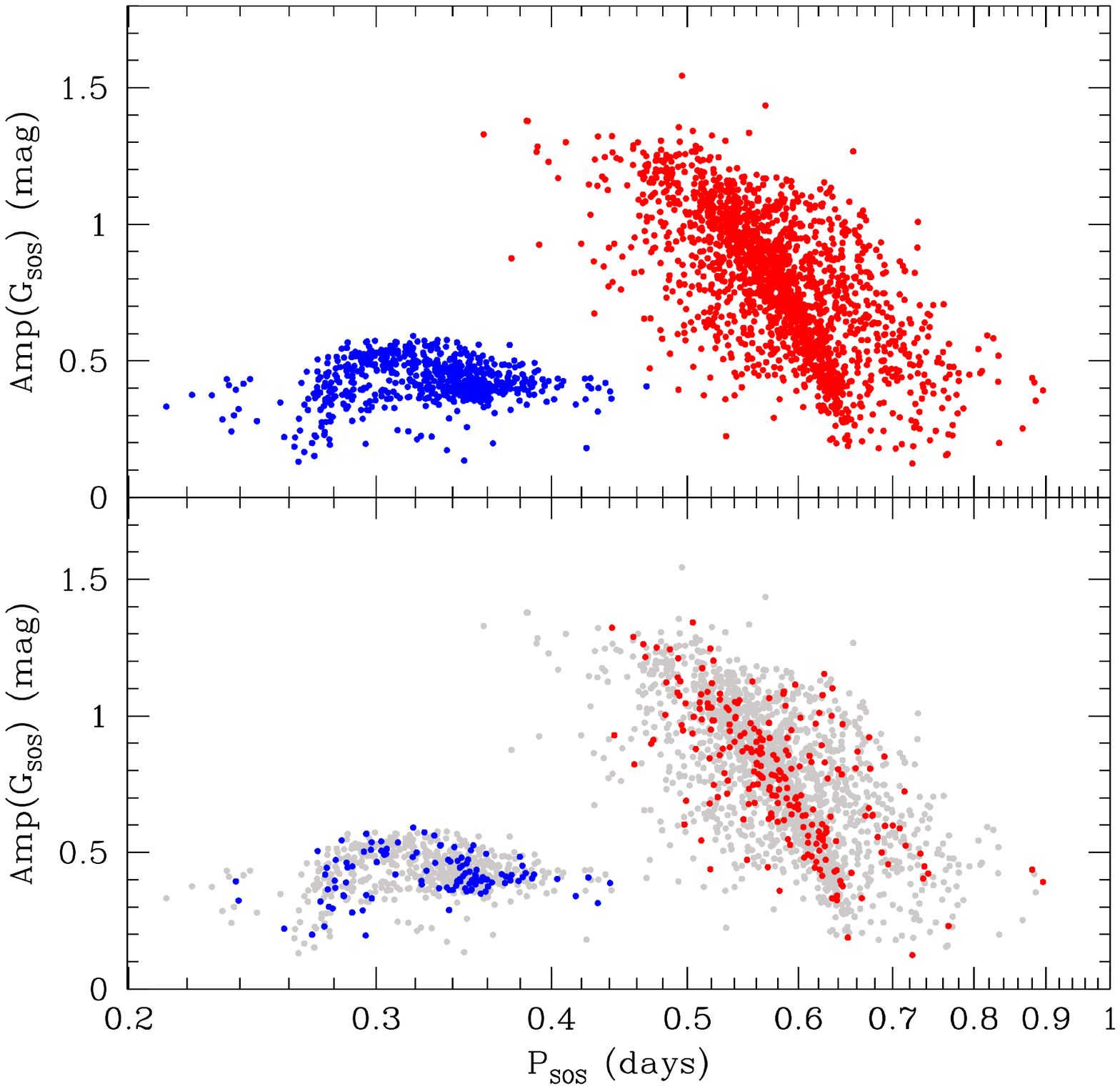}
  \caption{Period-$G$ amplitude diagram of the 2,595 SEP RR Lyrae stars published in {\it Gaia} DR1. 
    Blue and red filled circles indicate  RRc and RRab pulsators, respectively. The upper panel shows the total sample of 2,595 RR Lyrae stars. In the lower panel plotted in colour are  the new RRab (red) and RRc (blue) stars discovered by {\it Gaia} (343 in total), grey filled circles  are RR Lyrae  stars already known. 
}
              \label{pampRR}%
    \end{figure}

Figs.~\ref{PhistoRR} and ~\ref{PhistoCC} show, the period distributions of the RR Lyrae stars and Cepheids in {\it Gaia} DR1. 
 In both figures the light-blue shaded areas correspond to new discoveries by {\it Gaia}. New and known variables follow very similar distributions. The RR Lyrae stars show the typical bimodal distribution corresponding to RRc and RRab pulsators. The period distribution of the RRab stars peaks at $P \sim 0.59$ d and confirms a predominantly Oosterhoff-intermediate nature of the LMC RR Lyrae stars (\citealt{oo1939}, \citealt{catelan09} and references therein). However, a minor component of Oosterhoff type II RRab stars is also present, which defines a second less populated sequence in the $PA$ diagram shown in Fig.~\ref{pampRR}.
    
\begin{figure}
   \centering
   \includegraphics[width=9 cm, trim= 20 190 30 110, clip]{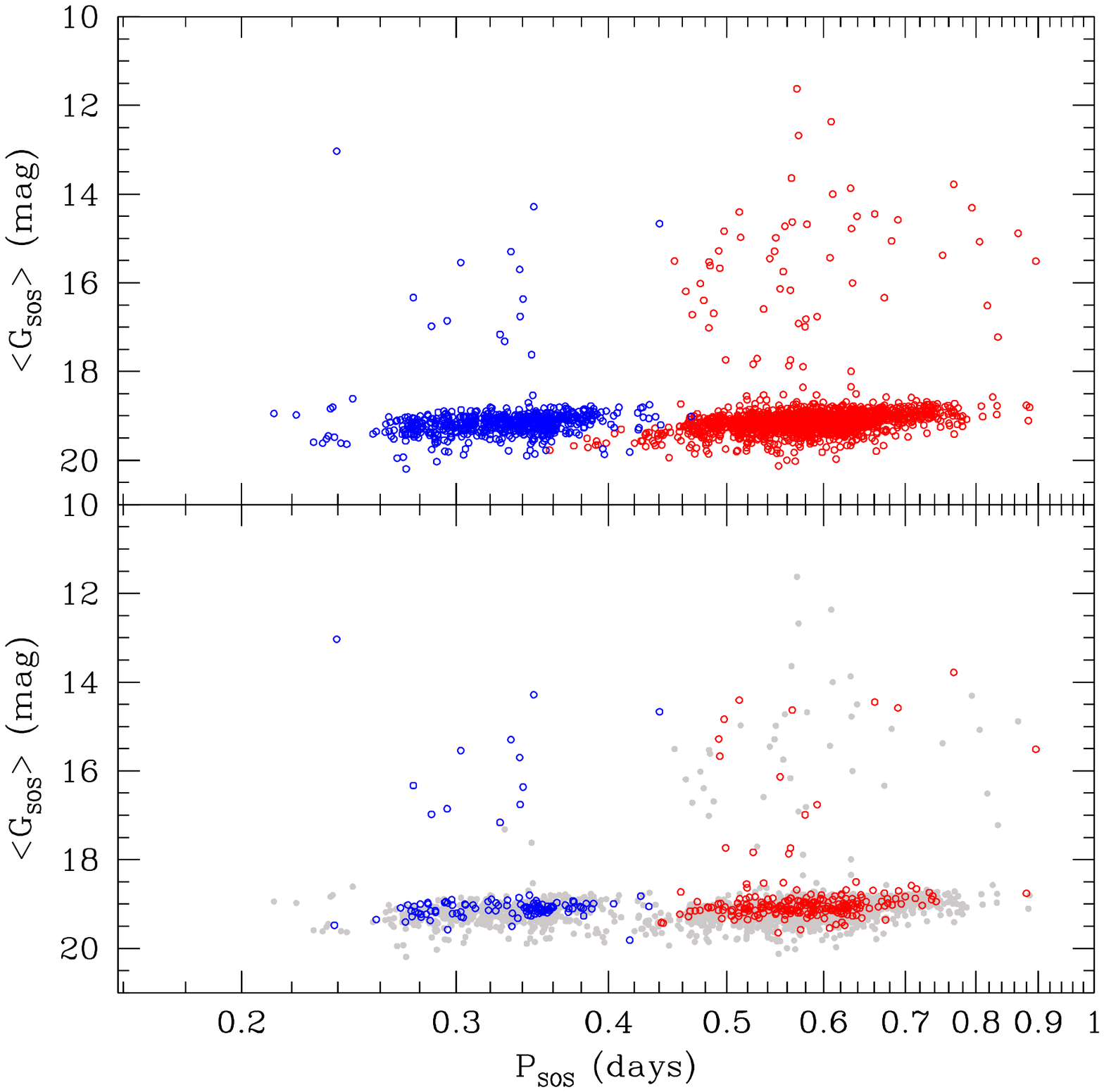}
   \caption{$PL$ distribution in the $G$-band of the SEP RR Lyrae stars published in {\it Gaia} DR1. 
    Blue and red open circles indicate  RRc and RRab pulsators, respectively. The upper panel shows the total sample of 2,595 RR Lyrae stars. In the lower panel marked in blu and red are new RR Lyrae stars discovered by {\it Gaia} (343 in total), grey filled circles are RR Lyrae  stars already known.     RR Lyrae stars brighter than 18- 18.5 mag likely belong to the MW halo in front of the LMC.}
              \label{ppl_RR}%
    \end{figure}

Fig.~\ref{ppl_RR} shows 
the $PL$ distribution of the  2,595  SEP RR Lyrae stars. 
 The LMC RR Lyrae stars in the {\it Gaia} SEP appear to typically have $\langle G \rangle \sim$19.0 mag with a dispersion of about  $\pm$ 0.5 mag. This rather large dispersion is  due to the 
combination of $PL$ intrinsic width, reddening, metallicity effects, LMC geometric depth and also to the extended throughput of {\it Gaia} $G$-band (330 - 1050 nm), which makes the  RR Lyrae star $PL$ relation 
to have a significant span in the $G$-band (\citealt{gaiacol-clementini}). On the other hand, the over  six magnitudes scatter seen in Fig.~\ref{ppl_RR} for $G \sim$ 18 mag is  largely due to Galactic RR Lyrae stars that fall within the {\it Gaia} SEP footprint and contaminate the LMC sample.

\begin{figure}
   \centering
   \includegraphics[width=9 cm,clip]{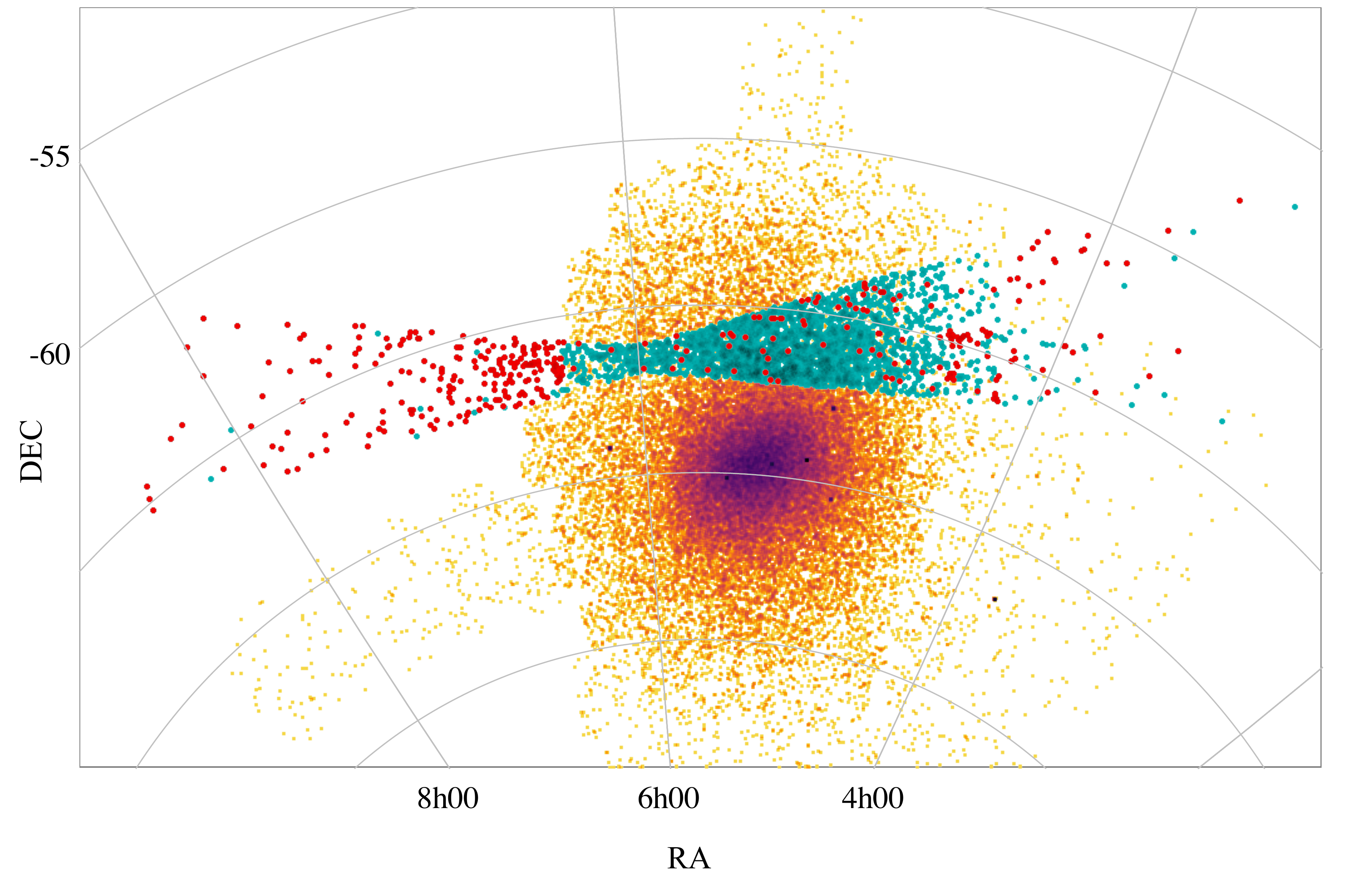}
   \caption{Spatial distribution of the 2,595 RR Lyrae stars released in {\it Gaia} DR1, compared with the distribution of LMC RR Lyrae stars from the OGLE-IV survey (orange dots; \citealt{soszynski16}).
   Green filled circles mark RR Lyrae stars observed by {\it Gaia}, in common with other surveys (see text for details). Red filled circles are 343 new RR Lyrae stars discovered by {\it Gaia}.}
              \label{mapRR}%
    \end{figure}

\begin{figure}
   \centering
    \includegraphics[width=9 cm, trim= 0 150 30 110, clip]{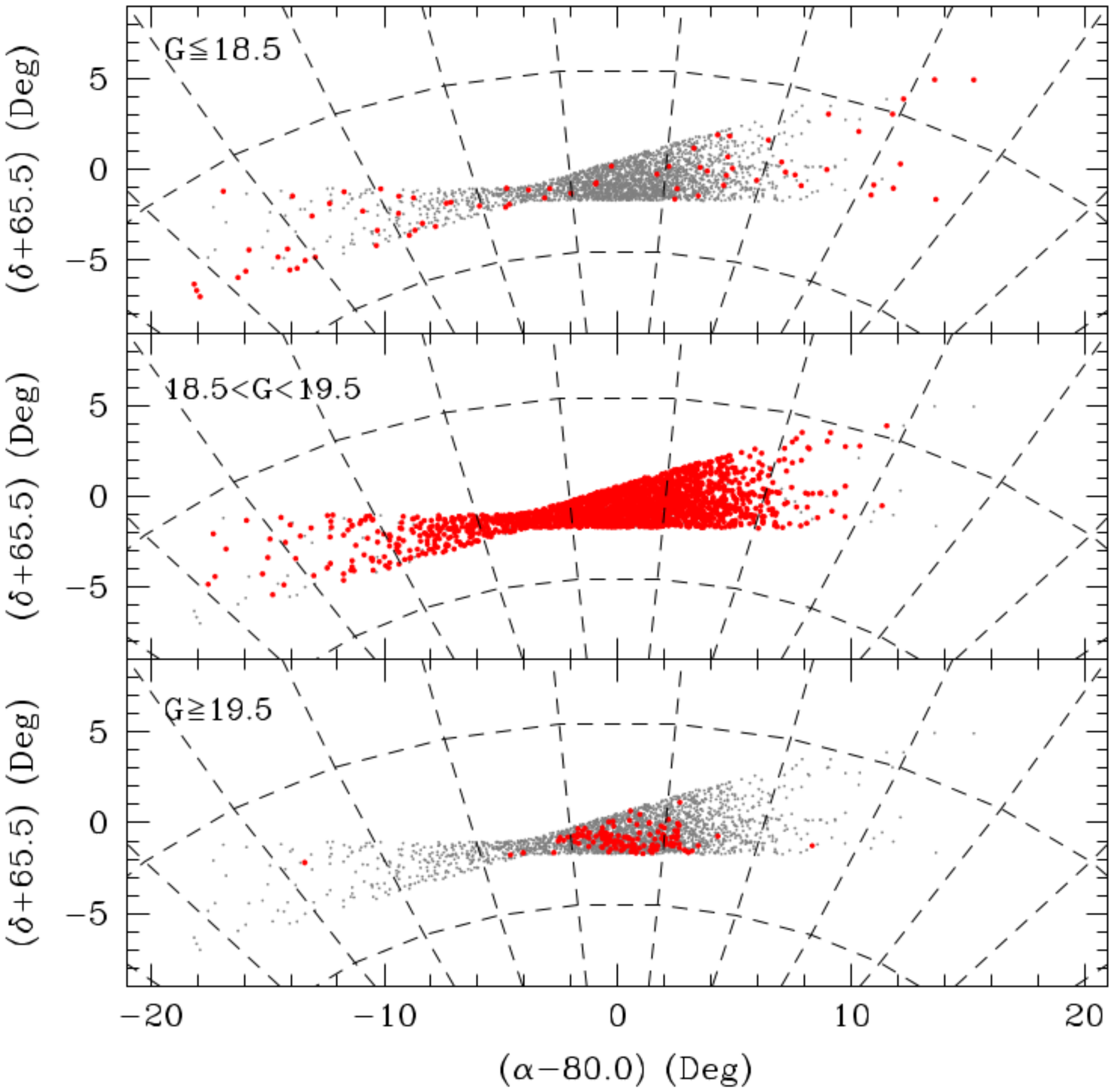}
   \caption{Distribution on sky of the SEP RR Lyrae stars in {\it Gaia} DR1 according to their mean apparent $G$-band  magnitude. 
   The stars have been divided in three different magnitude ranges that are labelled in the figure.  
   }
              \label{distrib}%
    \end{figure}

Fig.~\ref{mapRR} shows the spatial distribution of the 2,595 RR Lyrae stars in {\it Gaia} DR1  overlaid on the sample of LMC RR Lyrae stars recently released by OGLE-IV (\citealt{soszynski16}, orange filled circles).
Marked in green are RR Lyrae stars observed by {\it Gaia} that are in common with other surveys listed at the end of Section~\ref{s3}. About 700 of them were new discoveries by {\it Gaia} that were in the meantime identified also by OGLE (\citealt{soszynski16}).  Red filled circles are 343 new RR Lyrae stars discovered by {\it Gaia},  they mainly belong to the LMC and 
allow to cast a first glance on the very extended halo of this galaxy.
Also particularly interesting is the distribution in average magnitude of the 2,595 RR Lyrae stars in the {\it Gaia} SEP. This is shown in Fig.~\ref{distrib}, where we have sliced the sample into three 
magnitude bins. The top panel shows  73 RR Lyrae stars with $\langle G \rangle <$18.5 mag, sixty-three of them have $\langle G \rangle <$17.5 mag and 24 in this brighter sample are new discoveries by {\it Gaia}. All RR Lyrae stars in this panel likely belong to the MW halo in front of the LMC. The central panel shows 2,375 RR Lyrae stars with  18.5 $< \langle G \rangle <$19.5 mag. The average apparent magnitude of this sample is  $\langle G \rangle$ = 19.13 mag, with $\sigma$= 0.16 mag, this value is typical of the LMC RR Lyrae stars. They are quite homogeneously distributed and trace the LMC halo (see Fig.~\ref{mapRR}).  The bottom panel shows 147 RR Lyrae stars with $\langle G \rangle >$19.5 mag, they follow the clumped distribution of the LMC Cepheids in this area (see Fig.~\ref{mapCC}) and  are fainter than the average of LMC RR Lyrae stars likely because are more affected by reddening, which is expected to be higher in the dusty regions populated by Cepheids. This figure very nicely showcases  the potential of {\it Gaia} and variable stars to  study galactic structure.  

\begin{figure}
  \centering
   \includegraphics[width=9 cm,  trim= 10 180 10 130, clip]{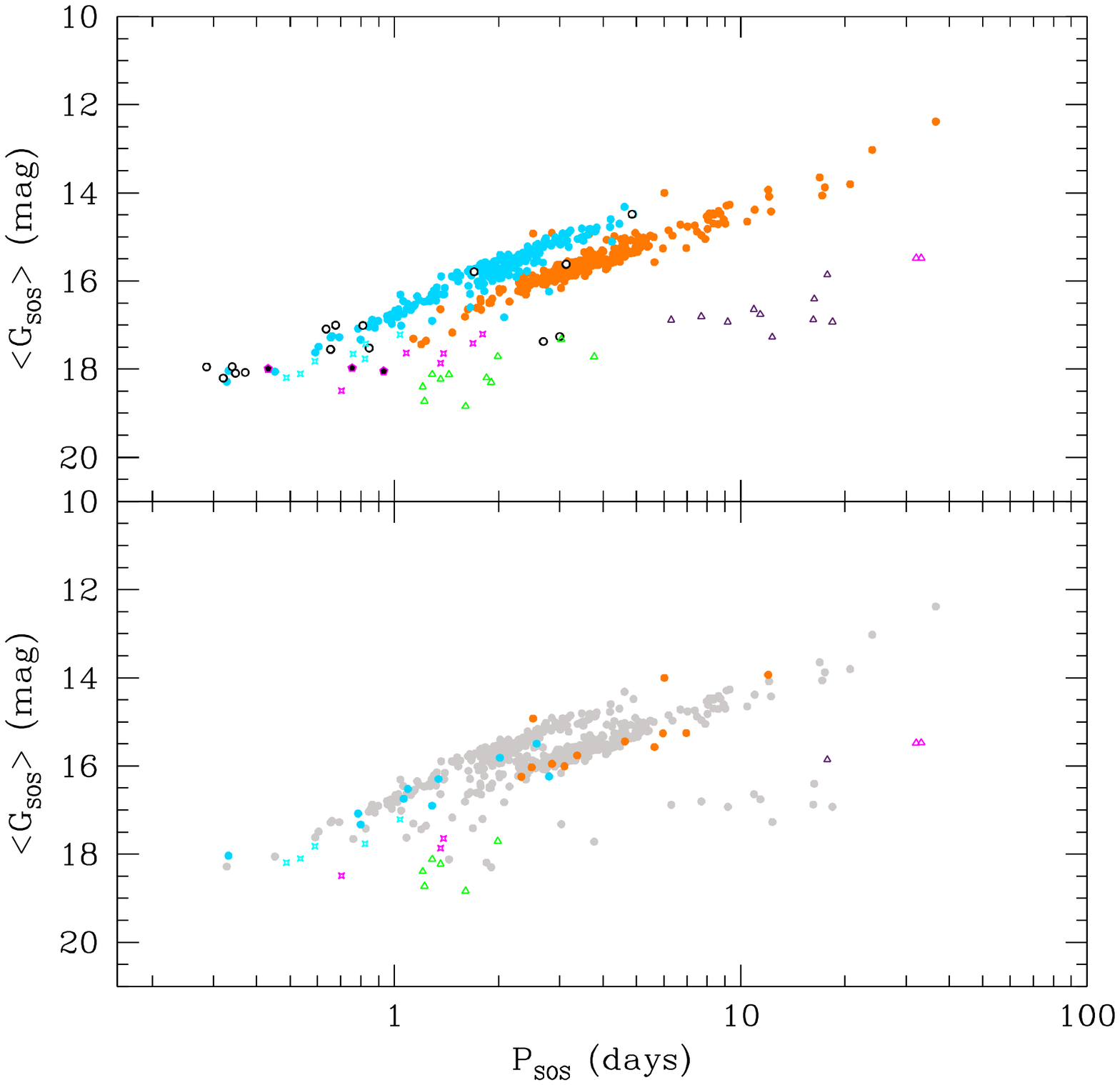}
   \caption{$PL$ distribution of the 599 Cepheids in {\it Gaia} DR1. 
   Symbols and colour-coding are the same as in Fig.~\ref{dp}. The upper panel shows the total sample of 599 Cepheids. In the lower panel marked in colour are 43 new Cepheids discovered by {\it Gaia}. Grey filled circles are Cepheids already known.}
              \label{ppl-2}%
    \end{figure}

\begin{figure}
   \centering
   \includegraphics[width=9 cm,clip]{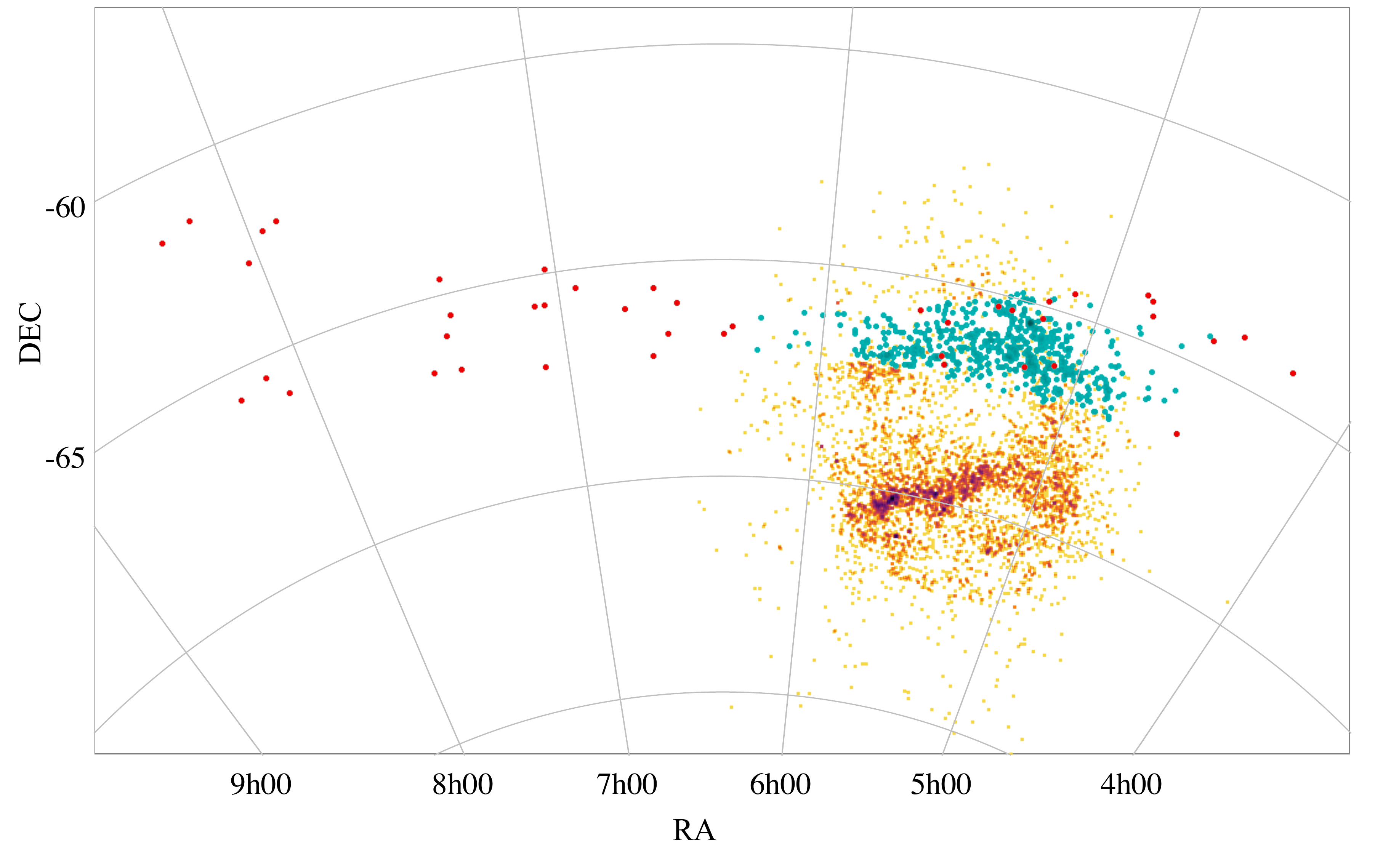}
  \caption{Spatial distribution of the 599 Cepheids released in {\it Gaia} DR1, compared with the distribution of LMC Cepheids from the OGLE-IV survey (orange dots; \citealt{soszynski15a,soszynski15b,soszynski15c}). Green filled circles mark Cepheids observed by {\it Gaia} and already known from the OGLE and EROS-2 surveys. Red filled circles are 43 new Cepheids  discovered by {\it Gaia}.}
              \label{mapCC}%
    \end{figure}

 Finally, Figs.~\ref{ppl-2} and ~\ref{mapCC} show, respectively, the $PL$ relation and the distribution on sky of the 599 Cepheids in {\it Gaia} DR1.  The various types of Cepheids in the sample 
 clearly define different $PL$ relations. They will be used to identify and classify different types of Cepheids in future {\it Gaia} data releases (see Section~\ref{conclusions}).
Red filled circles in Fig.~\ref{mapCC} mark 43 new Cepheids discovered by {\it Gaia}. This sample comprises 25 DCEPs, 10 ACEPs and 8 T2CEPs, which are located outside the OGLE-IV footprint and at increasing distance from the LMC.  According to the position on the $PL$ relation shown in Fig.~\ref{ppl-2} they belong to the LMC. However, it is premature to draw any conclusions on the LMC structure from this fairly small sample of new Cepheids, which might also be contaminated at 3\% level by ECLs (see \citealt{moretti14}), and  by other types of variables 
(e.g. ellipsoidals or spotted stars) that populate the same period, period-luminosity, period-amplitude, and Fourier parameter domain. 
Whether some of these misclassifications may have actually occurred, cannot be judged 
from Gaia DR1 data alone. This will become possible and checks will certainly be done, if/where appropriate, now that we  have started the re-processing of  all sources
 in preparation of Gaia DR2.  For Gaia second data release our analysis will  also be easier because other parts of the 
CU7 pipeline will be activated, those specifically devoted to binaries of different types 
(among which also ellipsoidals) and to spotted stars in particular. Furthermore, for Gaia DR2 we will also have  
$G_\mathrm{BP}$, $G_\mathrm{RP}$ time series available to help us disentangle Cepheids and RR Lyrae stars from other types 
of variables.  Hence, we leave any further discussion of this sample for {\it Gaia} DR2   when analysis of  Cepheid and RR Lyrae candidates will be extended beyond the SEP footprint and $G_\mathrm{BP}$,$G_\mathrm{RP}$ photometry will be used to help with the Cepheid classification. 

\section{Conclusions and future developments}\label{conclusions}

We have presented an overview of the Specific Objects Study 
(SOS) pipeline SOS Cep\&RRL, developed in the context of Coordination Unit 7 (CU7) 
of the {\it Gaia} Data Processing and Analysis Consortium (DPAC), to validate and fully characterise Cepheids and RR Lyrae variables 
observed by the spacecraft.  The SOS Cep\&RRL pipeline was  specifically 
tailored to analyse the {\it Gaia}  $G$-band  time-series photometry of sources in the South Ecliptic 
Pole (SEP) footprint, which covers an external region of the
LMC, and to produce results for confirmed RR 
Lyrae stars and Cepheids to be released with {\it Gaia} Data
Release 1 ({\it Gaia} DR1).

Results presented in this paper have been obtained applying 
the whole variable star analysis pipeline on the time-series photometry collected by {\it Gaia} 
during 28 days of Ecliptic Pole Scanning Law (EPSL) and 13 months  
 Nominal Scanning Law (NSL).  
 In addition to positions and $G$-band time-series photometry, 
for confirmed Cepheids and RR Lyrae stars in the {\it Gaia} SEP,  {\it Gaia} DR1 contains  the following outputs of the 
 SOS Cep\&RRL pipeline:
period of pulsation, classification in type and pulsation mode,  intensity-averaged mean magnitude,  
peak-to-peak amplitude and Fourier decomposition parameters
$R_{21}$ and $\phi_{21}$. All quantities are provided with related uncertainties.
The variable star inventory of {\it Gaia} DR1  includes 3,194 variables 
which comprise 599 Cepheids and 2,595 RR Lyrae stars,
386 of them (43 Cepheids and 343 RR Lyrae stars) are new discoveries by {\it Gaia}.

The published sources are distributed over an area  extending 38 degrees on either side from a point offset from the centre of the LMC by about 
3 degrees to the north and 4 degrees to the east.  
The vast majority, but not all, are located within the LMC.
The sample also includes 63  bright RR Lyrae stars that belong to the MW halo, of which 24 are new {\it Gaia} discoveries.

A number of  improvements of the SOS Cep\&RRL pipeline are planned in view of {\it Gaia}  forthcoming releases, 
of which the next one, {\it Gaia} DR2, is foreseen in  2017. They include:

\begin{enumerate}

\item
Check of the pass-band transformations. In preparation of {\it Gaia} DR2, all  tools 
and relations used by SOS Cep\&RRL to classify and characterise the {\it Gaia} 
sources will be re-derived directly from Cepheids and RR Lyrae stars 
released  in DR1, overcoming the need for transforming to {\it Gaia} pass-bands quantities that are generally known in 
the Johnson-Cousins system. This will definitely allow the $PL$  
relationships adopted to identify and classify different types of Cepheids to be refined. On the other hand, as 
shown by Fig.~\ref{color-equation} in Section~\ref{con-for}, the $G$-band transformation (eq. (A.1))  
worked rather well for the colour range spanned by the Cepheids and RR Lyrae stars in {\it Gaia} DR1.

\item
{\it Gaia} $G_\mathrm{BP}$, $G_\mathrm{RP}$ 
colours will become available with {\it Gaia} DR2. This will allow to use 
$PW$ relations, whose
reduced scatter, compared to the $PL$ relations, will allow  to further improve 
the classification of Cepheids.

\item
Double-mode RR Lyrae stars  and multimode classical Cepheids (F/1O, 
1O/2O, etc.) will be identified and fully characterised for {\it Gaia} DR2   by improving the detection algorithm to properly take  
into account the scatter in the folded light curve,  
 thus reducing the number of false positives.
 
\item
Estimate of the error in period, mean magnitude, peak-to-peak amplitude, etc.  will be refined. In particular, errors in the Fourier parameters $\phi_{ij}$ and $R_{ij}$, which are currently computed by propagation of the errors in  $\phi_i$, $\phi_j$, $R_i$ and  $R_j$,  will be entirely computed via Monte Carlo simulations.  

\item
A classifier will be developed to optimise the type and subtype classification of Cepheids and RR Lyrae stars performed by the SOS Cep\&RRL pipeline.  
\end{enumerate} 

The results for Cepheids and RR Lyrae stars shown in this paper demonstrate the excellent quality of {\it Gaia} photometry notwithstanding all limitations in the  dataset and processing 
for  {\it Gaia} Data Release 1. They nicely showcase the potential of {\it Gaia} and the promise of {\it Gaia} Cepheids and RR Lyrae stars for all areas of the sky in which an appropriate 
light curve sampling will be achieved.


\begin{acknowledgements}
This work has made use of data from the ESA space mission {\it Gaia}, processed by the {\it Gaia} Data
Processing and Analysis Consortium (DPAC). Funding for the DPAC has been
provided by national institutions participating in
the {\it Gaia} Multilateral Agreement. In particular, the Italian participation in DPAC has been supported by Istituto Nazionale di
Astrofisica (INAF) and the Agenzia Spaziale Italiana (ASI) through grants
I/037/08/0,  I/058/10/0,  2014-025-R.0, and 2014-025-R.1.2015 to INAF (PI M.G.
Lattanzi),  the Belgian participation by the BELgian federal Science Policy 
(BELSPO) through PRODEX grants, the Swiss participation by  the Swiss State Secretariat for Education, Research
and Innovation through the ESA Prodex program, the ``Mesures d'accompagnement", the
``Activit\'{e}s Nationales Compl\'{e}mentaires", the Swiss National Science
Foundation, and the Early Postdoc.Mobility fellowship, and the Spanish participation by the Spanish Ministry of Economy
MINECO-FEDER through grants AyA2014-55216,
AyA2011-24052. 
UK community participation in this work has been supported by funding
from the UK Space Agency, and from the UK Science and Technology
Research Council.
The {\it Gaia} mission website is:  \texttt{http://www.cosmos.esa.int/gaia}.

\end{acknowledgements}

%
   \bibliographystyle{aa} 
%

\begin{appendix}
\section{Conversion formula}\label{con-for}
\begin{figure}
   \centering
   \includegraphics[trim=20 180 0 120, width= 9.5 cm,clip]{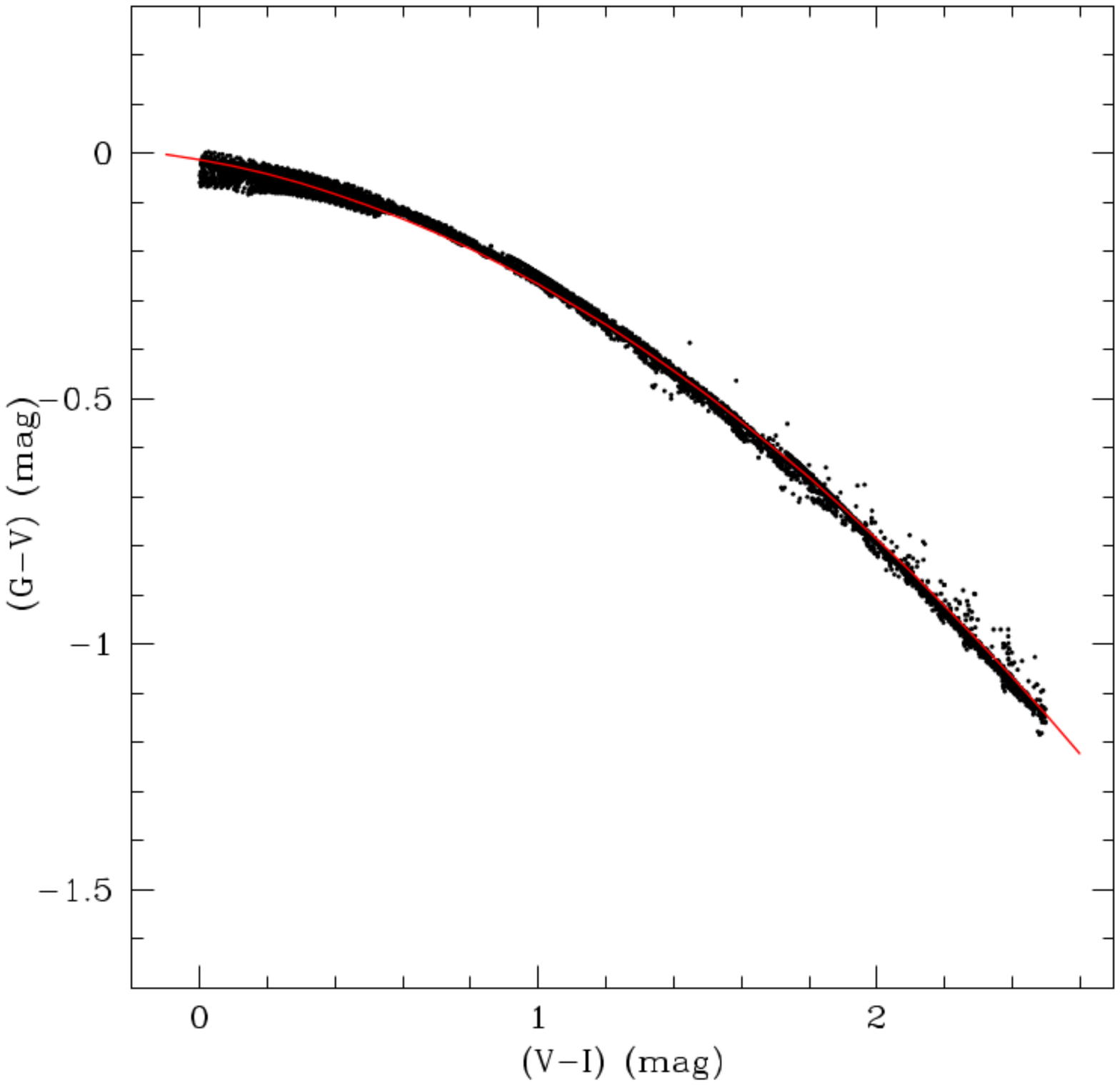}
  \caption{Conversion relation (red curve)  used in the present paper to transform Johnson $V$, $I$ to  the {\it Gaia} $G$-band 
    (eq. (A.1)). 
   The relation was computed by interpolating 
   in  the grid  of models by \citet{jordi10}  
   and subsequent updates (black solid points). New transformations have now been computed using {\it Gaia} real data (\citealt{gaiacol-van}), they supersedes the formula provided here.
  }
    \label{conversion}%
    \end{figure}

In this section we present the conversion formula, appropriate for the colour and metallicity ranges of Cepheids and RR Lyrae stars  ($V-I <$ 2.5 mag and   $-2.5 <$ [Fe/H] $<$0.5 dex),  we have used to transform the Johnson-Cousins $V$, $I$ to the {\it Gaia} $G$ 
pass-band. 
 This formula was computed using pass-band transformations provided in \citet{jordi10} 
and subsequent updates 
 (Jordi, personal communication).  
\begin{equation}
\begin{split}
G=V-(0.0128 \pm 0.0005)-(0.1218 \pm 0.0008)(V-I)\\
 -(0.1323 \pm 0.0003)(V-I)^{2}   ~~~~rms=0.017~{\rm mag}\\
\end{split}
\end{equation}
 The relation is shown by the red solid line in Fig.~\ref{conversion}. 

     \begin{figure}
   \centering
   \includegraphics[trim=20 180 0 120, width= 9.5 cm,clip]{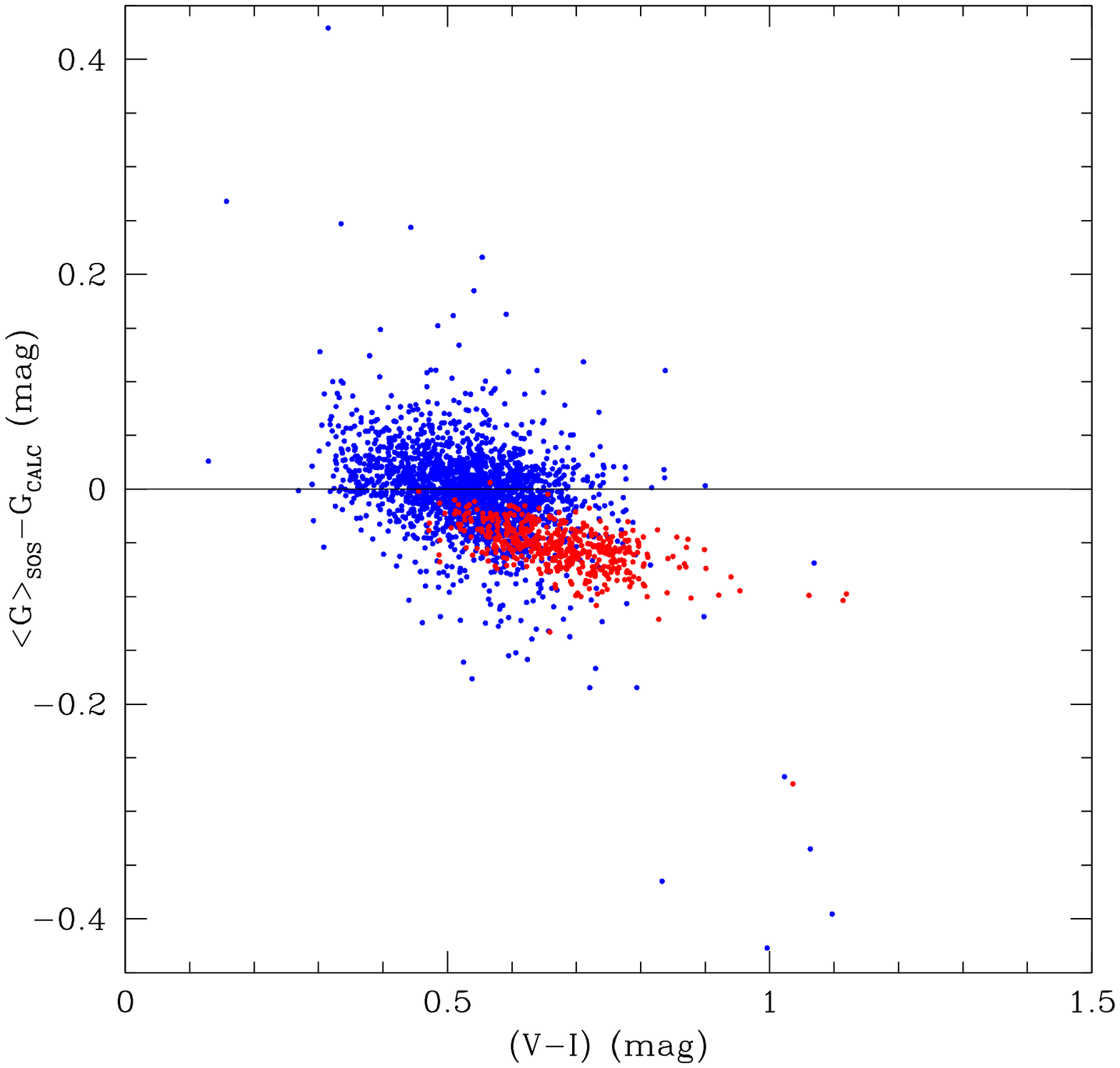}
   \caption{Comparison of the $G$-band  mean magnitudes computed by the SOS Cep\&RRL pipeline ( $\langle G_{\rm SOS} \rangle$) for  510 Cepheids (red filled circles) and 2,149 RR Lyrae stars (blue filled circles) in {\it Gaia} DR1   in common with the OGLE-IV survey and the  values obtained transforming  OGLE $V,I$ mean magnitudes to the $G$-band using Eq. A.1. 
    }
    \label{color-equation}%
    \end{figure}
It should be noted that the above conversion formula between {\it Gaia} $G$ 
 and Johnson-Cousins $V$,$I$ passbands is based on the pre-launch nominal {\it Gaia} $G$ passband,  hence it might not be perfect. Nevertheless, Fig.~\ref{color-equation} shows this transformation to have worked rather well for Cepheids and RR Lyrae stars that are published in {\it Gaia} DR1. In the figure 
we have plotted  the $\langle G \rangle$ mean magnitude computed by the SOS Cep\&RRL pipeline for 2,659 sources 
 (510 Cepheids and 2,149 RR Lyrae stars) in common with OGLE-IV  versus the  source $V-I$ colours from OGLE. 
Differences are, on average, within $\pm$0.1 mag.  

 New transformations have now been computed using {\it Gaia} real data (\citealt{gaiacol-van}). The reader is advised to use the new formulae to make pass-band transformations. 
 As far as future {\it Gaia} releases are concerned, Cepheids and RR Lyrae stars published in {\it Gaia} DR1  will be used to compute directly in the {\it Gaia} $G$-band the $PA$ and $PL$ relations used  in the SOS Cep\&RRL pipeline to classify these types of variable stars.

\section{Examples of \textit{\textbf{Gaia}} archive queries}
  \label{app:queries}


  \begin{table*}[!htbp]
    \caption{Queries to retrieve information on the Cepheids and RR Lyrae stars from the {\it Gaia} archive in the Astronomical Data Query Language (\citealt{osuna08}).}
    \label{tab:queries}
    \centering
    \begin{tabular}{p{\textwidth}}
      Query to retrieve time series of all Cepheids in the {\it Gaia} DR1.
      \begin{verbatim}
      select gaia.source_id, gaia.observation_time, gaia.g_flux, gaia.g_flux_error, 
        gaia.g_magnitude, gaia.rejected_by_variability_processing
      from gaiadr1.phot_variable_time_series_gfov as gaia
      inner join gaiadr1.cepheid as cep
        on gaia.source_id=cep.source_id
      \end{verbatim} \\

      Query to retrive time series of all RR Lyrae stars in the {\it Gaia} DR1.
      \begin{verbatim}
      select gaia.source_id, gaia.observation_time, gaia.g_flux, gaia.g_flux_error, 
        gaia.g_magnitude, gaia.rejected_by_variability_processing
      from gaiadr1.phot_variable_time_series_gfov as gaia
      inner join gaiadr1.rrlyrae as rr
        on gaia.source_id=rr.source_id
      \end{verbatim} \\

      Query to retrieve the number of processed observations and SOS Cep\&RRL-computed parameters of all Cepheids in the {\it Gaia} DR1.
      \begin{verbatim}
      select cep.source_id, num_observations_processed, type_best_classification, 
        mode_best_classification, type2_best_sub_classification, p1, p1_error, 
        num_harmonics_for_p1, epoch_g, epoch_g_error, int_average_g, int_average_g_error, 
        peak_to_peak_g, peak_to_peak_g_error, phi21_g, phi21_g_error, r21_g, r21_g_error
      from gaiadr1.phot_variable_time_series_gfov_statistical_parameters as stat
      inner join gaiadr1.cepheid as cep
       on stat.source_id=cep.source_id
      \end{verbatim} \\

      Query to retrieve the number of processed observations and SOS Cep\&RRL-computed parameters of all RR Lyrae in the {\it Gaia} DR1.
      \begin{verbatim}
      select rr.source_id, num_observations_processed, best_classification, p1, p1_error, 
        num_harmonics_for_p1, epoch_g, epoch_g_error, int_average_g, int_average_g_error, 
        peak_to_peak_g, peak_to_peak_g_error, phi21_g, phi21_g_error, r21_g, r21_g_error
      from gaiadr1.phot_variable_time_series_gfov_statistical_parameters as stat
      inner join gaiadr1.rrlyrae as rr
        on stat.source_id=rr.source_id
      \end{verbatim}
    \end{tabular}
  \end{table*}

\section{Atlas of the $G$-band light curves for Cepheids}\label{atlas-cep}
In this Section we present the $G$-band light curves of 599 Cepheids observed by {\it Gaia} in the SEP region. The light curves are folded according to the period and epoch of maximum light computed by the SOS Cep\&RRL pipeline and are plotted  grouping the Cepheids  by type and pulsation mode according to the following order:  DCEPs 1O (230 in total),  DECPs F (313), 
DCEPs NoMode (15 DCEPs for which we did not identify the pulsation mode);  ACEPs 1O (7 sources),  ACEPs F (6), ACEPs NoMode (3); T2CEPs BLHER (11 sources), T2CEPs WVIR (10), 
T2CEPs RVTAU (2) and T2CEPs NoMode (2). In each group the sources are desplayed in order of increasing period.  
\newpage
 \begin{figure*}
   \centering
   \includegraphics[width=16.5 cm, clip]{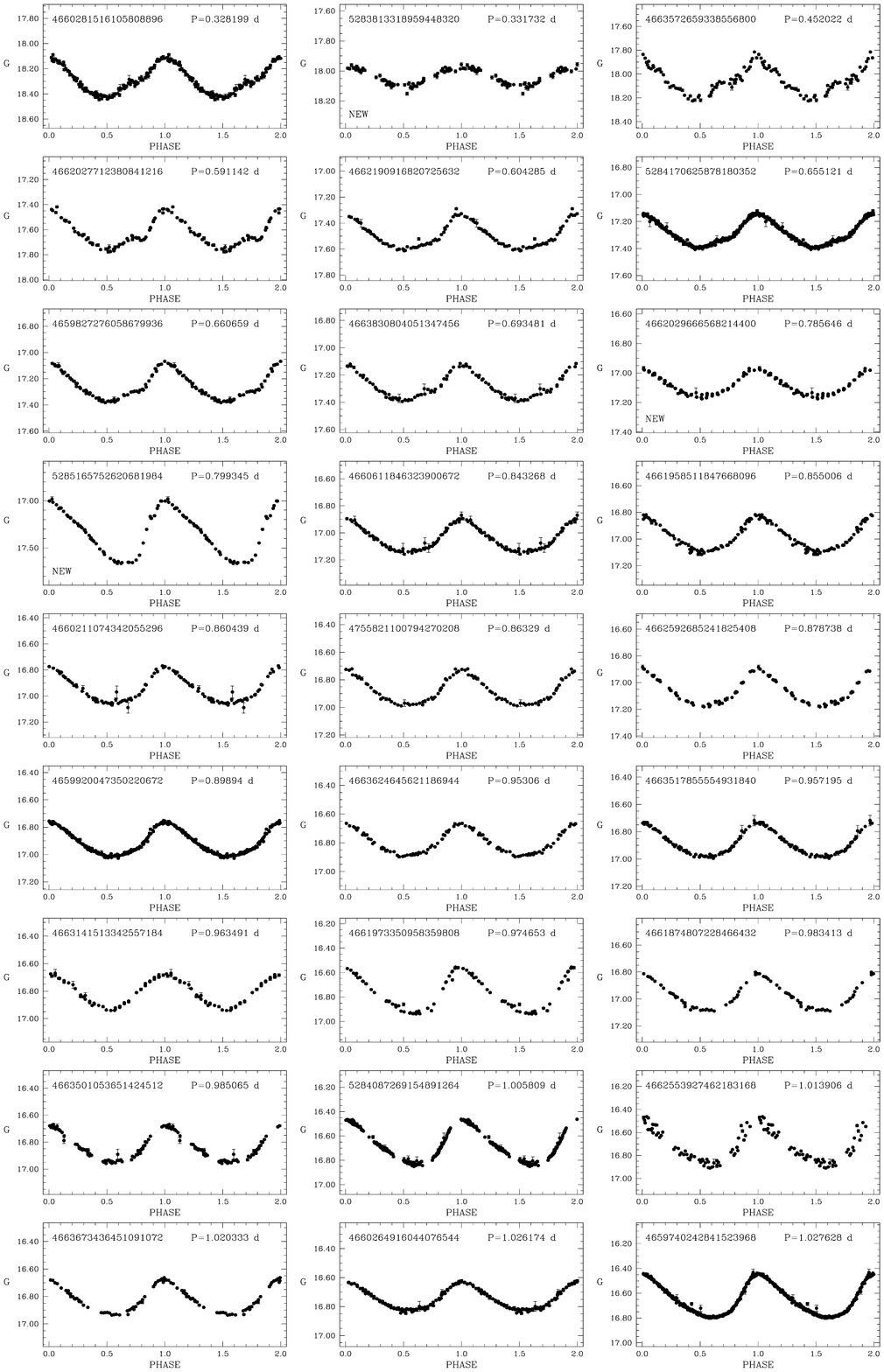}
  \caption{$G$-band light curves of first overtone DCEPs released in {\it Gaia} DR1 ordered by increasing period.    Error bars are comparable or smaller than the symbol size.  Measurements with errors larger than 0.05 mag are not displayed.  For each source we label the {\it Gaia} source ID and the pulsation period rounded to the last significant digit according to the error of the period  determination. New discoveries by Gaia are flagged as such.
   }
              \label{atlas-DcepFO}%
    \end{figure*}
\newpage
 \begin{figure*}
   \centering
   \includegraphics[width=16.5 cm, clip]{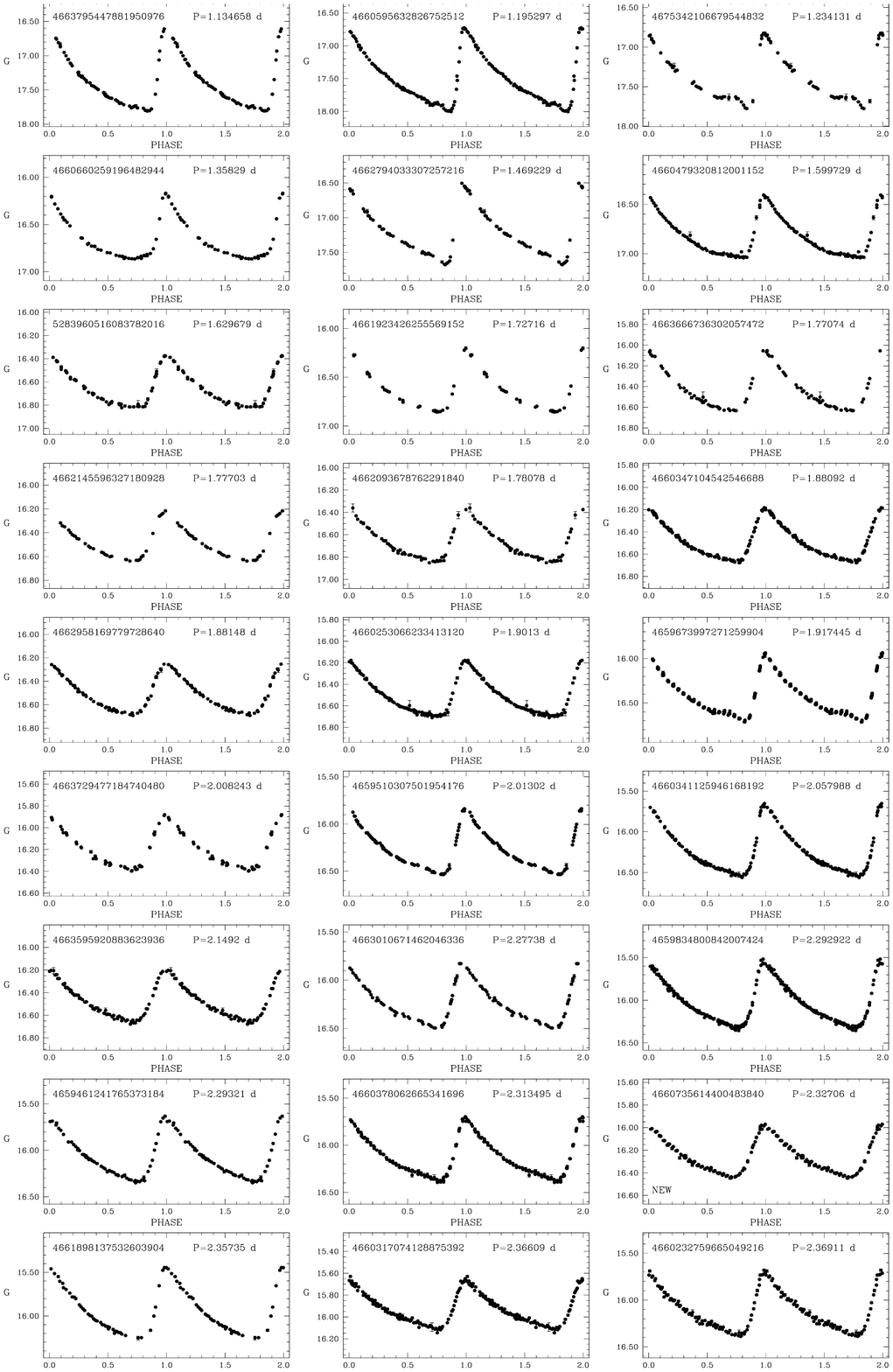}
   \caption{$G$-band light curves of  fundamental mode  DCEPs released in {\it Gaia} DR1 ordered by increasing period. Error bars are comparable or smaller than the symbol size. 
    Measurements with errors larger than 0.05 mag are not displayed.    
   For each source we label the {\it Gaia} source ID and  the pulsation period rounded to the last significant digit according to the error of the period  determination. New discoveries by Gaia are flagged as such.
   }
              \label{atlas-DcepF}%
    \end{figure*}
\newpage
 \begin{figure*}
   \centering
   \includegraphics[width=16.5 cm, clip]{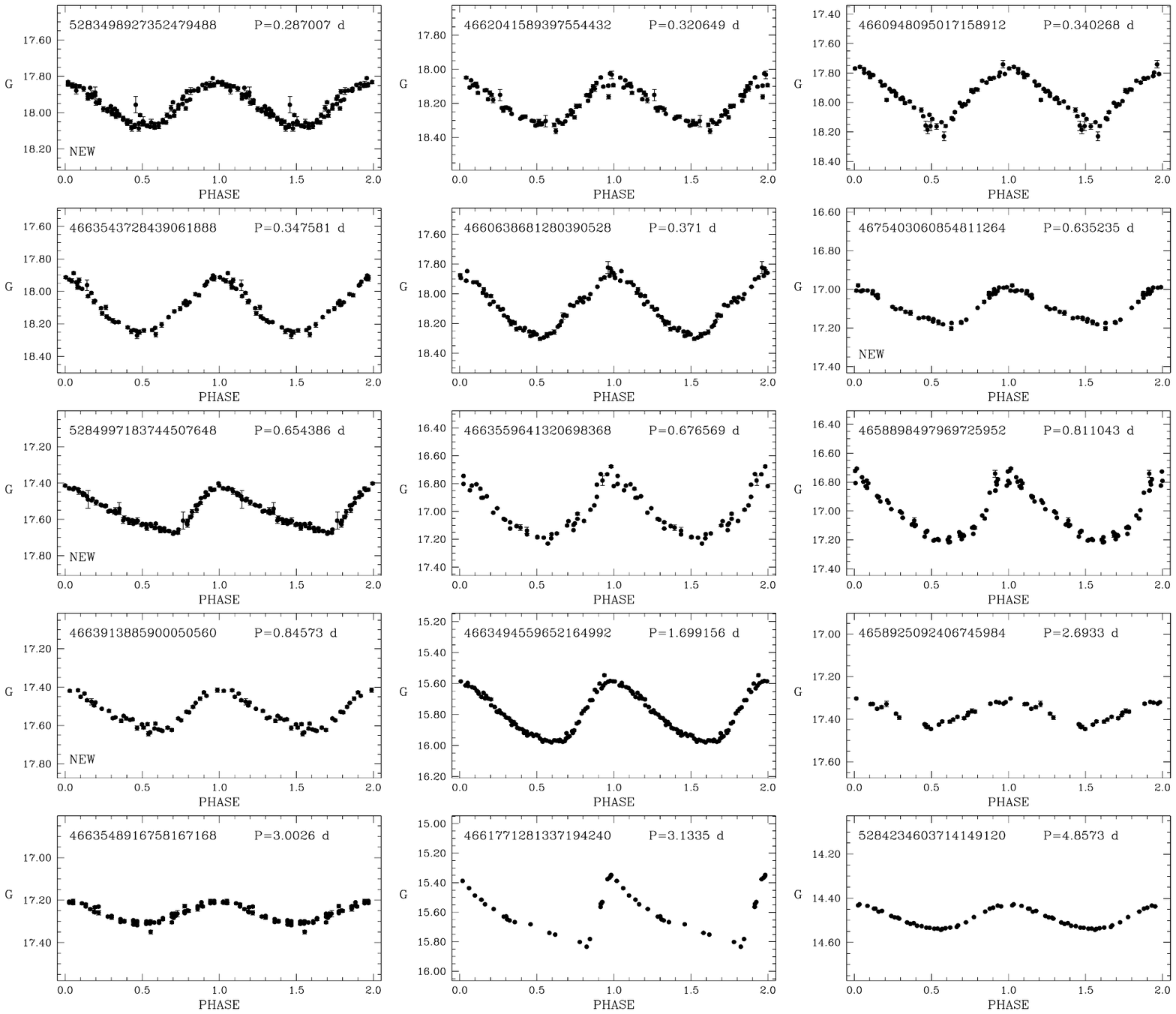}
   \caption{$G$-band light curves of DCEPs released in {\it Gaia} DR1 for which we did not identify the pulsation mode. Error bars are comparable or smaller than the symbol size. Measurements with errors larger than 0.05 mag are not displayed.    For each source we label the {\it Gaia} source ID and the pulsation period rounded to the last significant digit according to the error of the period  determination.  New discoveries by Gaia are flagged as such. 
   }
              \label{atlas-DcepTagged}%
    \end{figure*}
\newpage
 \begin{figure*}
   \centering
   \includegraphics[width=16.5 cm, clip]{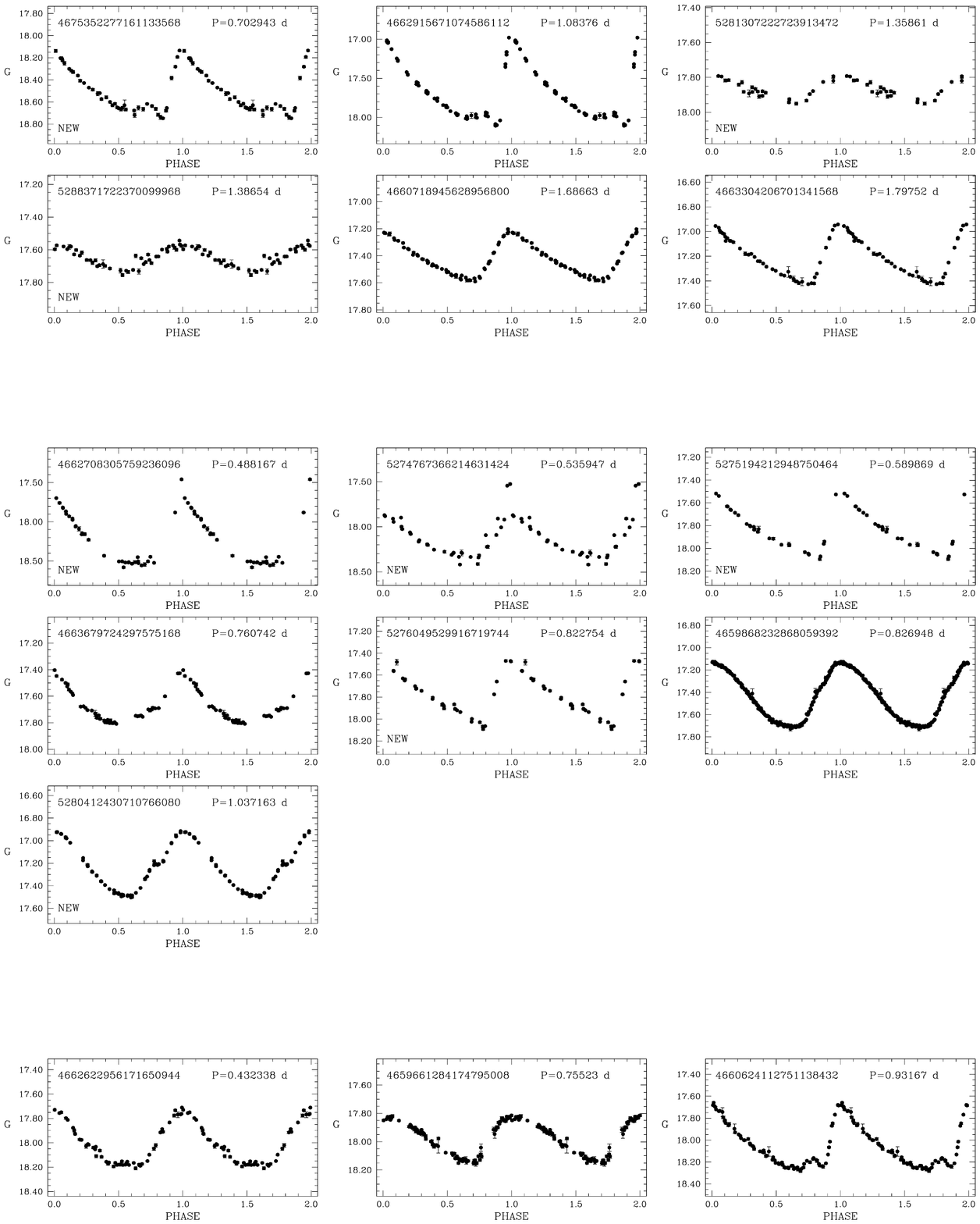}
   \caption{$G$-band light curves of ACEPs released in {\it Gaia} DR1 ordered by increasing period.   From top to bottom ACEPs F, ACEPs FO and ACEPs for whch we did not identify the 
   pulsation mode.  Error bars are comparable or smaller than the symbol size. Measurements with errors larger than 0.05 mag are not displayed.   For each source we label the {\it Gaia} source ID and the pulsation period rounded to the last significant digit according to the error of the period  determination. New discoveries by Gaia are flagged as such. 
   }
              \label{atlas-Acep}%
    \end{figure*}
    
 \newpage
 \begin{figure*}
   \centering
   \includegraphics[width=16.5 cm, clip]{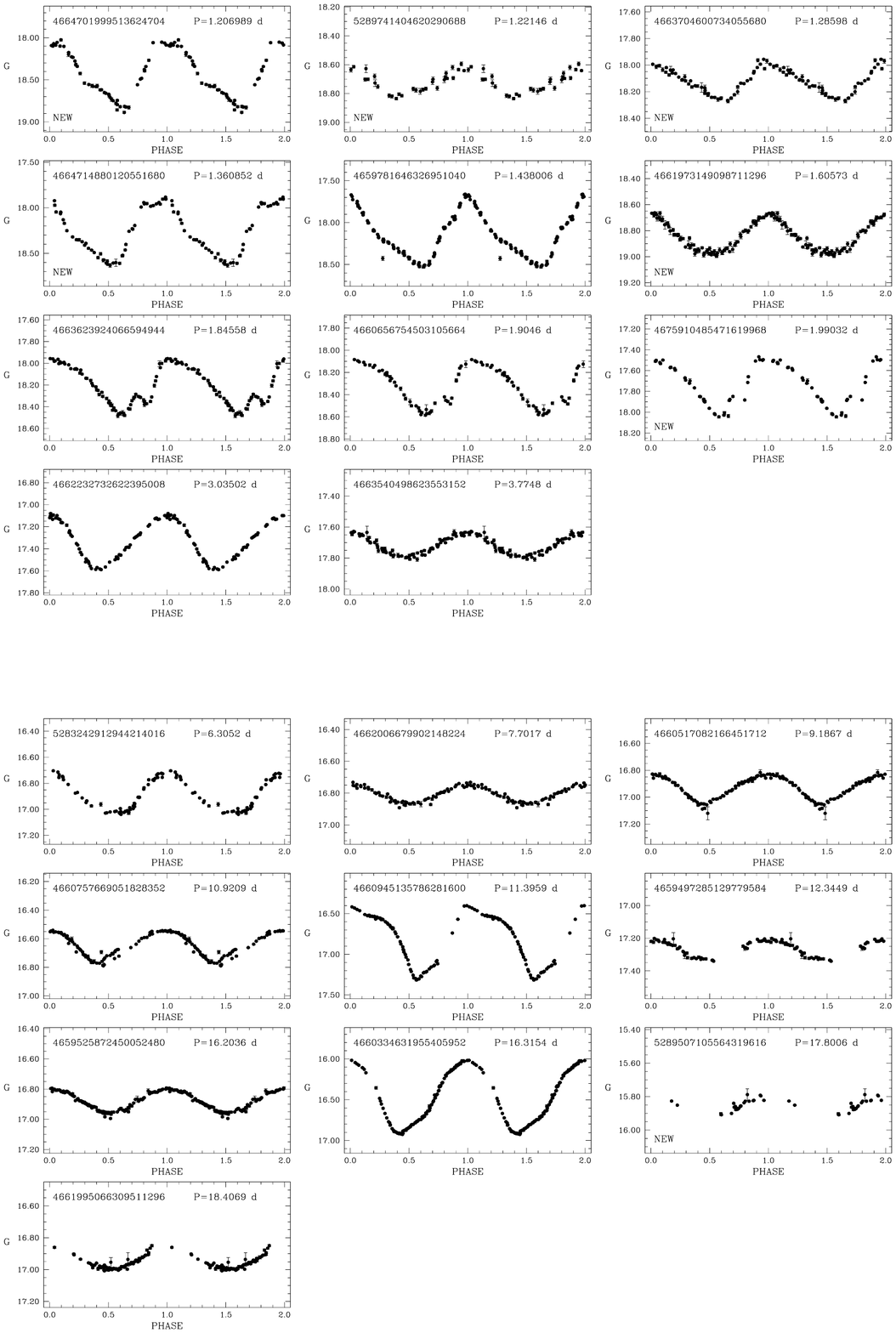}
   \caption{$G$-band light curves of T2CEPs released in {\it Gaia} DR1 ordered by increasing period.   Upper panels: T2CEPs BLHER; bottom panels: T2CEPs  WVIR.  Error bars are comparable or smaller than the symbol size. Measurements with errors larger than 0.05 mag are not displayed.  For each source we label the {\it Gaia} source ID and the pulsation period rounded to the last significant digit according to the error of the period  determination.  New discoveries by Gaia are flagged as such.
}
              \label{atlas-T2cep-p1}%
    \end{figure*}

 \newpage
 \begin{figure*}
   \centering
   \includegraphics[width=12.0 cm, trim=30 200 200 200, clip]{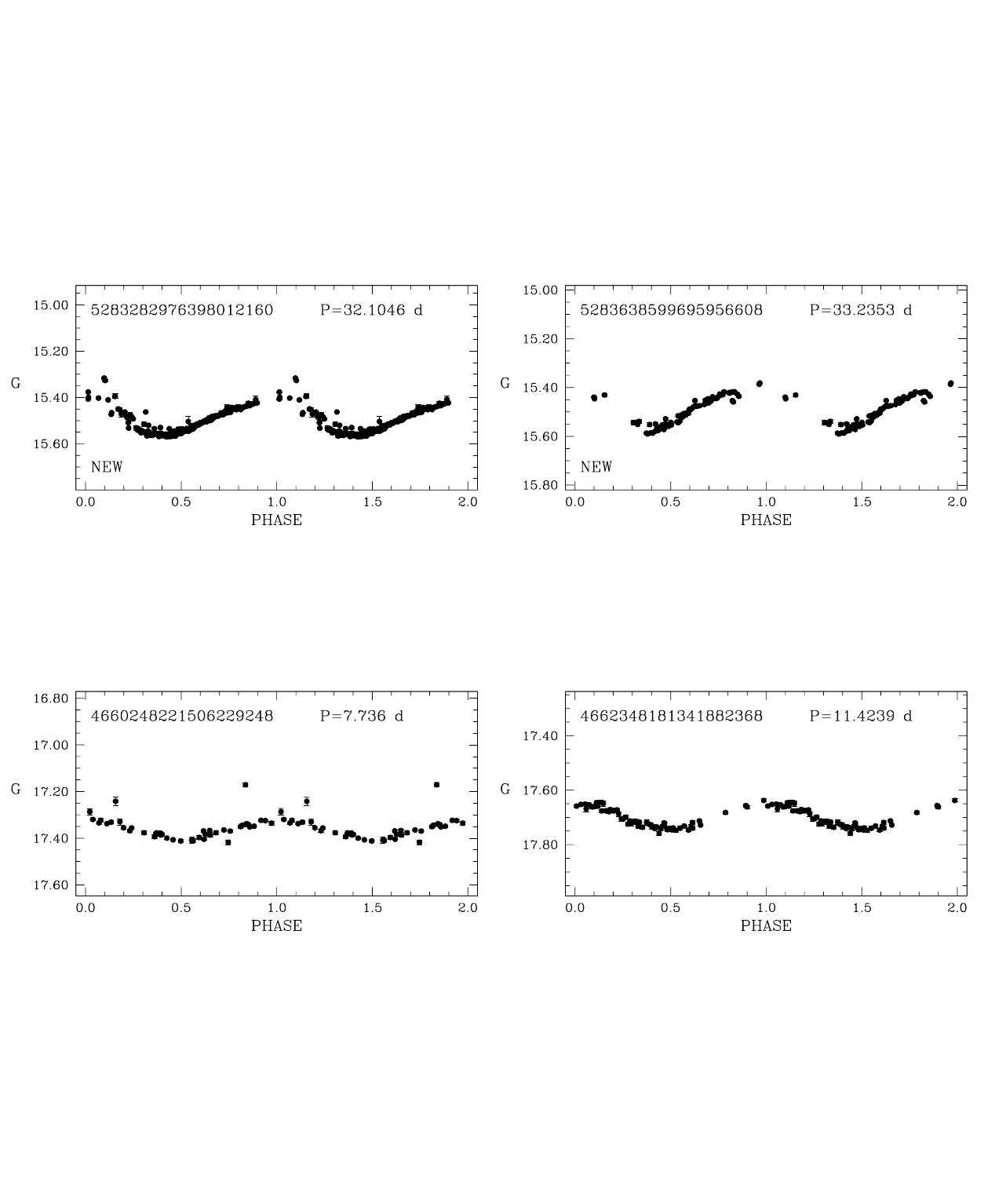}
   \caption{$G$-band light curves of T2CEPs released in {\it Gaia} DR1 ordered by increasing period.   Upper panels: T2CEPs RVTAU; bottom panels: T2CEPs  for which we did not indentify the subtype.  Error bars are comparable or smaller than the symbol size. Measurements with errors larger than 0.05 mag are not displayed.    For each source we label the {\it Gaia} source ID and the pulsation period rounded to the last significant digit according to the error of the period  determination. New discoveries by Gaia are flagged as such.
   }
              \label{atlas-T2cep-p2}%
    \end{figure*}
   
\section{Atlas of the $G$-band light curves for RR Lyrae stars}\label{atlas-rrl}
In this Section we present the $G$-band light curves of 2,595 RR Lyrae stars observed by {\it Gaia} in the SEP region. The light curves are folded according to the period and epoch of maximum light computed by the SOS Cep\&RRL pipeline and are plotted  grouping the RR Lyrae stars according to the pulsation mode, with 685 RRc stars first, followed by 1,910 RRab stars. In each group the sources are desplayed in order of increasing period.  
\newpage
 \begin{figure*}
   \centering
   \includegraphics[width=16.5 cm, clip]{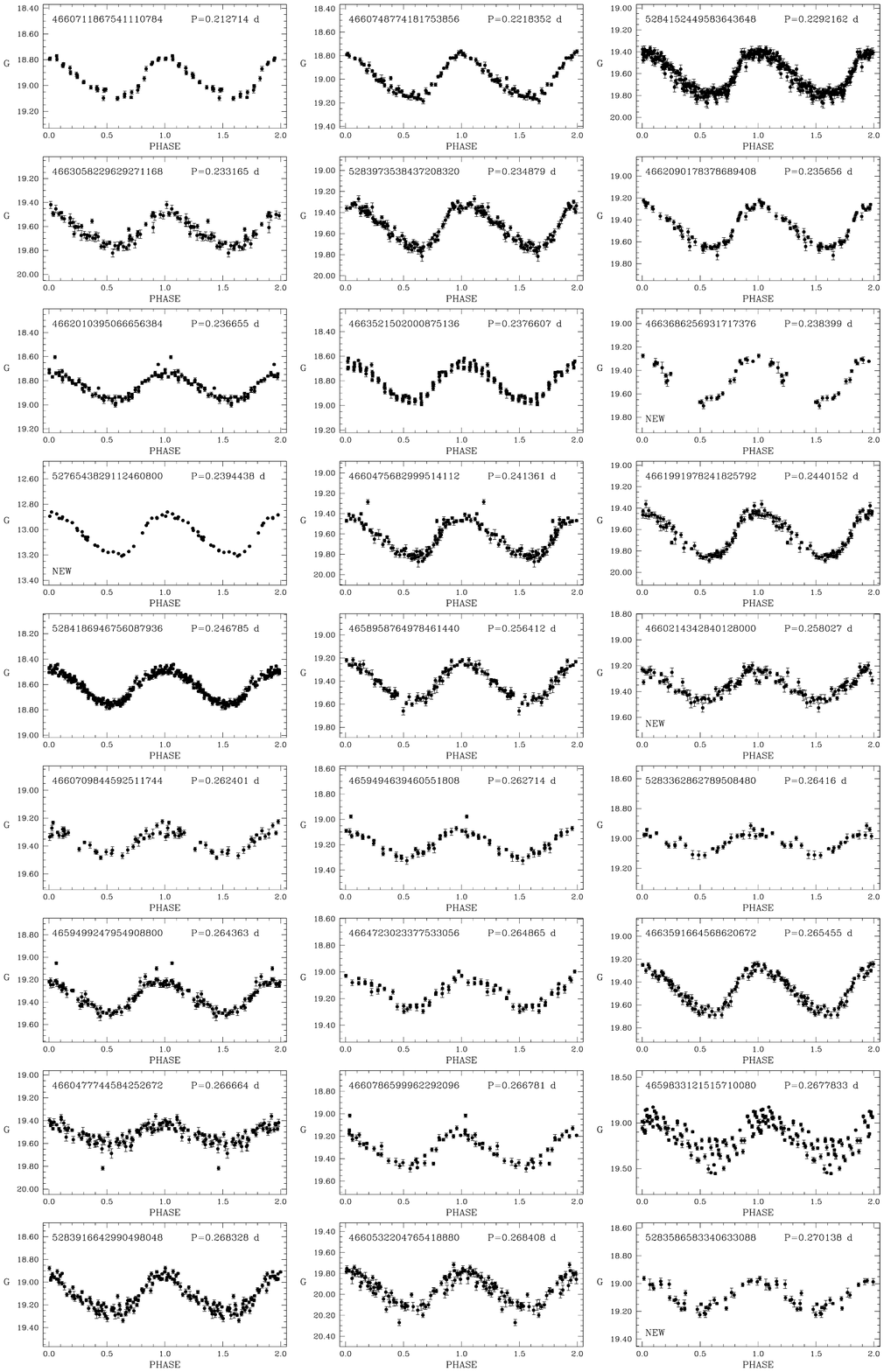}
   \caption{$G$-band light curves of RRc  stars released in {\it Gaia} DR1 ordered by increasing period. Measurements with errors larger than 0.05 mag are not displayed. For each source we label the {\it Gaia} source ID and the pulsation period rounded to the last significant digit according to the error of the period  determination.  New discoveries by Gaia are flagged as such.
   }
              \label{atlas-RRc}%
    \end{figure*}

\newpage
 \begin{figure*}
   \centering
   \includegraphics[width=16.5 cm, clip]{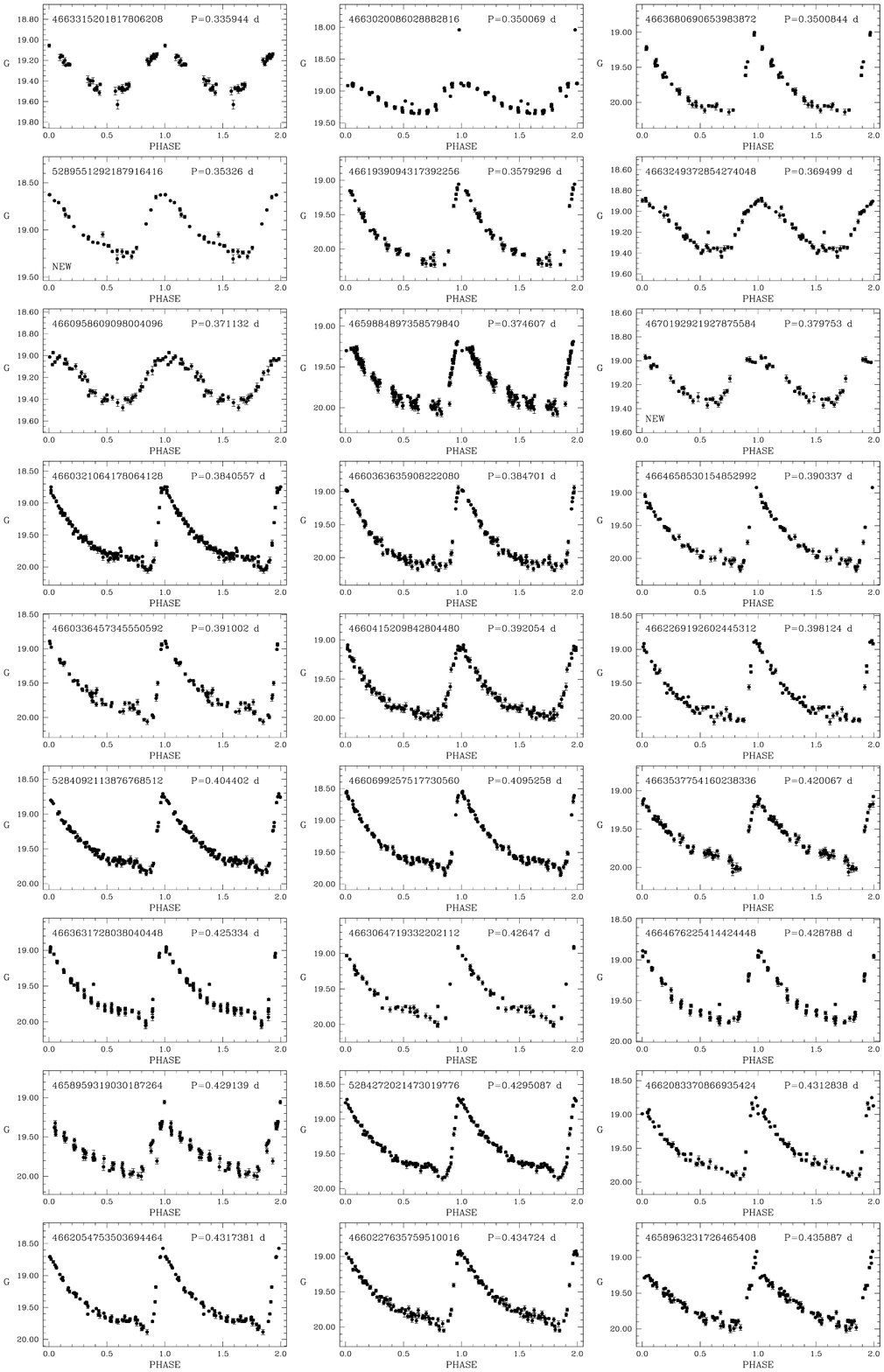}
   \caption{$G$-band light curves of RRab  stars released in {\it Gaia} DR1. Sources are ordered by increasing period. Measurements with errors larger than 0.05 mag are not displayed. For each source we label the {\it Gaia} source ID  and the pulsation period  rounded to the last significant digit according to the error of the  period determination. New discoveries by Gaia are flagged as such.
   }
              \label{atlas-RRab}%
    \end{figure*}

\end{appendix}

\end{document}